\documentclass[prb,aps,superscriptaddress,floatfix,tightenlines,showpacs,twocolumn]{revtex4-2}
\usepackage{graphicx}
\usepackage{dcolumn}   
\usepackage{bm}        
\usepackage{amssymb}   
\usepackage{amsmath}
\usepackage{url}
\usepackage{layout}
\usepackage{color}
\usepackage{epsfig}
\usepackage{graphicx}
\usepackage{mathrsfs}
\usepackage{bm}
\usepackage{appendix}
\usepackage{color}

\usepackage[thinlines]{easytable}
\usepackage{makecell}
\usepackage{tikz,pgf}
\usepackage{booktabs}
\usepackage{float}
\usepackage{hyperref}
\hypersetup{colorlinks,allcolors=blue}

\def\be{\begin{equation}}       \def\ee{\end{equation}}
\def\bea{\begin{eqnarray}}      \def\eea{\end{eqnarray}}
\def\bp{\begin{pmatrix}} \def\ep{\end{pmatrix}}
\def\beaa{\begin{equation}\begin{aligned}}
		\def\eeaa{\end{aligned}\end{equation}}

\begin{document}
\title{Superconductivity in bilayer La$_3$Ni$_2$O$_7$: A review focusing on the strong-coupling Hund's rule assisted pairing mechanism}

\author{Zhiming Pan}
\thanks{These two authors contributed equally to this work.}
\affiliation{Department of Physics, Xiamen University, Xiamen 361005, Fujian, China}
\author{Chen Lu}
\thanks{These two authors contributed equally to this work.}
\affiliation{School of Physics, Hangzhou Normal University, Hangzhou 311121, China}
\author{Fan Yang}
\email{yangfan\_blg@bit.edu.cn}
\affiliation{School of Physics, Beijing Institute of Technology, Beijing 100081, China}
\author{Congjun Wu}
\email{wucongjun@westlake.edu.cn}
\affiliation{New Cornerstone Science Laboratory, Department of Physics, School of Science, Westlake University, Hangzhou 310024, Zhejiang, China}
\affiliation{Institute for Theoretical Sciences, Westlake University, Hangzhou 310024, Zhejiang, China}
\affiliation{Key Laboratory for Quantum Materials of Zhejiang Province, School of Science, Westlake University, Hangzhou 310024, Zhejiang, China}
\affiliation{Institute of Natural Sciences, Westlake Institute for Advanced Study, Hangzhou 310024, Zhejiang, China}

\begin{abstract}
Discovery of high-$T_c$ superconductivity (SC) in the bilayer nickelate series La$_3$Ni$_2$O$_7$ have attracted substantial interest, providing a new platform for exploring unconventional SC.
Certain experimental evidence has pointed to a correlated electronic nature, which is the driving force responsible for its high critical temperature ($T_c$). 
This work reviews the SC in La$_3$Ni$_2$O$_7$, with a particular focus on theoretical understanding of its pairing mechanism driven by this strong-coupling, Hund-assisted scenario.
The electronic landscape is governed by two $E_g$-orbitals within the bilayer structure of NiO$_2$ planes.
The $3d_{z^2}$ orbital is nearly half-filled and exhibits a stronger localized character, while the $3d_{x^2-y^2}$ is approximately quarter-filled and remains highly itinerant.
The localized $3d_{z^2}$ orbitals experience robust interlayer hybridization, mediated by the $2p_z$ orbitals of the inner apical oxygen atoms.
This hybridization generates a strong interlayer antiferromagnetic (AFM) exchange. 
In the strong coupling regime, Hund's rule coupling aligns the spins of the two $E_g$ orbitals on the same nickel site.
The strong interlayer AFM exchange is effectively transferred to the itinerant $3d_{x^2-y^2}$ orbital, generating an effective coupling $J_{\perp}$ within this orbital. 
This mechanism is captured by a minimal strong-coupling bilayer $t$-$J$-$J_{\perp}$ model for the $3d_{x^2-y^2}$ band. 
Driven by $J_{\perp}$, $3d_{x^2-y^2}$ electrons can form interlayer Cooper pairs, leading to an extended $s$-wave pairing SC with high $T_c$. 
Meanwhile, the strongly localized $3d_{z^2}$ electrons tend to form interlayer rung singlets.
Due to a lack of phase coherence, these singlets do not directly participate in the SC condensate, but instead give rise to a pseudogap phase.
\end{abstract}
\maketitle

\section{Introduction}
The discovery of high-transition-temperature ($T_c$) superconductivity (SC) in cuprates marked a major milestone in condensed matter physics ~\cite{bednorz1986LBCO}.
This breakthrough has aroused considerable attentions, motivating decades of research aimed at enhancing $T_c$ and uncovering the underlying pairing mechanism responsible for this unconventional high-$T_c$ phenomena ~\cite{anderson1987resonating,tsuei2000pairing,damascelli2003angle,lee2006doping,armitage2010progress,taillefer2010scattering,keimer2015quantum,proust2019remarkable,zhang2022review}.
For a long time, however, cuprates remained the unique material family known to host SC with $T_c$ above the boiling point of liquid nitrogen.
This exclusivity posed a challenge to identify generic principles of unconventional high-$T_c$ SC:
is high-$T_c$ SC uniquely tied to the specific details of copper oxides, 
or is it a more universal consequence of strong electronic correlations?
Motivated by this, the search for nickelate analogs has been a primary focus in the community 
~\cite{zhang1994synthesis,anisimov1999electronic,lee2004infinite,pardo2011dft,nakata2017finite,li2019sc,botana2021nickelate,nomura2022sc,liu2023evidence}.
Nickelate series are compelling candidates because they share similar lattice structures and $3d$ electron configurations with cuprates. 
Researchers hoped that discovering SC in nickelate systems would bridge the gap between cuprates and other strongly correlated materials, ultimately providing a new perspective to unify the theories of unconventional high-$T_c$ SC.

This landscape has been transformed by the recent discovery of high-$T_c$ SC in the bilayer nickelate compound La$_3$Ni$_2$O$_7$ under high pressure ~\cite{sun2023lno,hou2023emergence,zhang2024high,wang2024pressure,wang2024bulk,ko2024sign}.
This series of material, with a high-$T_c$ reaching approximately $80$K under pressure, establishes a novel platform for exploring strongly correlated physics. 
While La$_3$Ni$_2$O$_7$ series shares some phenomenological similarities with cuprates, it also exhibits distinctly different electronic features.
ost notably, its low-energy physics involves two $E_g$ orbitals ($3d_{x^2-y^2}$ and $3d_{z^2}$) within a unique bilayer geometry,leading to a complex interplay between strong electronic correlations and magnetic exchange.
This break through has triggered a surge of experimental and theoretical efforts to characterize its electronic structure,magnetic correlations,and superconducting properties ~\cite{liu2024ele,yang2023arpes,zhang2023pressure,wang2023la2prnio7,wang2023structure,zhou2023evidence,cui2023strain,zhao2024pressure,abadi2024electronic,li2024ultrafast,li2024electronic,zhang2024doping,ren2025resolving,li2024dis,zhou2024revealing,su2024strongly,mijit2024local,chen2024unveiling,shi2025pre,huo2025low,liu2025evidence,li2025orbital,liu2025sc,li2025direct,zhong2025epitaxial,li2024pressure,zhou2025ambient,chen2024electronic,xie2024strong,wang2025electronic,shen2025nodeless,sun2025observation,bhatt2025resolving,cao2025direct,qiu2025interlayer,zhong2025evolution,li2025angle,zhang2025ident,hao2025sc,zhang2026spin,li2026enhanced,zhang2026strong,shu2026cont,li2026bulk,zhou2026sc,gao2026enhance,wang2026pauli,liu2026sc,gim2026spect,xu2026coex,he2026anisotropic,misawa2026polar,kumar2026non,zhan2026detecting,chen2026dissecting,han2026granular,li2026three,zhong2026doping,flavenot2026decoding,zhang2026interlayer,kriener2026control,luo2023bilayer,zhang2023electronic,yang2023possible,lechermann2023,sakakibara2024possible,gu2023effective,shen2023effective,christiansson2023correlated,shilenko2023correlated,wu2024superexchange,cao2023flat,chen2023critical,liu2023spm,lu2024interlayer,zhang2023structural,oh2023type2,liao2023electron,qu2024bilayer,yang2023interlayer,jiang2023high,zhang2023trends,huang2023impurity,qin2023high,tian2023correlation,lu2023sc,kitamine2023,luo2023high,zhang2023strong,pan2023rno,sakakibara2023La4Ni3O10,lange2023mixedtj,geisler2023structural,yang2023strong,rhodes2023structural,lange2023feshbach,labollita2024electronic,kumar2023softening,kaneko2023pair,ryee2023critical,zhang2023la3ni2o6,grusdt2023lno03349,chen2023iPEPS,liu2023dxy,ouyang2023hund,qu2023roles,sui2023rno,zheng2023twoorbital,kakoi2023pair,heier2023competing,jiang2024pressure,lu2024interplay,geisler2024fermi,talantsev2024debye,fan2024sc,ouyang2024absence,chen2024nonfermi,yi2024nature,yi2025unifying,xu2025com,bhatta2025structural,shi2025effect,yue2025correlated,wang2025mottness,shao2025band,liu2024origin,yang2025evolution,zhu2025quantum,cao2025strain,wang2025fermi,dong2025interstitial,wu2025ultrafast,ji2025signatures,ouyang2025phase,zhu2025magnetic,lu2025impact,ji2025strong,chen2025variation,haque2025dft,wang2025self,huo2025first,liu2025optimally,chen2025sc,chen2026electronic,sharma2026struct,geisler2026elect,shao2026possible,li2026machine,oh2026incom,lange2026simulating,korolev2026nearly,korolev2026threefold,liu2026triplon,qiu2026progress,watanabe2026hier,chen2026unified,wu2026sc,wu2026pressure,zhao2026pseudogap,wang2026jahn,cao2026tunable,wei2026perp,stepanov2026co,wang2026orbital,bejas2026raman,wu2026pair,zhang2026sc}.
Unraveling the pairing symmetry and underlying mechanism in La$_3$Ni$_2$O$_7$ is now a crucial task. 
Studying this system could enrich the understanding of the essential physics of strongly correlated SC, offering a critical step toward a unified framework for high-$T_c$ theories.

The emergence of SC in bulk La$_3$Ni$_2$O$_7$ samples is dictated by a delicate interplay between applied pressure and structural symmetry.
At ambient pressure (AP), the material crystallizes in an orthorhombic Amam structure ~\cite{sun2023lno}.
In this phase, high-quality stoichiometric single crystals exhibit metallic behavior at low temperatures,
though oxygen deficiency can drive the system into a weakly insulating state and explains why well-fabricated samples are necessary.
As external pressure is applied, La$_3$Ni$_2$O$_7$ could undergo a sequence of structural phase transitions ~\cite{sun2023lno,li2024pressure}.
The first transition occurs around $14$ GPa, shifting the lattice structure from the $Amam$ to an orthorhombic $Fmmm$ phase.
This transformation is considered to be physically crucial,
since it straightens the inner $c$-axis Ni-O-Ni bond angle within the bilayer,
increasing it from approximately $168^{\circ}$ towards $180^{\circ}$.
This structural straightening strongly enhances the electronic and magnetic coupling between the two NiO$_2$ layers in the bilayer unit.  
Upon further increasing the pressure, the material eventually transitions into a tetragonal $I4/mmm$ phase. 
SC is observed to emerge concurrently with the onset of the $Fmmm$ phase.
The critical temperature, $T_c$, reaches a maximum and subsequently declines as pressure continues to increase ~\cite{sun2023lno,li2024pressure}.

Bridging the gap between high-pressure physics and practical study, further experiments have successfully stabilized SC with a $T_c\approx 40$ K in La$_3$Ni$_2$O$_7$ thin-film series at ambient pressure ~\cite{ko2024sign,zhou2025ambient}. 
This breakthrough bypasses the need for complex high-pressure equipment by utilizing epitaxial strain engineering. 
The realization of this ambient-pressure superconducting state depends critically on the film's growth conditions, particularly the choice of substrate.
By selecting an appropriate substrate, an in-plane biaxial compressive strain is imposed on the La$_3$Ni$_2$O$_7$ film. 
Similar to the effect of hydrostatic pressure, this substrate-induced strain forces the inner Ni-O-Ni bond angle to straighten.
This geometric adjustment stabilizes the strongly coupled bilayer structure, thereby inducing the superconducting phase even without external pressure ~\cite{bhatt2025resolving,bhatta2025structural}. 
Ultimately, these thin-film results confirm that, a straightened Ni-O-Ni bond and the resulting robust interlayer coupling within the bilayer is the structural requisite for driving SC in this system.

While compelling experimental evidence confirms high-$T_c$ SC in La$_3$Ni$_2$O$_7$, its microscopic mechanism remains a subject of intense debate. 
The complexity of this material series stems from its unique two-orbital and bilayer electronic structure, 
which is widely believed to be the key to stabilizing the high-$T_c$ phase differing from cuprates. 
To build a complete physics picture, several fundamental questions must be addressed.
First is the role of the $\gamma$-pocket, a small electron-like pocket most relevant to the nearly half-filled $3d_{z^2}$-orbital, which is below the Fermi surface in the ambient pressure and predicted to cross the Fermi surface by the band structure calculation.
Determining whether this pocket survives in the superconducting phase and what critical role it might play is central to understanding the pairing mechanism.
Second, it is crucial to identify the dominant orbital character driving the SC: is it the itinerant $3d_{x^2-y^2}$ orbital, reminiscent of the cuprates, or the more localized $3d_{z^2}$ orbital, which facilitates the strong interlayer coupling?
Finally, establishing the SC pairing symmetry is essential.
Current proposals remain divided, ranging from a cuprate-like intralayer d-wave state ~\cite{cao2025direct} to an interlayer s-wave state driven by the unique bilayer geometry ~\cite{shen2025nodeless}.

One prominent theoretical framework addressing the SC and these issues in bilayer La$_3$Ni$_2$O$_7$ is the Hund's-assisted interlayer pairing mechanism ~\cite{lu2024interlayer,lu2024interplay}.
This theory suggests that strong Hund's rule coupling within the two-orbital bilayer system is a crucial ingredient.
It effectively links the localized magnetic interactions of the $3d_{z^2}$ orbital with the itinerant electrons of the $3d_{x^2-y^2}$ orbital.
This interplay facilitates a strong interlayer $s$-wave pairing state primarily hosted within the $3d_{x^2-y^2}$ band. 
Ultimately, this strong-coupling scenario offers an explanation for how the complex orbital physics and structural features of La$_3$Ni$_2$O$_7$ intertwined electronic correlation and Hund's rule to give rise to high-temperature SC.

This review is structured to provide theoretical Hund's-assisted pairing framework for SC in bilayer La$_3$Ni$_2$O$_7$.
Sect.~\ref{sec:elect} begins by summarizing the key experimental evidence for the material's strongly correlated nature.
In Sect.~\ref{sec:bilayer}, a bilayer two-orbital Hubbard model is introduced to explain its electronic behavior and detail how Hund's coupling assists in inter-orbital spin correlation.
Sect.~\ref{sec:strong} then presents a simplified $t$-$J$-$J_\perp$ model, 
which focus on the $3d_{x^2-y^2}$ orbital forms the basis for the proposed interlayer pairing mechanism.
Following this, Sect.~\ref{sec:extended} explores the interplay between the two orbitals clarifies the specific role of the $3d_{z^2}$ orbital. 
Finally, Sect.~\ref{sec:comp} compares the work with other theoretical frameworks, and Sect.~\ref{sec:disc} provides a summary and concluding discussion.

\section{Electronic properties and Experiments}
\label{sec:elect}

\begin{figure}[t!]
\centering
\includegraphics[width=0.95\linewidth]{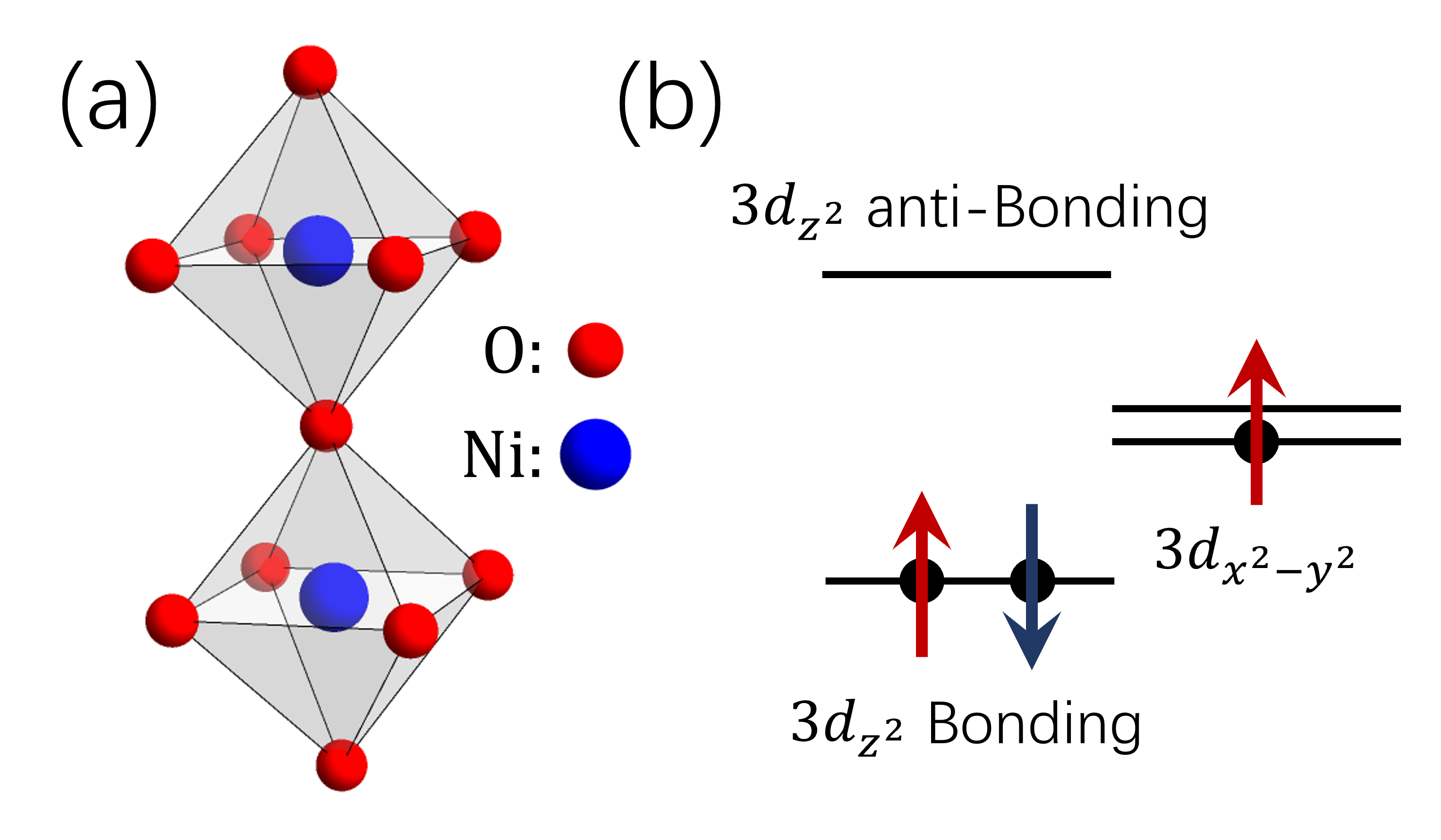}
\caption{(a). Schematic diagram of the bilayer NiO$_6$ octahedral structure, illustrating the local coordination environment of the nickel ions.
(b). Electronic configuration for the three $E_g$ electrons within the Ni$_2$O$_{11}$ bilayer unit.
The strong interlayer coupling of the $3d_{z^2}$ orbitals splits them into bonding and anti-bonding states.
The energetically favored bonding state is fully occupied by two electrons with opposite spins, leaving the remaining electron to occupy the in-plane $3d_{x^2-y^2}$ orbitals.}
\label{fig:NiOStruct}
\end{figure}

The electronic properties of La$_3$Ni$_2$O$_7$ series are primarily governed by the nickel centers, 
which possess a nominal valence of Ni$^{2.5+}$, 
corresponding to an averaged $3d^{7.5}$ electron configuration. 
As depicted in Fig.~\ref{fig:NiOStruct}(a), each nickel ion is octahedrally coordinated by six oxygen O$^{2-}$ ligands, 
forming a distorted NiO$_6$ unit ~\cite{sun2023lno}. 
This local crystalline environment lifts the degeneracy among the Ni $3d$ orbitals.
They split into a lower-energy, fully occupied $T_{2g}$ triplet and a higher-energy $E_g$ doublet, while the latter is responsible for the underlying electronic nature of the material.

The two $E_g$ orbitals exhibit distinct spatial orientations, leading to significantly different hopping and hybridization characteristics. 
The $3d_{x^2-y^2}$ orbitals are oriented within the NiO$_2$ planes, 
which promotes strong hybridization with the adjacent in-plane oxygen $2p_{x,y}$ orbitals, sharing similarity with cuprates.
In contrast, the $3d_{z^2}$ orbitals point perpendicular to the planes, aligning along the crystallographic $c$-axis.
Within the unique bilayer structure, the $3d_{z^2}$ orbitals of the top and bottom Ni sites in a rung can strongly interact. 
This hopping is mediated by the $2p_z$ orbital of the shared apical oxygen between the two NiO$_2$ layers. 
This robust interlayer coupling splits the $3d_{z^2}$ states into bonding and anti-bonding combinations of molecular orbitals, with the bonding state being significantly lower in energy.

Given the nominal Ni$^{2.5+}$ valence, the two Ni ions within a bilayer rung share a total of three $E-g$ electrons.
This configuration is distributed across the $E_g$ orbitals as follows:
These three electrons distribute themselves to minimize energy, as illustrated in Fig.~\ref{fig:NiOStruct}(b). 
Two electrons with opposite spins completely fill the lower-energy $3d_{z^2}$ bonding state.
The single remaining electron is shared between the two $3d_{x^2-y^2}$ orbitals.
This distribution yields an average of $\sim 0.5$ electrons per $3d_{x^2-y^2}$ orbital on each Ni Site, resulting in a highly itinerant, approximately quarter-filled band.

\begin{figure}[t!]
\centering
\includegraphics[width=0.4\linewidth]{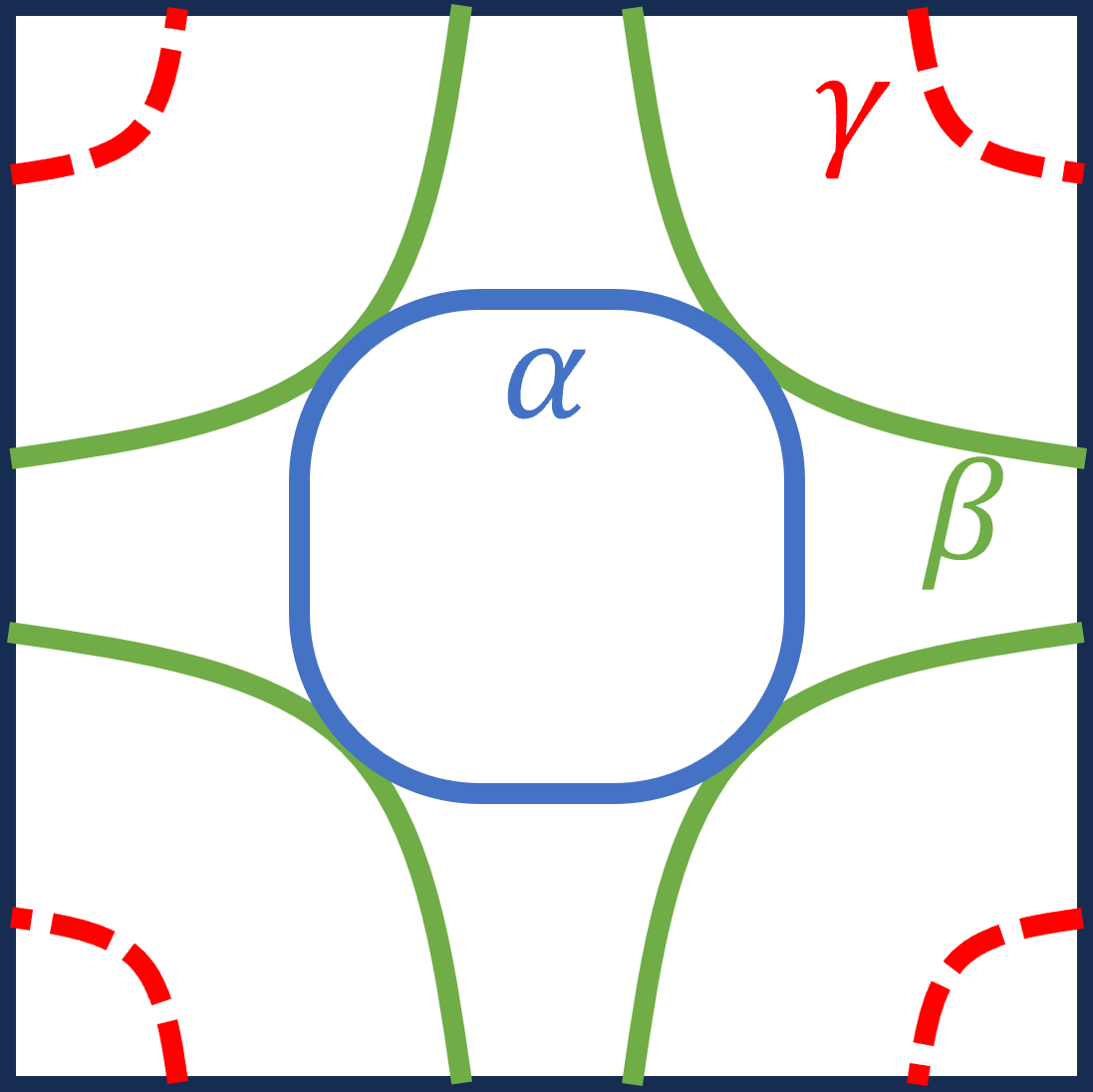}
\caption{Schematic diagram of the Fermi surface for bilayer La$_3$Ni$_2$O$_7$.
Under ambient pressure, the Fermi surface consists solely of the $\alpha$- and $\beta$-pockets.
The $\gamma$-pocket, indicated by dashed lines, is predicted to cross the Fermi surface (metallize) when the material is subjected to high pressure.}
\label{fig:BareFermiSurface}
\end{figure}

The Fermi surface topology of bilayer La$_3$Ni$_2$O$_7$ exhibits both parallels and distinct differences when compared to cuprates, as illustrated in Fig.~\ref{fig:BareFermiSurface}.
At ambient pressure, the Fermi surface consists solely of an electron-like $\alpha$-pocket and a hole-like $\beta$-pocket.
Both of these are derived primarily from the in-plane $3d_{x^2-y^2}$ orbitals.
The $\beta$-pocket is driven by intralayer hopping and closely resembles the characteristic Fermi surface features found in cuprate superconductors.
However, a major distinguishing feature of the nickelate system is the bonding $\gamma$-band, which originates from the out-of-plane $3d_{z^2}$ orbitals.
Angle-resolved photoemission spectroscopy (ARPES) measurements on bulk crystals confirm that at ambient pressure, 
this $\gamma$-band lies completely below the Fermi energy ($E_F$) and exhibits strong band renormalization ~\cite{luo2023bilayer,yang2023arpes}.
Under high pressure, density functional theory (DFT) predicts a metallization of the $\gamma$-band, causing it to cross the Fermi level ~\cite{luo2023bilayer}.
However, direct experimental verification of this high-pressure Fermi surface remains elusive due to the severe technical challenges of performing ARPES under high pressure.

Interestingly, ARPES investigations on superconducting La$_3$Ni$_2$O$_7$ thin-film samples have revealed a complex and controversial electronic landscape ~\cite{li2025angle,yue2025correlated,shen2025nodeless,wang2025electronic,sun2025observation}.
Unlike bulk samples, these epitaxially grown films can exhibit SC at ambient pressure with a $T_c$ of approximately $40$ K.
Due to the controlled condition, directly determining their electronic structure is experimentally reachable, which is pivotal for understanding the underlying pairing mechanism.
A central point of debate in the literature concerns the precise position of the $\gamma$-band relative to the Fermi level ($E_F$).
Some experimental studies report that the $\gamma$-band crosses $E_F$, becoming metallic and explicitly contributing a pocket to the Fermi surface ~\cite{li2025angle,yue2025correlated,shen2025nodeless}.
This observation aligns with high-pressure DFT predictions for bulk material, implying that substrate-induced strain effect acts as an effective pressure that  metallizes the $3d_{z^2}$ band and drive the emergence of SC.
In sharp contrast, other ARPES reports contend that the $\gamma$-band remains fully occupied and located entirely below $E_g$, much like the ambient-pressure  bulk electronic structure ~\cite{wang2025electronic,sun2025observation}. 
Although these latter studies do observe that the band shifts closer to the Fermi level compared to bulk samples, the strict absence of a Fermi crossing suggests that the $\gamma$ electrons may not directly participate as itinerant carriers in the superconducting condensate.
Resolving this ongoing experimental discrepancy is absolutely essential for clarifying whether the $\gamma$-band plays an active or passive role in the SC of bilayer nickelates.

Experimental investigations consistently underscore the crucial role of electron correlation effects in understanding the SC of bilayer La$_3$Ni$_2$O$_7$.
Specifically, ARPES measurements reveal a substantial renormalization of the electronic band structure compared to DFT predictions ~\cite{yang2023arpes,li2025angle,yue2025correlated,shen2025nodeless,wang2025electronic,sun2025observation}.
This strong renormalization is particularly evident in the $\gamma$-band, regardless of its proximity to the Fermi surface.
Such pronounced band narrowing is a key signature of strong electronic correlations, suggesting a highly localized character for the $3d_{z^2}$ orbital. 
Further supporting this correlated picture, optical conductivity measurements demonstrate a marked reduction in the kinetic energy of charge carries upon cooling to low temperatures ~\cite{liu2024ele}.

Furthermore, further studies effectively rule out the conventional electron-phonon-coupling Bardeen-Cooper-Schrieffer (BCS) framework as the dominant mechanism ~\cite{ouyang2024absence,talantsev2024debye,zhu2025magnetic}. 
While phonon dispersion experiments have determined the system's characteristic Debye energy, theoretical calculations based solely on electron-phonon coupling yield critical temperatures far below the high $T_c$ observed in La$_3$Ni$_2$O$_7$. 
Taken together, these findings strongly suggest that conventional electron-phonon-mediated pairing alone is insufficient to explain the observed SC.
Instead, they strongly point toward an unconventional pairing mechanism driven primarily by strong electronic correlations.

\begin{figure}[t!]
\centering
\includegraphics[width=0.95\linewidth]{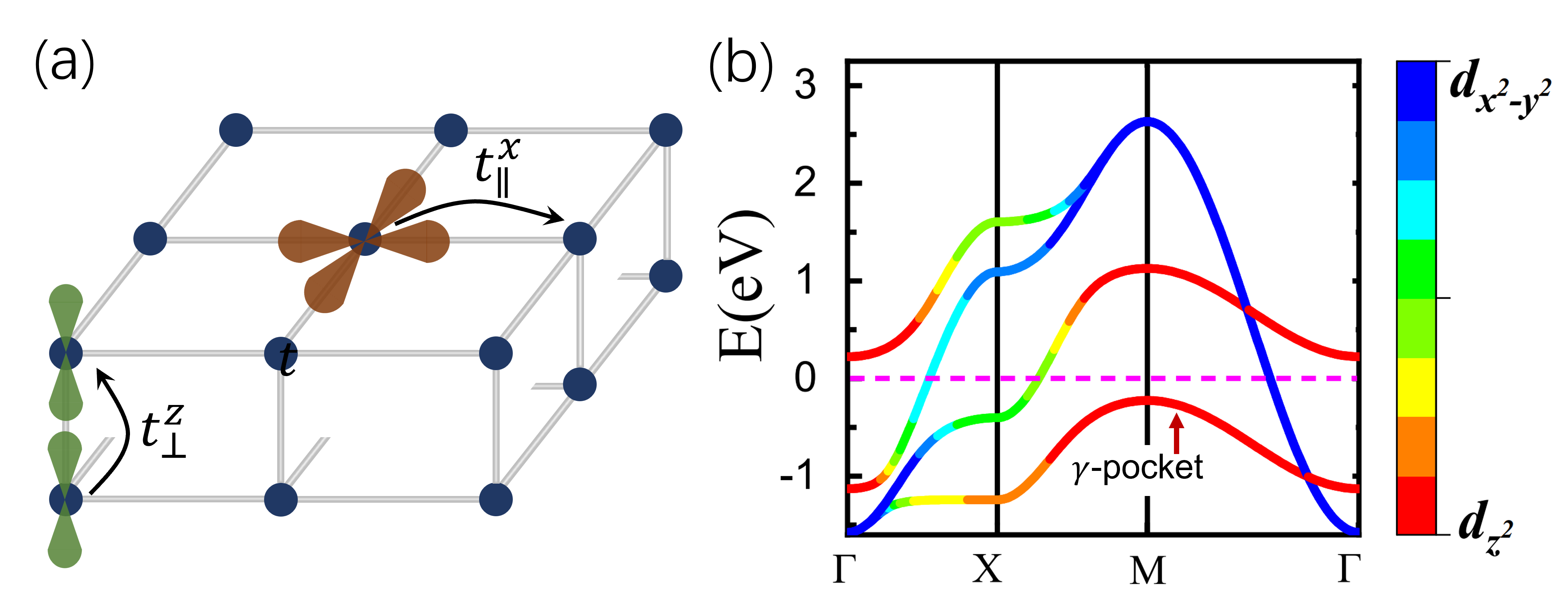}
\caption{(a). Schematic diagram of the bilayer lattice structure. The key hopping parameters for the two relevant $E_g$ orbitals are highlighter: 
the in-plane $3d_{x^2-y^2}$-orbital is dominant by intralayer hopping $t_{\parallel}^{x}$, 
whereas the out-of-plane $3d_{z^2}$ orbital governs the principal interlayer hopping $t_{\perp}$.
(b). Bare band structure derived from the two-orbital model at ambient pressure. 
The Fermi level ($E_F$) is marked by the horizontal dashed line.
Note that the $\gamma$-band, which originates primarily from the $3d_{z^2}$ orbital, lies entirely below $E_F$.
This indicates that it is fully occupied and does not form a Fermi surface pocket at ambient pressure.}
\label{fig:BilayerLatticeBands}
\end{figure}

\section{Bilayer correlated model}
\label{sec:bilayer}

The electronic structure and superconducting properties of La$_3$Ni$_2$O$_7$ are governed by the following three essential ingredients: 
the unique bilayer NiO$_2$ lattice geometry, 
the active $E_g$ orbital degree of freedom (specifically, the $3d_{z^2}$ and $3d_{x^2-y^2}$ orbitals), 
and the strong electron correlations. 
To distill the essential physics from this complex system, the material's electronic properties are typically captured by a two-orbital Hubbard-Kanamori model.
This total Hamiltonian is given by,
\begin{align*}
{H} ={H}_0 +{H}_{\text{int}}.
\end{align*}
Here, ${H}_0$ represents the non-interacting tight-binding term describes the charge dynamics of the two $E_g$ orbitals.
Meanwhile, ${H}_{\text{int}}$ captures the local electron correlation effects,
explicitly incorporating the standard Slater-Kanamori terms for on-site Coulomb  repulsion and Hund's rule coupling.

The tight-binding Hamiltonian ${H}_0$ describes the kinetic energy of electrons as they hop between different sites and orbitals.
This term is central to generating the bare lattice band structure shown in Fig.~\ref{fig:BilayerLatticeBands}(a) and (b).
Explicitly, the relevant term of the bare kinetic Hamiltonian is formulated as:
\begin{equation}
\begin{aligned}
H_0&=-t^{x}_{\parallel} \sum_{\langle i,j\rangle \alpha\sigma} \big(d_{x\alpha\sigma}^{\dagger} (i) d_{x\alpha\sigma} (j) +\text{h.c.}\big)     \\
&+\Delta_g \sum_{i\alpha\sigma} d_{x\alpha\sigma}^{\dagger} 
(i)
d_{x\alpha\sigma} (i)
\\ 
&-t^{z}_{\perp} \sum_{i\sigma} \big(d_{z1\sigma}^{\dagger} (i) d_{z2\sigma} (i)
+\text{h.c.}\big)   \\
&-t^{z}_{\parallel} \sum_{\langle i,j\rangle\alpha\sigma} \big(d_{z\alpha\sigma}^{\dagger} (i)
d_{z\alpha\sigma} (j)
+\text{h.c.}\big)   \\
&-\sum_{\langle i,j\rangle \alpha\sigma} t_{\parallel,j-i}^{xz} \big(d_{x\alpha\sigma}^{\dagger} (i)
d_{z\alpha\sigma} (j) +\text{h.c.}\big).
\end{aligned}
\label{eq:ham0}
\end{equation}
In this expression, $d_{a \alpha\sigma}^{\dagger}(i)$ is the creation operator for an electron with spin $\sigma$ in orbital $a$ at site $i$ in layer $\alpha$.
Here, the orbital index $a=x$ corresponds to the in-plane $3d_{x^2-y^2}$ orbital,
while $a=z$ denotes the out-of-plane $3d_{z^2}$ orbital.

Physically, the band topology of the kinetic ${H}_0$ is characterized by several essential hopping parameters, generating the bare band structure shown in Fig.~\ref{fig:BilayerLatticeBands} (b).
First, the intralayer hopping amplitudes $t_{\parallel}^x$ and $t_{\parallel}^z$ describe the itinerant motion of the $3d_{x^2-y^2}$ and $3d_{z^2}$ electrons within the Ni$O_2$ planes, respectively.
The presence of a dominant $t_{\parallel}^x$ generates a a highly dispersive $3d_{x^2-y^2}$ band that closely resembles the electronic structure of cuprates.
Conversely, $t_{\parallel}^z$ is typically small due to the highly localized spatial distribution of $3d_{z^2}$ orbital within the plane, 
though it remains necessary for establishing the slight in-plane dispersion of this band.
Second, the vertical hopping $t_{\perp}^z$ captures the crucial interlayer coupling of the $3d_{z^2}$ orbital along the rung, 
which is strongly mediated by the $2p_z$ orbital of the shared inner apical oxygen between the two layers.
Third, the crystal-field splitting term $\Delta_g$ lowers the on-site energy of the $3d_{z^2}$ orbital relative to the $3d_{x^2-y^2}$ orbital, ensuring the $3d_{z^2}$ states are preferentially occupied.
Finally, the inter-orbital hopping $t_{\parallel,j-i}^{xz}$ introduces essential in-plane hybridization between the two active $E_g$ orbitals.
Due to the distinct orbital symmetries of the two orbitals, this hybridization exhibits a strict sign alternation in momentum space (i.e., $t^{xz}_{\parallel,x}=-t^{xz}_{\parallel,y}\equiv t^{xz}_{\parallel}$).

The specific physical values of these hopping parameters could be extracted from DFT calculations, e.g., in the Ref.~\cite{luo2023bilayer}. 
DFT calculations ~\cite{luo2023bilayer} reveal typical values of these parameters.
Typical intralayer hopping amplitudes are $t_{\parallel}^{z}=0.110$ eV, $t_\parallel^{xz}=0.239$ eV, and $t_\parallel^{x}=0.483$ eV. 
The interlayer hopping for the $3d_{z^2}$ orbital is $t^{z}_\perp=0.635$ eV, whereas the interlayer hopping involving the $3d_{x^2-y^2}$ orbital is negligible and effectively vanishes.
The onsite energies are typically evaluated as $\epsilon_z = 0.409$ eV and $\epsilon_x = 0.776$ eV, establishing the crystal-field splitting $\Delta_g=\epsilon_x-\epsilon_z>0$.
This tight-binding parameterization accurately reproduces the bare band structure shown in Fig.~\ref{fig:BilayerLatticeBands} (b).
It successfully captures the system's Fermi surface topology, including the metallic electron pockets and the fully occupied $\gamma$-band that lies entirely below the Fermi level at ambient pressure.

Beyond the kinetic terms, the strongly correlated physics is captured by the multi-orbital Hubbard interaction Hamiltonian, which reads:
\begin{equation}
\begin{aligned}
&{H}_{\text{int}}= U\sum_{ia\alpha} n_{a\alpha\uparrow}(i)
n_{a\alpha\downarrow}(i) 
+V \sum_{i\alpha} n_{z\alpha}(i) n_{x\alpha}(i)
\\
&+P \sum_{i\alpha} \left( d^\dagger_{x\alpha\uparrow}(i) d^\dagger_{x\alpha \downarrow}(i)
d_{z\alpha \downarrow}(i) d_{z\alpha 
\uparrow}(i) +\text{h.c}\right) \\
&-J_{H} \sum_{i\alpha} \bm{S}_{z\alpha}(i) \cdot \bm{S}_{x\alpha}(i),
\end{aligned}
\label{eq:ham1UVJh}
\end{equation}
Here, $n_{a\alpha\sigma}(i)=d_{a\alpha\sigma}^{\dagger}(i)d_{a\alpha\sigma}(i)$ and $\bm{S}_{a\alpha}(i)=\frac{1}{2}d_{a\alpha}^{\dagger}(i)[\bm{\sigma}]d_{a\alpha}(i)$ represents the standard particle number and spin operators, respectively, for an electron in orbital $a$ at site $i$ within layer $\alpha$.
The interaction terms account for the intra-orbital Coulomb repulsion $U$, the
inter-orbital repulsion $V$, the pair-hopping amplitude $P$, and the ferromagnetic Hund's rule coupling $J_H$.
To preserve the rotational symmetry of the atomic orbitals in real space, these parameters are typically constrained by the standard Kanamori relations: $V=U-J_H$ and $P=J_H$.

Physically, in La$_3$Ni$_2$O$_7$ series, the on-site Coulomb repulsion is estimated to be approximately $U=5$ eV.
This value places La$_3$Ni$_2$O$_7$ in a moderately correlated regime, situating it between the weaker correlations of iron-based superconductors and the very strong correlations of cuprates.
Additionally, the Hund's coupling $J_H$ typically ranges from $\frac{U}{8}$ to $\frac{U}{4}$.
Since the electronic configuration is switched between $3d^7$ and $3d^8$, $J_H$ establishes as crucially important energy scale that aligns the spins of the two $E_g$ orbitals and influences the low-energy magnetic physics.

\subsection{Hierarchy of interactions}
The low-energy physics characterized by both $3d_{z^2}$ and $3d_{x^2-y^2}$ orbitals is fundamentally restricted by the hierarchy of the interaction scales. 
First, the largest intra-orbital Coulomb repulsion $U$ effectively suppresses the double occupancy of any single orbital, thereby strongly localizing the electrons and resulting Mott physics.
However, more profound physics arises from the substantial Hund's rule coupling $J_H$ due to the partial occupied of the two orbitals.
This ferromagnetic exchange interaction strongly favors the parallel alignment of spins between the $3d_{z^2}$ and $3d_{x^2-y^2}$ orbitals on the same nickel.
When two electrons for the two orbitals occupy on a single $Ni^{2+}$ site with $3d^8$ configuration, 
they form a local spin-$1$ triplet state to minimize energy. 
This triplet formation driven by strong Hund's first rule, serves as a robust local constraint when deriving the low-energy effective model.

The interplay between the inter-orbital Coulomb repulsion $V$ and the Hund's coupling $J_H$ is another critical ingredient dictating the local electronic structure. 
While $J_H$ energetically favors the alignment of electron spins to form local triplets, 
the $V$ term imposes a severe energy penalty on the coexistence of two electrons in different orbitals on the very same site. 
This competitions is conceptually crucial.
A simplified theoretical analysis that artificially neglects $V$ might erroneously predict an on-site triplet Cooper pairing state, driven solely by the attractive Hund's channel.
In reality, the physically realistic $V$ term is highly repulsive and easily large enough to suppress this unphysical on-site inter-orbital pairing channel. 
Therefore, an accurate and proper description of the pairing mechanism must explicitly incorporate this delicate balance between the ferromagnetic Hund's coupling $J_H$ and the inter-orbital repulsion $V$.

In the strong coupling limit, the magnetic landscape is further characterized by a pronounced anisotropy, directly stemming from the distinct geometries of the two $E_g$ orbitals.
The $3d_{z^2}$ orbitals, with their lobes oriented along the $c$-axis, exhibit strong hybridization with the shared apical oxygen $2p_z$ orbitals.
This rung-like structural unit geometry mediates a dominant interlayer AFM superexchange interaction, denoted as $J^{z}_{\perp}$.
From conventional analysis, this term approximately scales as $J^{z}_{\perp}\sim 4(t_{\perp}^z)^2/U$
In contrast, the $3d_{x^2-y^2}$ orbitals are confined to the in-plane NiO$_2$ square lattice, mediating a significantly weaker intralayer AFM exchange, $J^{x}_{\parallel}\sim 4(t_{\parallel}^x)^2/U$.
The low-energy magnetic physics is thus overwhelmingly dominated by the strong interlayer AFM coupling $J_{\perp}^z$.
In theoretical models, this robust interlayer interaction could act as the primary magnetic glue for the proposed superconducting pairing, whereas the weaker intralayer correlations play only a secondary role.

This theoretically predicted magnetic anisotropy in La$_3$Ni$_2$O$_7$ is strongly substantiated by a convergence of experimental evidence.
Both DFT calculations and electronic structure measurements via ARPES consistently reveal a pronounced interlayer hybridization specifically for the $3d_{z^2}$ orbitals, laying a firm electronic foundation for this magnetic picture.
More directly, the existence of a robust interlayer AFM coupling has been unambiguously confirmed.
Independent experimental studies utilizing resonant inelastic X-ray scattering (RIXS) \cite{chen2024electronic,zhong2025epitaxial} and inelastic neutron scattering (INS) \cite{xie2024strong} have  consistently detected a dominant interlayer spin interaction.
These measurements reveal a remarkably large energy scale for the interlayer exchange, roughly $S J_{\perp} \approx 60-70$ meV. 
This massive, experimentally established energy scale firmly validates the strong-coupling bilayer framework, confirming that the interlayer AFM exchange is the principal driving force behind the material's magnetic ground state and possible relevant for its emergent SC.

\subsection{Hund assisted magnetic exchange}

\begin{figure}[t!]
\centering
\includegraphics[width=0.8\linewidth]{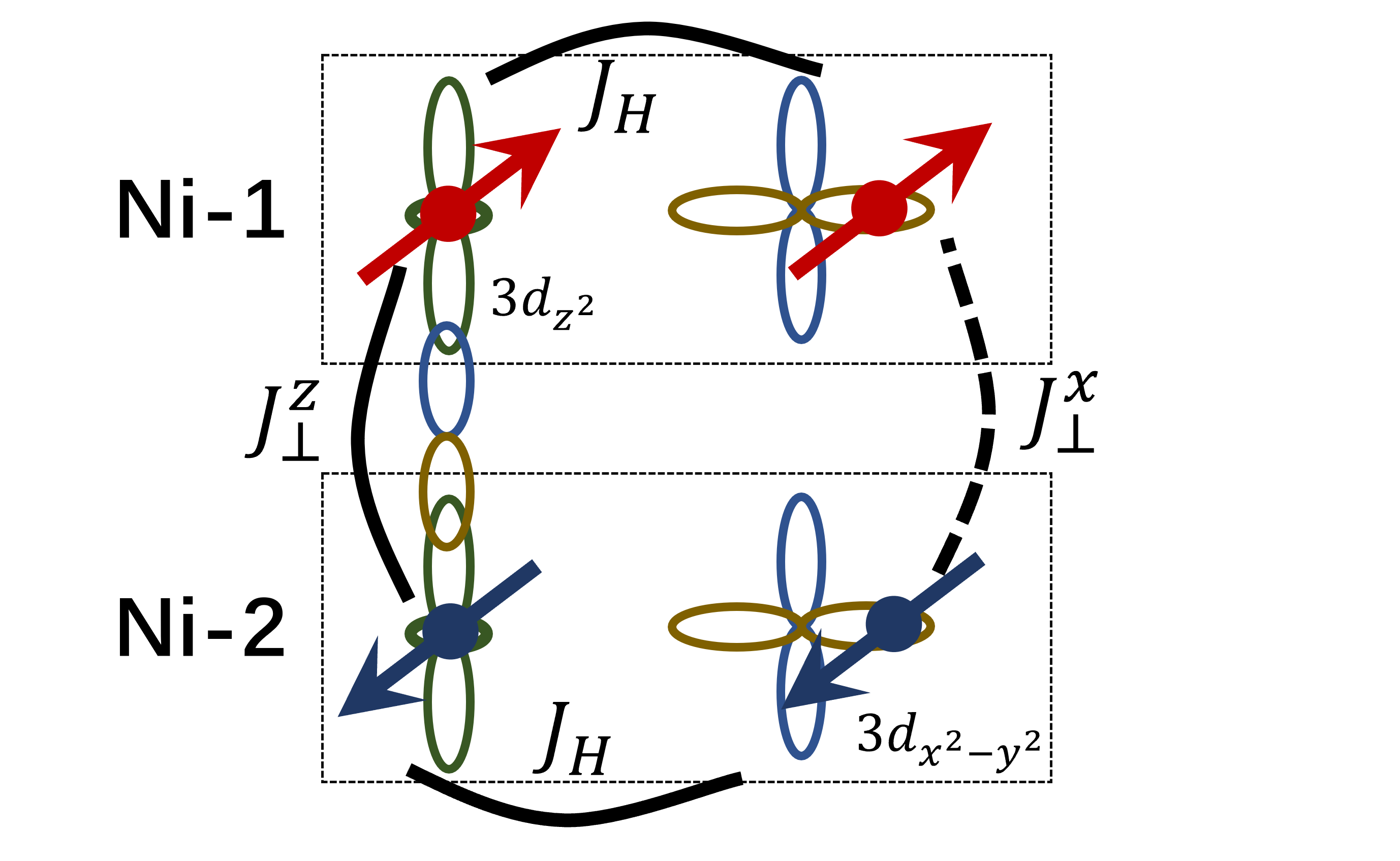}
\caption{Schematic diagram of the effective interlayer antiferromagnetic exchange $J_{\perp}^x$ driven by the interplay of Hund's rule coupling and bilayer geometry. 
The strong superexchange $J_{\perp}^z$ energetically favors an antiparallel alignment of the two $3d_{z^2}$ spins across the bilayer rung.
Simultaneously, the on-site ferromagnetic Hund's coupling tends to align the $3d_{x^2-y^2}$ and $3d_{z^2}$ spins within each individual Ni sites. 
As a result, the antiparallel correlation of the $3d_{z^2}$ spins is transferred to the $3d_{x^2-y^2}$ spins, generating an effective antiferromagnetic coupling $J_{\perp}^x$ between them.}
\label{fig:HundAssisted}
\end{figure}

More profound physics emerges from the interplay between the two $E_g$ orbitals.
Although the direct interlayer hopping of the $3d_{x^2-y^2}$ electrons is negligible, 
these itinerant electrons still experience a significant effective interlayer AFM exchange, denoted as $J_{\perp}^{x}$,
This induced interaction is primarily generated via the on-site Hund's coupling ~\cite{lu2024interlayer}, as illustrated in Fig.~\ref{fig:HundAssisted}.

To demonstrate this Hund-assisted mechanism, one can employ a semi-classical spin-coherent-state path integral approach. 
Consider an isolated vertical rung connecting two Ni sites (labeled by $i=1, 2$) across the bilayer, as shown in Fig.~\ref{fig:HundAssisted}. 
The relevant spin Hamiltonian governing this single rung consists of the on-site Hund's coupling $J_H$ and the direct interlayer AFM superexchange $J_{\perp}^z$ between the $3d_{z^2}$ orbitals: 
\begin{align}
\hat{H}_{12} = -J_H \sum_{i=1,2} \bm{S}_{z^2,i}\cdot\bm{S}_{x^2,i} 
+ J_{\perp}^z \bm{S}_{z^2,1}\cdot\bm{S}_{z^2,2}
\end{align}
In the path integral formalism, the quantum spin operators are mapped to classical unit vectors, $\bm{S}_{\alpha,i} \rightarrow S\hat{\Omega}_{\alpha,i}$.
The resulting imaginary-time action $\tilde{S}[\hat{\Omega}_{z^2},\hat{\Omega}_{x^2}]$ encompasses the topological Berry phase terms, the Hund's coupling $\tilde{S}_H$, and the interlayer $3d_{z^2}$ exchange $\tilde{S}_{\perp} = S^2 J_{\perp}^z \int_0^{\beta} d\tau \hat{\Omega}_{z^2,1} \cdot \hat{\Omega}_{z^2,2}$.

In the strong Hund's coupling regime ($J_H \gg J_{\perp}^z$), the large ferromagnetic $J_H$ imposes a strict energy penalty against local spin misalignment.
This creates a strong tendency for the electrons in the two $E_g$ orbitals to form an on-site spin triplet state. 
Consequently, the orientation of the $3d_{z^2}$ spin heavily dynamically locks to the local $3d_{x^2-y^2}$ spin, 
acting as a dynamical constraint: 
$\hat{\Omega}_{z^2,i}(\tau) \rightarrow \hat{\Omega}_{x^2,i}(\tau)$. 
Under this semi-classical approximation, we can integrate out the localized $3d_{z^2}$ spin degrees of freedom.
The unit vectors $\hat{\Omega}_{z^2,i}$ in the partition function are then effectively substituted by $\hat{\Omega}_{x^2,i}$. 
As a result, the original interlayer action $\tilde{S}_{\perp}[\hat{\Omega}_{z^2}]$ directly transforms into an effective action for the itinerant orbitals: $\tilde{S}_{\perp}[\hat{\Omega}_{x^2}] = J_{\perp}^z S^2 \int_0^{\beta} d\tau \hat{\Omega}_{x^2,1} \cdot \hat{\Omega}_{x^2,2}$.
This equivalently yields an emergent interlayer AFM spin-exchange $J_{\perp}^x \approx J_{\perp}^z$ directly between the two $3d_{x^2-y^2}$ orbitals.

The physical picture can be intuitively understood as follows:
Consider the two nickel sites along a single vertical rung.
The dominant superexchange $J_{\perp}^z$ strongly couples the localized $3d_{z^2}$ moments, locking them into a robust antiparallel AFM configuration across the bilayer.
The on-site physics within each nickel site is governed by the strong ferromagnetic Hund's coupling $J_H$.
This coupling energetically enforces a parallel alignment between the $3d_{z^2}$ and $3d_{x^2-y^2}$ spins on the exact same site. 
The antiparallel correlation already established among the $3d_{z^2}$ orbitals is imprinted onto the itinerant $3d_{x^2-y^2}$ orbitals.
This chain of interactions induces a potent interlayer AFM coupling $J_{\perp}^x$ for the $3d_{x^2-y^2}$ moments,
effectively bridging the localized magnetic and itinerant electronic sectors of the material.

\subsection{Effective single orbital model}

The physics of the $3d_{z^2}$ orbitals is dominated by their proximity to half-filled and the exceptionally strong interlayer superexchange.
Driven by a large $J_{\perp}^z$, the magnetic moments of the two Ni atoms along a vertical rung lock into robust spin-singlet bonds.
The formation of local singlets effectively opens a spin gap and freezes out the charge degrees of freedom in the $3d_{z^2}$ sectors, and these electrons become electrically silent at low energies.
Therefore, the $3d_{z^2}$ electrons do not participate directly as itinerant carriers in the formation of the superconducting condensate in La$_3$Ni$_2$O$_7$.

However, these localized $3d_{z^2}$ are far from passive; 
they play a decisive role in shaping the interactions of the itinerant $3d_{x^2-y^2}$ electrons as analyzed in previously.
By integrating out the high-energy $3d_{z^2}$ spin degrees of freedom, the system can be reduced to an effective single-orbital model based solely on the metallic $3d_{x^2-y^2}$ bands.
The resulting low-energy effective Hamiltonian is thus described by the following bilayer $t$-$J_{\perp}$-$J_{\parallel}$ model ~\cite{lu2024interlayer}:
\begin{equation}
\begin{aligned}
& H_{\text{eff}} =-t^{x}_{\parallel} \sum_{\langle i,j\rangle \alpha\sigma}
\mathcal{P} \big(d_{x\alpha\sigma}^{\dagger} (i) d_{x\alpha\sigma} (j) +\text{h.c.}\big)  \mathcal{P}   \\
+&J_{\parallel}^x \sum_{\langle i,j\rangle \alpha} \bm{S}_{x\alpha}(i)\cdot  \bm{S}_{x\alpha}(j)
+J_{\perp}^x \sum_{i} \bm{S}_{x 1}(i)\cdot  \bm{S}_{x 2}(i)
\end{aligned}
\end{equation}
where $\mathcal{P}$ represents the Gutzwiller projection operator, which enforces the strong-coupling constraint by strictly forbidding double occupancy of the $3d_{x^2-y^2}$ orbitals on any given site.

This effective single-orbital Hamiltonian captures the essential ingredients of the low-energy physics.
The first term, characterized by in-plane hopping $t_{\parallel}^x$, represents the kinetic energy of the itinerant $3d_{x^2-y^2}$ electrons moving the NiO$_2$ planes.
The second term describes the residual intralayer AFM exchange $J_{\parallel}^x$.
The third, and most physically crucial, term is the effective interlayer exchange $J_{\perp}^x$. 
Even though the direct interlayer hopping of the $3d_{x^2-y^2}$ electrons is negligible,
this strong AFM coupling is dynamically generated via the Hund's rule coupling to the $3d_{z^2}$ rung singlets.
Ultimately, this induced $J_{\perp}^x$ serves as the primary magnetic glue that drives the formation of Cooper pairs in the system. 
Physically, this effective bilayer $t$-$J$ model can be understood as the strong-coupling limit of a more fundamental bilayer Hubbard model. 
While the subsequent sections will deeply explore how SC naturally emerges directly from this $t$-$J$ Hamiltonian, it is highly instructive to first seek exact numerical verification of this underlying pairing mechanism.

\subsection{Exact numerical insights from sign-problem-free QMC simulations}
\label{sec:qmc}

Solving strongly correlated systems, particularly the Hubbard model and its variants like $t$-$J$ model, remains a central challenge in condensed matter physics. 
A key outstanding question is determining exactly whether and how SC can emerge in these systems. 
Quantum Monte Carlo (QMC) methods can rigorously solve the half-filled case, which typically exhibits an AFM Mott insulating ground state. 
However, simulating the physically crucial finite-doping regime is notoriously difficult. 
Away from half-filling, QMC simulations are severely plagued by the fermion sign problem, which causes numerical errors to grow exponentially, leading the results to be unjustified.

To overcome this critical barrier, it is highly desirable to identify or construct specific models that remain sign-problem-free under doping. 
Recently, a series of studies have successfully designed such sign-problem-free bilayer Hubbard models by utilizing a Kramers-invariant decomposition \cite{ma2022doping,ma2025parameter}. 
The underlying lattice geometry of these specially designed models naturally mirrors the physical bilayer structure of La$_{3}$Ni$_{2}$O$_{7}$ as mentioned above.
This structural alignment provides a reliable and numerically exact platform to investigate the emergence of unconventional SC. 
It allows researchers to explore broad parameter regimes directly, entirely avoiding the biases and approximations inherent in mean-field or perturbative theories.

\begin{figure}[t!]
\centering
\includegraphics[width=0.9\linewidth]{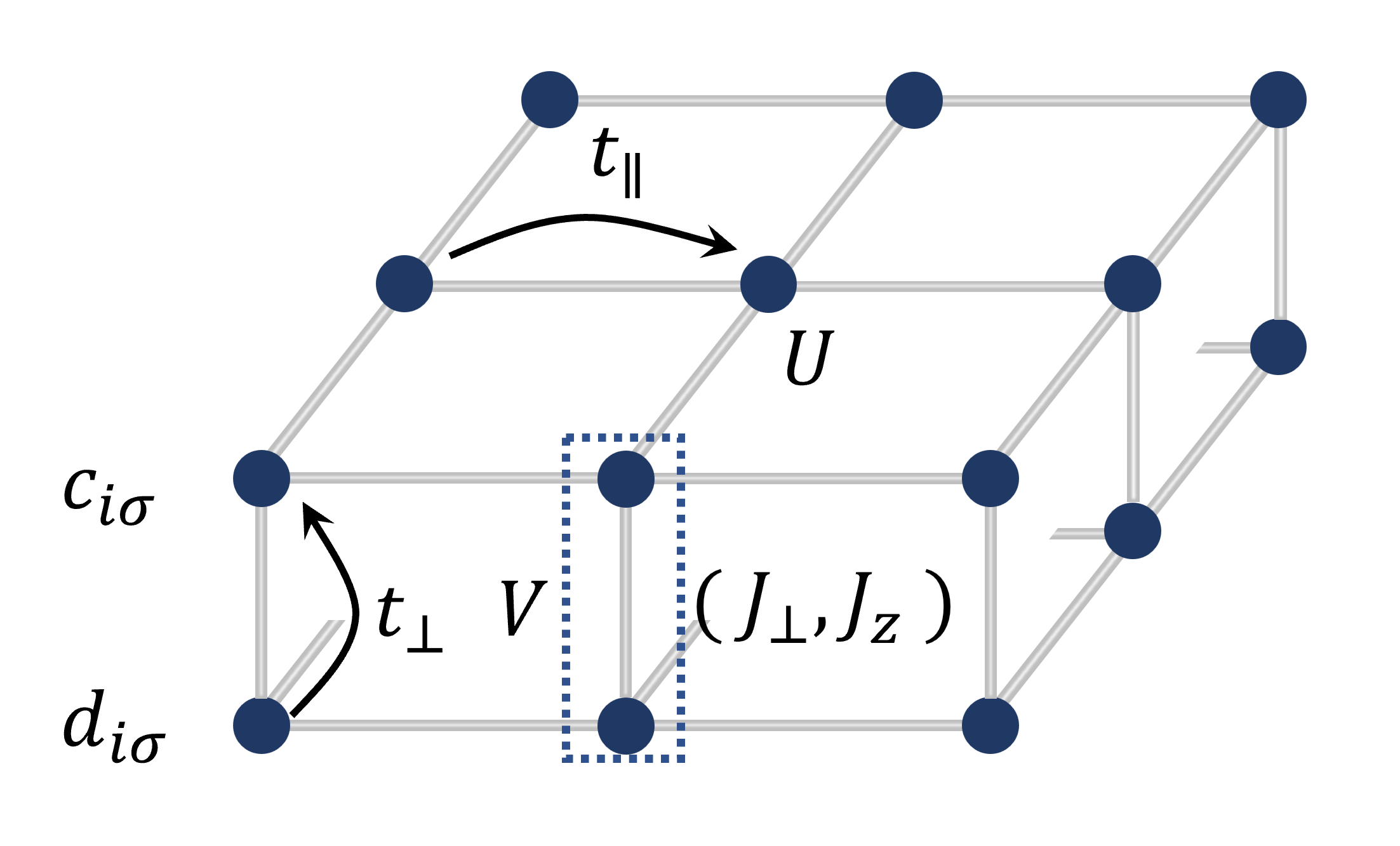}
\caption{Schematic diagram of the generalized SZH model on a bilayer square lattice. The system exhibits intralayer hopping $t_{\parallel}$, interlayer hopping $t_{\perp}$, on-site Coulomb $U$, interlayer Coulomb repulsive $V$ and interlayer AFM superexchange $(J_{\perp},J_z)$.}
\label{fig:bilayerHubbardQMC}
\end{figure}

The microscopic Hamiltonian of the bilayer version of the Scalapino-Zhang-Hanke (SZH) model \cite{scalapino1998so},
which is defined on a bilayer square lattice given by the following Hamiltonian \cite{ma2022doping,ma2025parameter}, as depicted in Fig.~\ref{fig:bilayerHubbardQMC}:
\begin{equation}
\begin{aligned}
H = &-t_{\parallel} \sum_{\langle ij\rangle\sigma} (c_{i\sigma}^{\dagger}c_{j\sigma} + d_{i\sigma}^{\dagger}d_{j\sigma} + \text{h.c.})  \\
&+t_{\perp} \sum_{i\sigma} (c_{i\sigma}^{\dagger}d_{i\sigma} + \text{h.c.}) - \mu \sum_{i\sigma} (n_{ic\sigma} + n_{id\sigma}) \\
& +\frac{J_{\perp}}{2} \sum_{i} (S_{ic}^+ S_{id}^- + \text{h.c.}) + J_z \sum_{i} S_{ic}^z S_{id}^z  \\
& +U \sum_{i} \Big[ \Big(n_{ic\uparrow}-\frac{1}{2}\Big)\Big(n_{ic\downarrow}-\frac{1}{2}\Big)  \\
& \qquad +\Big(n_{id\uparrow}-\frac{1}{2}\Big)\Big(n_{id\downarrow}-\frac{1}{2}\Big) \Big]     \\ 
& +V \sum_{i} (n_{ic}-1)(n_{id}-1) 
\end{aligned}
\label{eq:bilayerHubbard}
\end{equation}
Here, $c_{j\sigma} $ and $d_{j\sigma}$ denote the electron annihilation operators in the upper and lower layers, respectively. 
The first two kinetic terms describe the intralayer nearest-neighbor hopping ($t_{\parallel}$) and the vertical interlayer hopping ($t_{\perp}$). 
For the interaction terms, the parameter $U$ represents the strong on-site Hubbard repulsion, which effectively drives the Mott physics at half-filling. 
The $V$ term captures the charge-channel interaction between the two vertically aligned sites along a bilayer rung. 
$J_{\perp}$ and $J_z$ explicitly define the transverse and longitudinal AFM superexchange interactions across the vertical rung. 
In the context of La$_3$Ni$_2$O$_7$, this explicit interlayer magnetic coupling essentially serves as the primary pairing glue for SC.

The most remarkable feature of this generalized bilayer model is that it remains completely free of the sign problem at arbitrary doping levels in a large parameter region, opening up possibility to obtain exact results from QMC. 
This is mathematically achieved through a Kramers-invariant decomposition framework \cite{ma2022doping,ma2025parameter}, where the time-reversal symmetry is ensured \cite{wu2005sufficient}. 
By introducing a four-component spinor representation, the model in Eq.~\ref{eq:bilayerHubbard} can be mapped to a spin-$\frac{3}{2}$ fermionic Hubbard model \cite{wu2003exact,capponi2004current,wu2005sufficient,wu2006hidden,ma2022doping,ma2025parameter}:
\begin{equation}
\begin{aligned}
H=& -t_{\parallel} \sum_{\langle ij\rangle} \big( \psi_i^{\dagger}\psi_j +\text{h.c.} \big) -t_{\perp} \psi_i^{\dagger} \Gamma^5 \psi -\mu \sum_i n_i \\
&-\sum_i \frac{g_c}{2} (n_i-2)^2 -\sum_{i,a=1-5} \frac{g_a}{2} ((n_i^a)^2,
\end{aligned}
\label{eq:spin32hubbard}
\end{equation}
where $\psi_i=(c_{i\uparrow},c_{i\downarrow},d_{i\uparrow},d_{i\downarrow})$ is the spin-$\frac{3}{2}$ operator.
Here, $n_i=\psi_i^{\dagger}\psi_i$ and $n_i^a=\frac{1}{2}\psi_i^{\dagger}\Gamma^a\psi_i$ and the five Dirac $\Gamma^a$ ($a=1,\cdots,5$) matrices form a Clifford algebra $\{\Gamma^a,\Gamma^b\}=2\delta_{ab}$.
The mapping of the parameters between the two Hamiltonian in Eq.~\ref{eq:bilayerHubbard} and Eq.~\ref{eq:spin32hubbard} is given explicitly in the following,
\begin{equation}
\begin{aligned}
4g_c=& \frac{J_{\perp}}{2} +\frac{J_z}{4} -U -4V,    \\
g_{1,5}=& \frac{J_{\perp}}{2} +\frac{J_z}{4} -U +V,    \\
g_{2,3}=& \frac{J_{\perp}}{2} -\frac{J_z}{4} +U -V,    \\
g_4=& -\frac{J_{\perp}}{2} +\frac{3J_z}{4} +U -V.
\end{aligned}
\end{equation}
During the QMC simulations, the interaction terms are decoupled using a discrete Hubbard-Stratonovich (HS) transformation. 
The reason the sign problem is eliminated is that this specific HS decomposition is formulated to strictly preserve a generalized Kramers symmetry ($\mathcal{T}$). 
Physically, this $\mathcal{T}$ operator represents a combination of the standard time-reversal transformation and a spatial layer-exchange transformation. 
Because this Kramers symmetry remains unbroken for every single auxiliary field configuration, the eigenvalues of the fermion matrix naturally appear in complex conjugate pairs. 
The product of their determinants, which serves as the statistical sampling weight in the QMC algorithm, is guaranteed to be strictly positive definite. 
This elegant mathematical property ensures that the QMC algorithm can simulate the heavily doped regime with exact numerical precision, providing a highly reliable platform to observe the underlying pairing mechanisms.

The ground-state phase diagram of this bilayer model exhibits a remarkably rich structure. 
Depending on the specific interaction parameters, the system can host a variety of correlated phases, including charge-density-wave, bond-current, AFM, and SC orders \cite{wu2003exact,capponi2004current,wu2005sufficient,wu2006hidden,ma2022doping,ma2025parameter}. 
This model serves as a powerful and reliable platform for exploring the complex competition among these unconventional quantum states.

At exact half-filling, the strong on-site Hubbard repulsion drives the system into a robust Mott insulating state. 
Depending on the spatial anisotropy of the defined spin-exchange interactions, this parent insulator manifests either as an anisotropic Ising AFM phase or an $SU(2)$-invariant Mott phase. 
When hole doping is introduced into the system, the injected mobile charge carriers progressively weaken the static AFM background. 
Interestingly, before the magnetic order is completely suppressed, there exists a critical underdoped regime. 
In this specific region, the residual AFM order is found to stably coexist with the emerging macroscopic singlet SC state.

The established superconducting pairing occurs across the vertical rungs and possesses an extended $s$-wave symmetry \cite{ma2022doping}. 
This confirms the theoretical expectation that the strong interlayer exchange $J_{\perp}$ acts as the dominant magnetic glue, stabilizing an interlayer $s$-wave pairing state. 
Furthermore, symmetry principles dictate a coupling among the AFM order, the superconducting order, and a triplet pair-density wave (tPDW). 
In the coexistence regime of the AFM and superconducting orders, an enhanced tPDW correlation is induced as a secondary effect, highlighting the rich interplay of competing orders upon doping the parent Mott state.

Beyond the basic phase diagram, unbiased determinant QMC (DQMC) simulations shed light on the parameter dependence of the superconducting transition temperature ($T_c$), revealing a delicate interplay between the local Hubbard interaction $U$, the interlayer hopping $t_{\perp}$, and the spin-exchange $J_{\perp}$ \cite{ma2022doping,ma2025parameter}.
It is uncovered that the on-site interaction $U$ exerts a distinctly non-monotonic effect on SC. 
At low doping levels, increasing $U$ actually suppresses $T_c$, indicating that strong local Coulomb repulsion can be detrimental when hole mobility is restricted. 
However, at higher doping levels, increasing $U$ significantly enhances $T_c$. 
Because of this intricate dependence, the optimal doping level shifts toward $\langle n \rangle = 0.5$ as $U$ increases, which directly corresponds to the quarter-filled itinerant $3d_{x^2-y^2}$ bands in the actual La$_{3}$Ni$_{2}$O$_{7}$ material.

Furthermore, while a robust interlayer spin-exchange $J_{\perp}$ is unequivocally critical to forming local pairs, excessively large interlayer hopping $t_{\perp}$ is shown to inhibit macroscopic SC. 
A large $t_{\perp}$ modifies the Fermi surface topology by pushing the bonding and anti-bonding states away from the Fermi level, thereby suppressing essential many-body correlation effects. 
This numerical finding imposes critical constraints on theoretical proposals aiming to elevate $T_c$ via structural tuning or pressure: while enhancing interlayer magnetic coupling is beneficial, amplifying the bare single-particle interlayer hopping beyond a certain threshold can destabilize the superconducting condensate.

Finally, beyond the superconducting state itself, calculating the regular part of the optical resistivity within this sign-problem-free model unveils a linear-in-$T$ scattering rate at elevated temperatures \cite{ma2025parameter}. 
This linear relationship between temperature and resistivity indicates the potential emergence of a strange-metal-like phase in the normal state of the bilayer system. 
The manifestation of strange metallicity not only echoes the well-known phenomenology of cuprates but also aligns seamlessly with recent transport measurements of pressurized La$_{3}$Ni$_{2}$O$_{7}$ crystals. 
This agreement solidifies the proposed bilayer strong-coupling framework as a sound driver of both the anomalous normal state and the high-temperature superconducting behaviors in this novel nickelate compound.

\section{Strong coupling pairing scenario}
\label{sec:strong}

\begin{figure}[t!]
\centering
\includegraphics[width=0.6\linewidth]{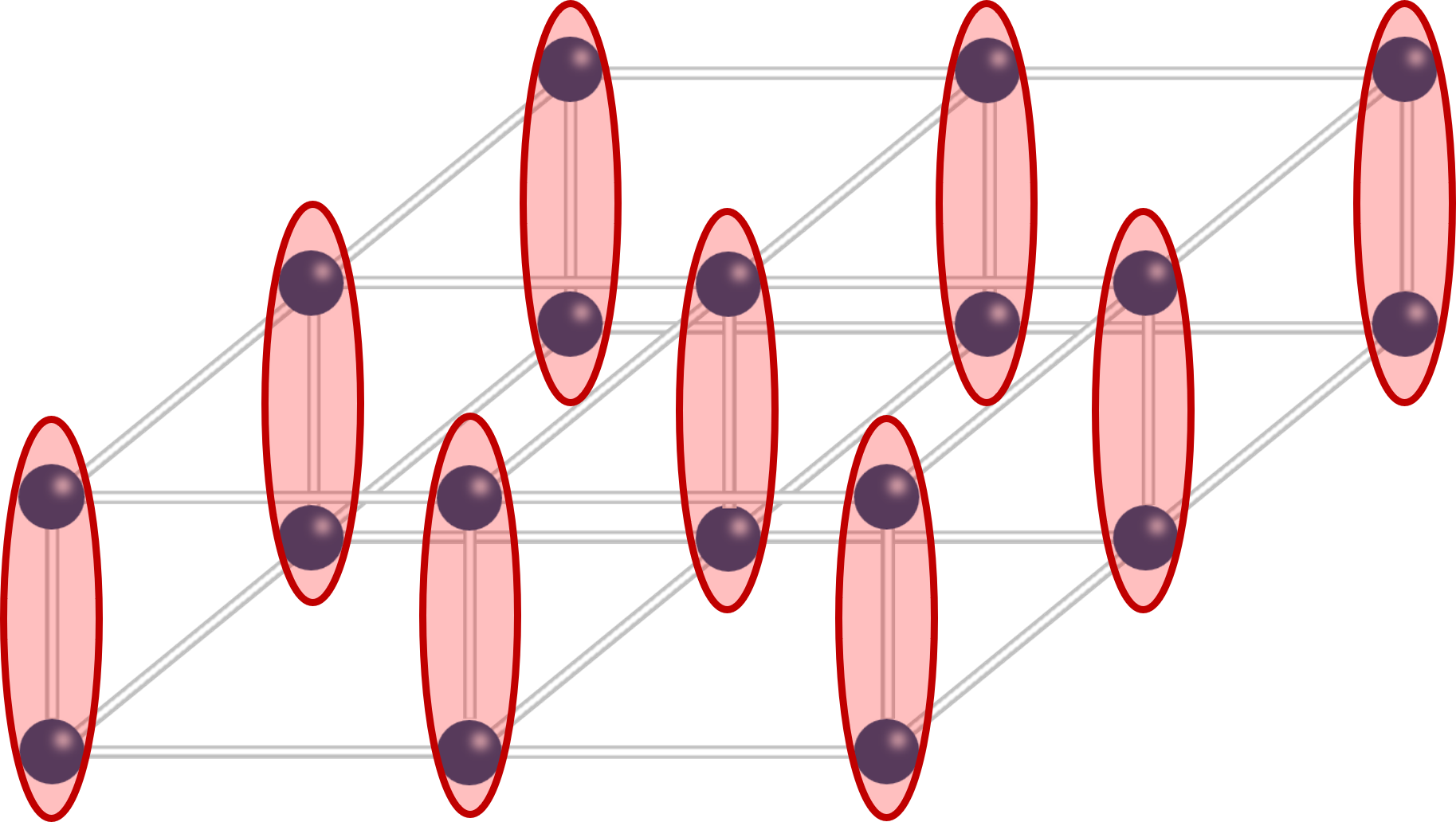}
\caption{Schematic diagram of the interlayer $s$-wave singlet pairing.
The local singlet pairs are formed along the vertical rungs connecting the two layers.}
\label{fig:swavePairing}
\end{figure}

The superconducting behavior is primarily driven by the itinerant $3d_{x^2-y^2}$ electrons, 
governed by the single-orbital bilayer $t$-$J_{\perp}$-$J_{\parallel}$ model.
Within this framework, the pairing mechanism is dominated by the effective interlayer magnetic correlations ~\cite{lu2024interlayer}.
The strong interlayer exchange interaction, $J_{\perp}^x$, which is dynamically inherent from the localized $3d_{z^2}$ rung singlets via the Hund's coupling,
acts as a robust attractive potential.
This interaction also induces the formation of local interlayer singlet pairs within the $3d_{x^2-y^2}$ sector,
as illustrated in Fig.~\ref{fig:swavePairing}.
Unlike the frozen and strictly localized $3d_{z^2}$ rung singlets, these $3d_{x^2-y^2}$ singlet rung-pairs exhibit high mobility.
Because the $3d_{x^2-y^2}$ band is approximately quarter-filled, effectively placing it deep in the heavily overdoped regime, these pairs can readily hop between neighboring lattice sites.
Their coherent motion across the lattice establishes the global phase coherence required for macroscopic, long-range SC.
Furthermore, in this heavily overdoped regime, phase coherence temperature is expected to significantly exceed the mean-field pairing temperature.
Therefore, the physical onset of SC (i.e., the critical temperature $T_c$) is primarily governed by the pairing energy scale determined by $J_{\perp}^x$.

The competition between intralayer and interlayer pairing tendencies enriches the superconducting phase diagram.
In the weak-interlayer situation where $J_{\perp}^x\ll J_{\parallel}^x$, 
the system effectively decouples into two independent single-layer $t$-$J$ models.
In this limit, the intralayer AFM fluctuations strongly favor a conventional intralayer $d$-wave pairing state, akin to the physics of cuprates.
However, as $J_{\perp}^x$ gradually increases and eventually dominates the interactions, the ground state must evolve into an interlayer $s$-wave symmetry.
A natural physical consequence of bridging these distinct $d$-wave and $s$-wave limits is the emergence of an exotic intermediate phase.
In this crossover regime, a mixed pairing state, most likely possessing an $s+id$ pairing symmetry, becomes energetically favored ~\cite{lu2024interlayer}.
This complex order parameter spontaneously breaks time-reversal symmetry, highlighting the remarkably rich physics dictated by the dimensional crossover in this strongly correlated bilayer system.

This intuitively physical proposal is substantiated by further numerical simulations.
For instance, slave-boson mean-field (SBMF) calculations ~\cite{lu2024interlayer} demonstrate that robust interlayer pairing can indeed emerge even in the highly overdoped regime.
This occurs provided that the interlayer superexchange $J_{\perp}^x$ dominates over the intralayer component $J_{\parallel}^x$. 
This conclusion is further corroborated by state-of-art DMRG studies.
These numerical methods unambiguously reveal a realization of high-$T_c$ SC and pronounced tendency toward interlayer pairing rather than intralayer pairing within the physically relevant parameter space ~\cite{qu2024bilayer}.

\subsection{Interlayer $s$-wave pairing}

To capture the strongly correlated nature of the system, the effective single-orbital bilayer $t$-$J_{\perp}$-$J_{\parallel}$ model for the $3d_{x^2-y^2}$ orbital can be analyzed using the SBMF theory. 
In this framework, the physical electron operator is fractionized into a neutral spin-$1/2$ fermion (spinon), $f_{i\alpha\sigma}^{\dagger}$, and a spinless charged boson (holon), $b_{i\alpha}$.
The electron creation operator is thus expressed as $d_{x\alpha\sigma}^{\dagger}(i)=f_{i\alpha\sigma}^{\dagger}b_{i\alpha}$. 
This representation could inherently enforces the Gutzwiller constraint through the local constraint $\sum_{\sigma}f_{i\alpha\sigma}^{\dagger}f_{i\alpha\sigma} +b_{i\alpha}^{\dagger}b_{i\alpha}=1$, forbidding double occupancy on any local site.

The ground state formulation assumes that the bosonic holons undergo Bose-Einstein condensation (BEC) at low temperatures. 
They can be treated as a coherent macroscopic background with an expectation value $b_{i\alpha}=\sqrt{\delta}$, where $\delta=1-2x$ represents the hole-doping concentration relative to the half-filled state. 
With the charge degrees of freedom condensed, the low-energy physics is dictated entirely by the spinon dynamics.
These dynamics are characterized by the intralayer and interlayer spinon pairing mean-field order parameters decoupled from the respective superexchange channels:
\begin{equation}
\begin{aligned}
\Delta_{ij}^{(\alpha)} =& \langle f_{j\alpha\downarrow} f_{i\alpha\uparrow } -f_{j\alpha\uparrow} f_{i\alpha\downarrow } \rangle, \\ 
\Delta_{z} =& \langle f_{i2\downarrow } f_{i1\uparrow} -f_{i2\uparrow} f_{i1\downarrow} \rangle
\end{aligned}
\label{eq:spinonpairing}
\end{equation}
where $\Delta_{ij}^{(\alpha)}$ and $\Delta_{z}$ denote the in-plane and vertical rung pairing amplitudes, respectively.

By self-consistently solving the mean-field Hamiltonian, a rich ground-state phase diagram emerges ~\cite{lu2024interlayer}.
The resulting phases are primarily governed by the electron filling fraction $x$ and the magnetic interaction ratio $J_\perp^x/J_\parallel^x$. 
As expected, three distinct superconducting phases are numerically identified. 
In the regime of weak interlayer coupling and high electron filling approaching half-filling, the system exhibits a standard intralayer $d$-wave pairing.
This state seamlessly connects to the well-known physics of single-layer cuprates superconductor. 
Conversely, when the interlayer superexchange dominates ($J_\perp^x > J_\parallel^x$),
the system transitions into a robust interlayer $s$-wave pairing state driven by the vertical rung singlets.
Interestingly, wedged directly between these two distinct limits is a narrow, intermediate phase with a mixed $s+id$ pairing symmetry. 
This exotic phase spontaneously breaks time-reversal symmetry, physically bridging the dimensional crossover between the purely two-dimensional $d$-wave regime and the strongly coupled bilayer $s$-wave regime.

\begin{figure}[t!]
\centering
\includegraphics[width=0.48\textwidth]{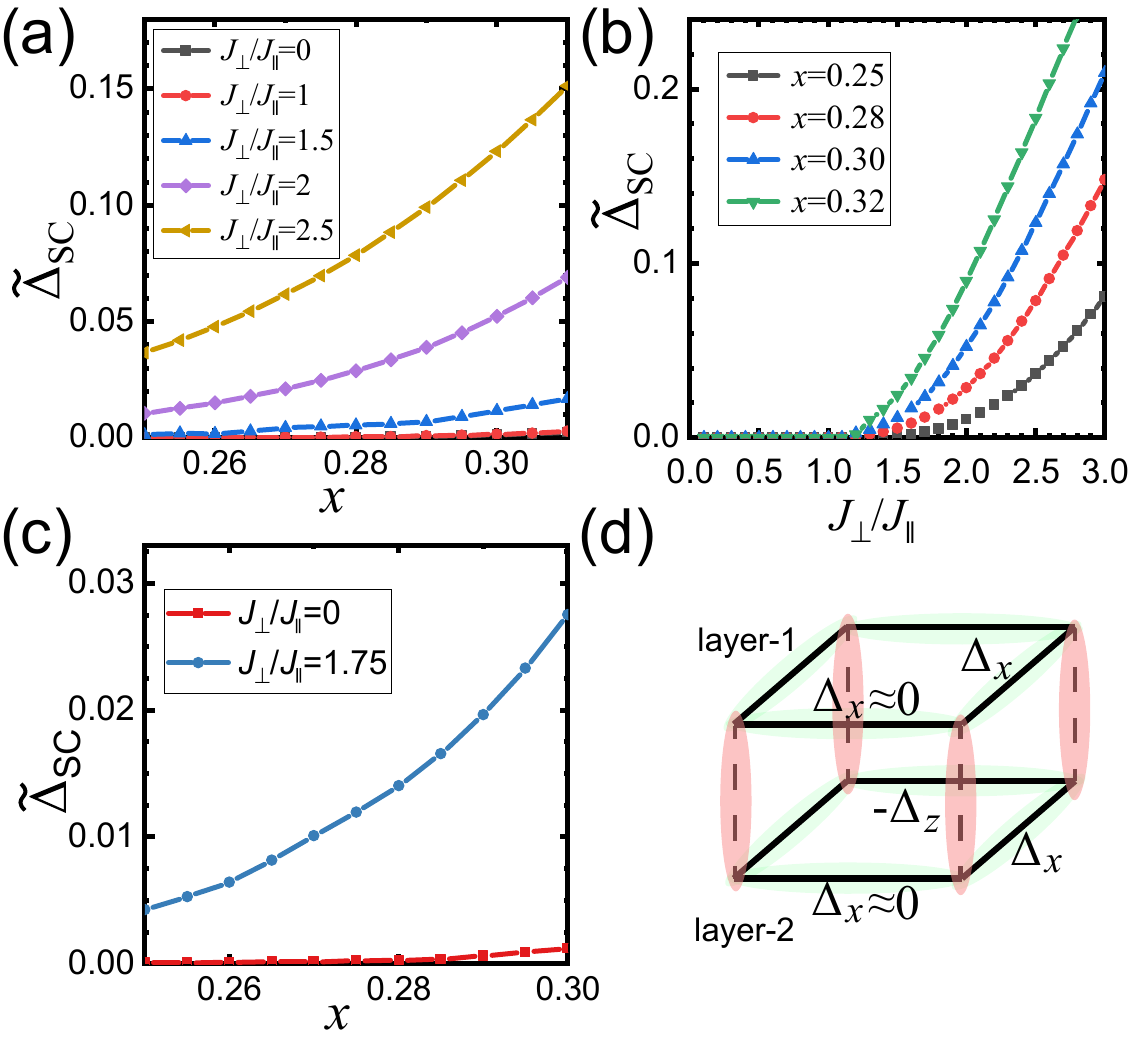}
\caption{(a). Ground state superconducting order parameter $\tilde{\Delta}_{\text{SC}}$ as a function of the filling fraction x $x$ at $J_{\parallel}^x=0.4t_{\parallel}^x$ for various coupling ratios $J_{\perp}^x/J_{\parallel}^x=0, 1, 1.5, 2, 2.5$ derived from slave-boson mean-field calculation.
(b). Simulated $\tilde{\Delta}_{\text{SC}}$ versus $J_{\perp}^x/J_{\parallel}^x$ for different filling levels $x=0.25, 0.28, 0.3, 0.32$.
(c). Comparison of $\tilde{\Delta}_{\text{SC}}$ versus $x$ between $J_{\perp}^x/J_{\parallel}^x=0$ and physical relevant $J_{\perp}^x/J_{\parallel}^x=1.75$.
(d). Pairing configuration of the obtained interlayer $s$-wave state for $J_{\perp}^x/J_{\parallel}^x=1.75$. 
Results adapted from Ref.~\cite{lu2024interlayer}.}
\label{fig:JperpTcFilling}
\end{figure}

The behavior of the physical pairing order parameter is crucial for understanding the robustness of SC in this system.
Within the SBMF theory, the realization of a true macroscopic superconducting state requires both the BEC of the holons and the Cooper pairing of the spinons.
Therefore, the physical superconducting order parameter is defined as $\tilde{\Delta}_{\text{SC}}=\delta \Delta_{\text{pair}}$.
Here, $\delta=1-2x$ represents the condensed holon density which is directly tied to the hole-doping concentration. 
$\Delta_{\text{pair}}$ represents the spinon pairing amplitude, which typical takes the form defined in Eq.~(\ref{eq:spinonpairing}).

The numerical simulated dominant superconducting order parameter $\tilde{\Delta}_{\text{SC}}$ is plotted in Fig.~\ref{fig:JperpTcFilling} ($a$) as a function of the filling level $x$ for various interlayer-to-intralayer superexchange ratios $J_{\perp}^x/J_{\parallel}^x$. 
Clearly, the physical pairing amplitude $\tilde{\Delta}_{\text{SC}}$ rises as the filling $x$ increases for all the simulated values of $J_{\perp}^x/J_{\parallel}^x$.
This general feature shares a similar trend with the purely two-dimensional single-layer $t$-$J$ model.
For instance, in the completely decoupled limit $J_\perp^x=0$, $\tilde{\Delta}_{\text{SC}}$ drops rapidly as $\delta$ approaches $0.5$ (or equivalently, as $x \to 0.25$), which reflects the system moving away from its optimal doping regime.

To isolate the effect of the interlayer coupling, Fig.~\ref{fig:JperpTcFilling}(b) shows the evolution of $\tilde{\Delta}_{\text{SC}}$ as a function of $J_{\perp}^x/J_{\parallel}^x$ for specific values of $x$.
Notably, once the interlayer exchange surpasses the intralayer exchange, $J_{\perp}>J_{\parallel}$, the physical order parameter $\tilde{\Delta}_{\text{SC}}$ increases monotonically and significantly across all experimentally relevant fillings.
In this strong interlayer coupling regime, the pairing configuration is overwhelmingly dominated by an interlayer $s$-wave pattern, with the intralayer pairing becoming physically negligible, as depicted in Fig.~\ref{fig:JperpTcFilling} ($d$).

The emergence of the robust interlayer $s$-wave state, and its correspondingly high transition temperature ($T_c$), can be intuitively understood through the concept of pairing frustration. 
In a pure intralayer pairing scenario, a given electron must symmetrically choose to pair with one of its four in-plane neighbors, leading to strong spatial competition and geometric frustration. 
In sharp contrast, for interlayer pairing, an electron uniquely and unambiguously couples with the single adjacent site located directly across the vertical rung.
This strict lack of geometrical frustration allows the strong $J_\perp^x$ to easily establish a dominant, uniform vertical pairing state ($\Delta_z \gg \Delta_{x,y}$).

The macroscopic superconducting order parameter, $\tilde{\Delta}_{\text{SC}} = \delta \Delta_{\text{pair}}$, is dramatically enhanced by interlayer magnetic correlation $J_\perp^x$. 
This theoretical picture provides a natural, microscopic explanation for key experimental observations in La$_3$Ni$_2$O$_7$. 
When external pressure is applied, it structurally straightens the inner $c$-axis Ni-O-Ni bond.
This geometric straightening strongly boosts the interlayer magnetic coupling $J_\perp^x$, thereby triggering the on-site high-$T_c$ SC. 
Conversely, if structural defects such as inner apical oxygen vacancies are present, they physically sever this critical vertical bond. 
This destroys the local $J_{\perp}^x$ exchange, directly and immediately suppressing the superconducting state.

Within the strong coupling scenario, the intrinsic doping level of the system places the itinerant $3d_{x^2-y^2}$ band deep in the overdoped regime.
The bare electronic filling of this orbital corresponds to $0.5$ electrons per site (a quarter-filled band.
Relative to a half-filled Mott insulating state, this equates to an effective hole-doping concentration of $\delta=0.5$.
This concentration is significantly higher than the typical optimal hole-doping level of $0.16$ observed in cuprate superconductors.
As a result, introducing electrons into the system (electron doping) would reduce the effective hole concentration, pushing the system closer to optimal doping.
This analysis reveals an inverse relationship between the hole-doping level and the pairing strength, namely, as hole concentration is reduced, the pairing strength is strongly enhanced ~\cite{lu2024interlayer,chen2026unified}. 
This physical picture leads to a clear and testable prediction that the SC critical temperature of La$_2$Ni$_2$O$_7$ should increase under conditions of electron doping, a predication which might be experimentally verified.

To capture the strong correlation effects beyond mean-field approximations, recent theoretical efforts have also employed the VMC method ~\cite{chen2025variation}. 
These VMC simulations confirm that the interlayer $s$-wave pairing remains the overwhelmingly dominant pairing channel. 
Compared to conventional mean-field theories, the rigorous VMC approach reveals a drastic enhancement in the calculated SC order parameters. 
This enhancement is of great physical significance for realistic materials. 
In actual La$_3$Ni$_2$O$_7$ samples, the finite nature of Hund's rule coupling inevitably reduces the effective interlayer superexchange $J_{\perp}^x$ transmitted to the $3d_{x^2-y^2}$ orbital. 
While mean-field theories struggle to predict robust SC in this reduced-$J_{\perp}^x$ regime, the VMC results demonstrate that the strict treatment of strong electron correlations effectively compensates for this reduction, namely, predicting a larger $T_c$ for smaller $J_{\perp}^x$ compared to the SBMF analysis. 
This numerical study establishes that the Hund-assisted interlayer pairing mechanism is resilient under realistic physical constraints.

Several other research groups have also investigated the bilayer La$_3$Ni$_2$O$_7$ based on this proposed single-orbital framework for the $3d_{x^2-y^2}$ electrons.
For example, the authors of Ref.~\cite{zhang2023strong} studied the pairing mechanism in the presence of an explicit on-site Hubbard repulsion $U$, rather than relying solely on effective spin-exchange interactions.
By analyzing a simplified effective single-orbital $t$-$J_{\perp}$-$U$, they found that introducing a large $U$ could suppress the kinetic energy of the itinerant electrons, thereby enhancing the electronic correlations.
As a result of this correlation-induced bandwidth narrowing, the relative importance of the magnetic exchange is amplified, allowing a robust superconducting state to be stabilized even for smaller values of the interlayer coupling $J_{\perp}$.

\subsection{Strong interlayer limit}
To deeply understand the physics governing $T_c$, it is instructive to examine the asymptotic limit of extremely strong interlayer exchange ($J_{\perp}\rightarrow \infty$) ~\cite{lu2023sc}.
In this regime, the infinitely large $J_{\perp}$ forces the system decouple into a collection of isolated vertical rungs.
The physics thus reduces to an effective Hamiltonian describing the competition between the intralayer kinetic hopping and the formation of interlayer spin-singlets driven by $J_{\perp}$:
\begin{equation}
\begin{aligned}
H_{\text{eff}} =-& t_{\parallel} \sum_{\langle i,j\rangle \alpha\sigma}
\mathcal{P} \big(d_{\alpha\sigma}^{\dagger} (i) d_{\alpha\sigma} (j) +\text{h.c.}\big)  \mathcal{P}   \\
+&J_{\perp} \sum_{i} \bm{S}_{1}(i)\cdot  \bm{S}_{2}(i).
\end{aligned}
\end{equation}
This simplified theoretical framework can be applied to both the $3d_{z^2}$ orbital and $3d_{x^2-y^2}$, albeit at their vastly different effective doping levels.
Energetically, the massive $J_{\perp}$ unambiguously drives the formation of tightly bound interlayer spin-singlets as illustrated previously in Fig.~\ref{fig:swavePairing}.
At exact half-filling, this rung-singlet formation results in a symmetric-mass-generation (SMG) insulator.
This is a featureless Mott insulator unique to bilayer systems, arising because the strict bilayer geometry completely cancels the Lieb-Schultz-Mattis (LSM) anomaly.
In this limit, the electronic degrees of freedom are effectively frozen into local rungs, which can be mapped to hard-core bosons, effectively leading to a singlet-layer bosonic model.
SC then naturally emerges upon doping this SMG insulator, as these preformed Cooper pairs gain the kinetic mobility necessary to establish long-range phase coherence across the lattice and form BEC condensation.

Physically, the effective $T_c$ is typically dictated by the lower of two characteristic scales: the pairing temperature ($T_{\text{pair}}$), driven by the interlayer superexchange, and the phase-coherence temperature ($T_{\text{BEC}}$), governed by the kinetic hopping of those preformed pairs.
The evolution between the BEC and BCS limits is explicitly controlled by the dimensionless ratio of interlayer interaction to the intralayer hopping ($J_{\perp}/t_{\parallel}$) ~\cite{lu2023sc}.

In the strong-coupling BEC limit, which is physically relevant to the nearly half-filled $3d_{z^2}$ orbitals, the pairing energy is exceptionally large.
This leads to the emergence of preformed interlayer Cooper pairs at a very high pairing temperature $T_{\text{pair}}$.
However, due to the smaller intralayer hopping, these pairs remain phase-incoherent and do not immediately condense into a superfluid.
In this regime, the actual superconducting transition temperature $T_c$ is determined entirely by the phase-ordering BEC temperature $T_{\text{BEC}}$.
In this simplified perturbative picture, this $T_{\text{BEC}}$ is proportional to the square of the effective kinetic hopping amplitude of the preformed Cooper pairs, given by $\tilde{t} = 8t^2/3J_{\perp}$. 
Thus, $T_c$ paradoxically decreases as the interlayer coupling $J_{\perp}$ increases, because the tightly bound pairs become too heavy to acquire macroscopic phase coherence.

Conversely, in the weak-coupling BCS limit, which is more representative of the highly overdoped $3d_{x^2-y^2}$ orbitals, the pairing formation and the establishment of phase coherence occur nearly simultaneously.
In this regime, $T_c$ is primarily limited by the pairing gap itself, which roughly scales positively with the interaction strength $J$.
Therefore, the itinerant $3d_{x^2-y^2}$ band resides on the side of the BEC-BCS crossover where an enhanced $J_{\perp}$ directly boosts $T_c$, consistently with previous analysis.

\begin{table}[t!]
\centering
\begin{tabular}{lcc}
\hline\hline
Parameters & $3d_{x^2-y^2}$ electron & $3d_{z^2}$ electron \\ \hline
$x$ & $1/4$ & $1/2 \rightarrow 0.483$ \\
$t$ & $0.48$~eV & $\sim 0.1$~eV \\
$J$ & $\sim 0.66$~eV & $\sim 0.66$~eV \\
$J/t$ & $1.4$ & $6.6$ \\
$T_c/t$ & $0.014$ & $0 \rightarrow 0.067$ \\
$T_c$ & $79$~K & $0 \rightarrow 78$~K \\
SC order & \multicolumn{2}{c}{s-wave interlayer spin-singlet} \\
Mechanism & BCS & BEC \\ \hline\hline
\end{tabular}
\caption{Parameter estimation and theoretically predicted $T_c$ for the Ni $3d_{x^2-y^2}$ and $3d_{z^2}$ electrons. 
Symbol definitions: $x$ denotes the electron filling fraction, $t$ is the hopping intralayer parameter, $J$ is the interlayer AFM Heisenberg coupling strength, and $T_c$ is the estimated superconducting transition temperature.
Table adapted from Ref.~\cite{lu2023sc}}
\label{tab:parameterestimation}
\end{table}

As summarized in Tab.~\ref{tab:parameterestimation}, this dual kinetic and magnetic nature results in a non-monotonic dependence of $T_c$ on both interaction strength and doping levels.
The $3d_{x^2-y^2}$ and $3d_{z^2}$ orbitals orbitals exhibit starkly contrasting superconducting behaviors ~\cite{lu2023sc}. 
The global maximum of $T_c$ is achieved exclusively within the intermediate BCS-BEC crossover region, where the pairing and coherence scales are optimally balanced.
In terms of band fillings, $T_c$ is theoretically maximized near a quarter-filled configuration ($x=1/4$).
At this specific filling, the interlayer Cooper pairs acquire substantial kinetic mobility while maintaining a robust pairing amplitude.

This framework reveals a fundamental dichotomy between the two orbitals. 
Because the highly itinerant $3d_{x^2-y^2}$ electrons reside on the BCS side of the crossover, their $T_c$ is expected to increase upon electron doping, which reduces the hole concentration and pushes the system toward the optimal crossover regime.
In sharp contrast, the nearly half-filled $3d_{z^2}$ electrons are situated deep on the strong-coupling BEC side.
Therefore, their $T_c$ would be enhanced by hole doping, which provides the necessary kinetic mobility for the otherwise localized rung pairs to establish phase coherence.
This theoretically predicted dichotomy provides a crucial and testable experimental signature. By carefully measuring how $T_c$ responds to chemical doping, future experiments could determine which orbital predominantly drives the physical SC in La$_3$Ni$_2$O$_7$.

\subsection{Element substitution}
The discovery of SC near $80$K in pressurized La$_3$Ni$_2$O$_7$ naturally prompts the search for physical or chemical avenues to further enhance the transition temperature $T_c$. 
One promising candidate route is rare-earth (RE) element substitution.
Specifically, this involves replacing lanthanum with smaller RE elements to synthesize the isostructural R$_3$Ni$_2$O$_7$ (where R $=$ La to Sm) series. 
In the superconducting $Fmmm$ phase, progressing from La to Sm systematically decreases the lattice constant ~\cite{zhang2023trends} . 
This intrinsic "chemical pressure" physically forces the Ni atoms closer together, directly enhancing the spatial overlap of the adjacent Ni $3d$ wavefunctions.

\begin{figure}[t!]
\includegraphics[width=1.0\linewidth]{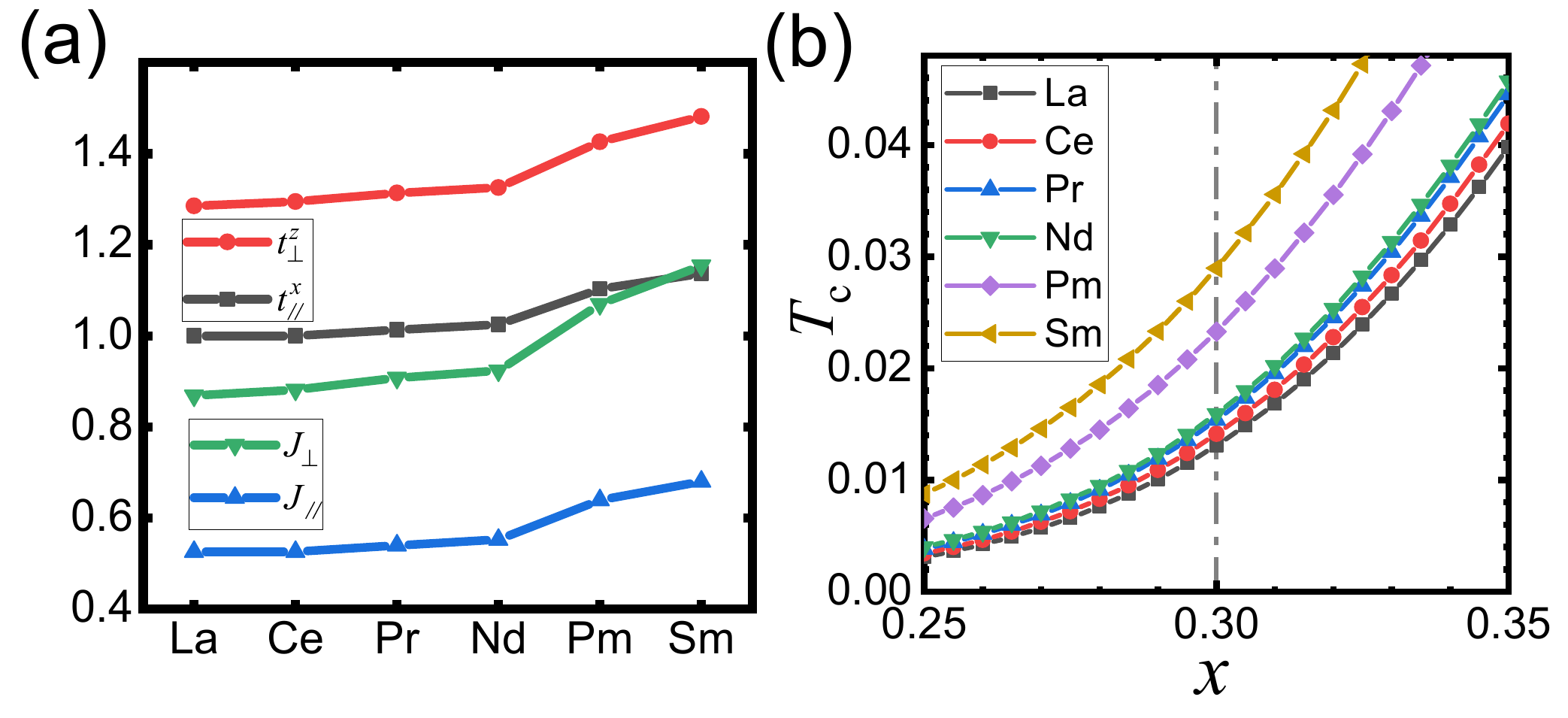}
\caption{(a). Relevant hopping integrals and effective spin-exchange parameters for different RE elements from La to Sm in the R$_3$Ni$_2$O$_7$ Fmmm phase under pressure.
As the rare-earth element changes from La to Sm, both the interlayer $3d_{z^2}$ hopping $t^{z}_{\perp}$ and the intralayer $3d_{x^2-y^2}$ hopping $t_{\parallel}^x$ gradually increase \cite{zhang2023trends}.
The effective interlayer $3d_{z^2}$ spin exchange $J_{\perp}$ and intralayer $3d_{x^2-y^2}$ spin exchange $J_{\parallel}$ follow a similar monotonically increasing trend.
(b). Superconducting critical temperature $T_c$ versus filling level $x$ under element substitution for the Fmmm phase.
Progressing from La to Sm, the interlayer pairing strength $\Delta_{\perp}$ and the superconducting $T_c$ increase simultaneously. 
Results adapted from Ref.~\cite{pan2023rno}.}
\label{fig:REDeltaTc}
\end{figure}

Both the intralayer hopping ($t_{\parallel}^x$) of the $3d_{x^2-y^2}$ orbitals and the crucial interlayer hopping ($t_{\perp}^z$) of the $3d_{z^2}$ orbitals undergo a significant and monotonic increase, as depicted in Fig.~\ref{fig:REDeltaTc} (a).
Because the magnetic superexchange interactions approximately scale as $J_{\perp}\sim (t_{\perp}^z)^2/U$ and $J_{\parallel}\sim (t_{\parallel}^x)^2/U$, they are also strongly amplified by this substitution.
Numerical evaluations confirm that while the interlayer $s$-wave pairing symmetry remains entirely robust across the RE series, both the macroscopic pairing gap ($\Delta_{\perp}$) and $T_c$ are elevated ~\cite{pan2023rno}. 
As clearly shown in Fig.~\ref{fig:REDeltaTc}(b), the predicted $T_c$ increases systematically as the RE ionic radius decreases.

The evolution of $T_c$ across the RE series highlights an important difference between weak- and strong-coupling pairing mechanisms ~\cite{pan2023rno,chen2026unified}. 
In a conventional weak-coupling framework, such as BCS or the random-phase approximation (RPA) theories, the transition temperature is primarily governed by the density of states (DOS) at the Fermi level, $N_F$, and the effective interaction strength, $g$.
The dependence of $T_c$ roughly follows the BCS-like relation $T_c\propto e^{-1/gN_F}$.
Increasing the hopping integral $t$ through element substitution (from La to Sm) broadens the electronic bands, which inherently reduces $N_F$.
Furthermore, within the RPA framework, the effective coupling $g$ is proportional to the bare susceptibility, which is inversely related to the bandwidth.
As the hopping increases and the bands broaden, the bare susceptibility and DOS $N_F$ are both strongly suppressed.
Consequently, weak-coupling theories typically predict a decrease in $T_c$ as one moves toward smaller RE ions.

However, bilayer nickelates are strongly correlated systems characterized by a large on-site Hubbard repulsion, with $U$ estimated at approximately $\approx 4\sim 5$ eV.
While weak-coupling methods often employ an artificially reduced $U$ (e.g., $\sim 1$eV) due to methodological constraints, they approach severely underestimates the actual correlation effects.
In the strong-coupling limit, captured by the effective $t$-$J_{\parallel}$-$J_{\perp}$ model, the pairing mechanism shifts from a DOS-driven process to one governed by local magnetic superexchange.
In this regime, the interlayer superexchange interaction $J_{\perp}$ is the dominant energy scale for pairing and scales proportionally with $t^2/U$ ~\cite{lu2024interlayer,pan2023rno}.
Because the local Coulomb repulsion $U$ remains relatively constant across the RE series, the substitution-induced increase in the hopping integrals $t$ leads to a direct and significant enhancement of the magnetic superexchange $J_{\perp}$.
This amplified magnetic coupling provides a clear, physical rationale for the rise in $T_c$ observed in substituted systems.

This strong-coupling amplification seamlessly integrates with the Hund-assisted multi-orbital physical picture.
The enhanced vertical spatial overlap of the $3d_{z^2}$ orbitals strengthens the localized interlayer superexchange $J_{\perp}^z$. 
Through the robust on-site ferromagnetic Hund's coupling, the enhanced magnetic pairing force is efficiently transmitted to the itinerant $3d_{x^2-y^2}$ electrons. 
In fact, replacing La with Sm, particularly when accompanied by a slight substitution-induced increase in the carrier filling fraction,
is found to largely increase the $T_c$ as shown in Fig.~\ref{fig:REDeltaTc} ~\cite{pan2023rno}.
This establishes Sm$_3$Ni$_2$O$_7$ and similar substituted compounds as prime candidates for realizing even higher temperature SC, providing experimental attachable route.
Actually, a $T_c$ exceeding $90$K has been observed in the partially element substituted sample La$_2$SmNi$_2$O$_{7-\delta}$ \cite{li2026bulk}, which is consistent with this theoretical proposal.

\subsection{Effect of Pressure and Apical Oxygen Vacancies}

Beyond the fundamental pairing symmetry, the delicate interplay between structural geometry, oxygen stoichiometry, and SC provides crucial insights into the realization of high-$T_c$ in La$_3$Ni$_2$O$_7$. 
The specific influences of external pressure and apical oxygen vacancies have been systematically investigated using the effective bilayer $t$-$J_{\parallel}$-$J_{\perp}$ model within a real-space SBMF framework ~\cite{lu2025impact}.

Theoretical investigations reveal a stark contrast in the superconducting properties between the ambient-pressure and high-pressure phases. 
In the ambient-pressure orthorhombic $Amam$ phase, the interlayer $c$-axis Ni-O-Ni bond angle significantly deviates from $180^{\circ}$. 
This geometric distortion heavily suppresses the interlayer hopping integral ($t_{\perp}^z$) of the $3d_{z^2}$ orbitals, leading to a weaker effective Hund's assisted interlayer superexchange $J_{\perp}^x\sim J_{\perp}^z$. 
As external pressure is applied, the system undergoes a structural transition to the more symmetric tetragonal $I4/mmm$ phase.
This transition is characterized by the physical straightening of the vertical Ni-O-Ni bond.
This structural straightening is an essential ingredient for the pairing mechanism because it drastically enhances the interlayer magnetic coupling. 
Numerical simulations ~\cite{lu2025impact} demonstrate that this enhanced $J_{\perp}^x$ can elevate the theoretical $T_c$ from a weak bulk tendency of roughly $15$K at ambient pressure to nearly $87$K at $30$GPa, echoing the experimentally observed pressure-induced emergence of robust high-$T_c$ states.

Interestingly, experimental phase diagrams also indicate that the superconducting $T_c$ eventually decreases upon further increasing the pressure beyond an optimal point. 
This non-monotonic behavior can also be captured within the strong-coupling framework by examining the ratio of the interlayer to intralayer exchange couplings, $J_{\perp}^x/J_{\parallel}^x$. 
DFT calculations suggest that extreme compression disproportionately increases the in-plane intralayer hopping compared to the out-of-plane hopping. 
This differential scaling leads to a steady reduction in the $J_{\perp}/J_{\parallel}$ ratio. 
Model calculations ~\cite{lu2025impact} confirm that the superconducting $T_c$ drops rapidly as this ratio decreases.
This drop underscores a delicate physical balance between the mutually frustrating interlayer and intralayer pairing channels under extreme physical compression, which suppresses the macroscopic superconducting state.

\begin{figure}[t!]
\centering
\includegraphics[width=0.45\textwidth]{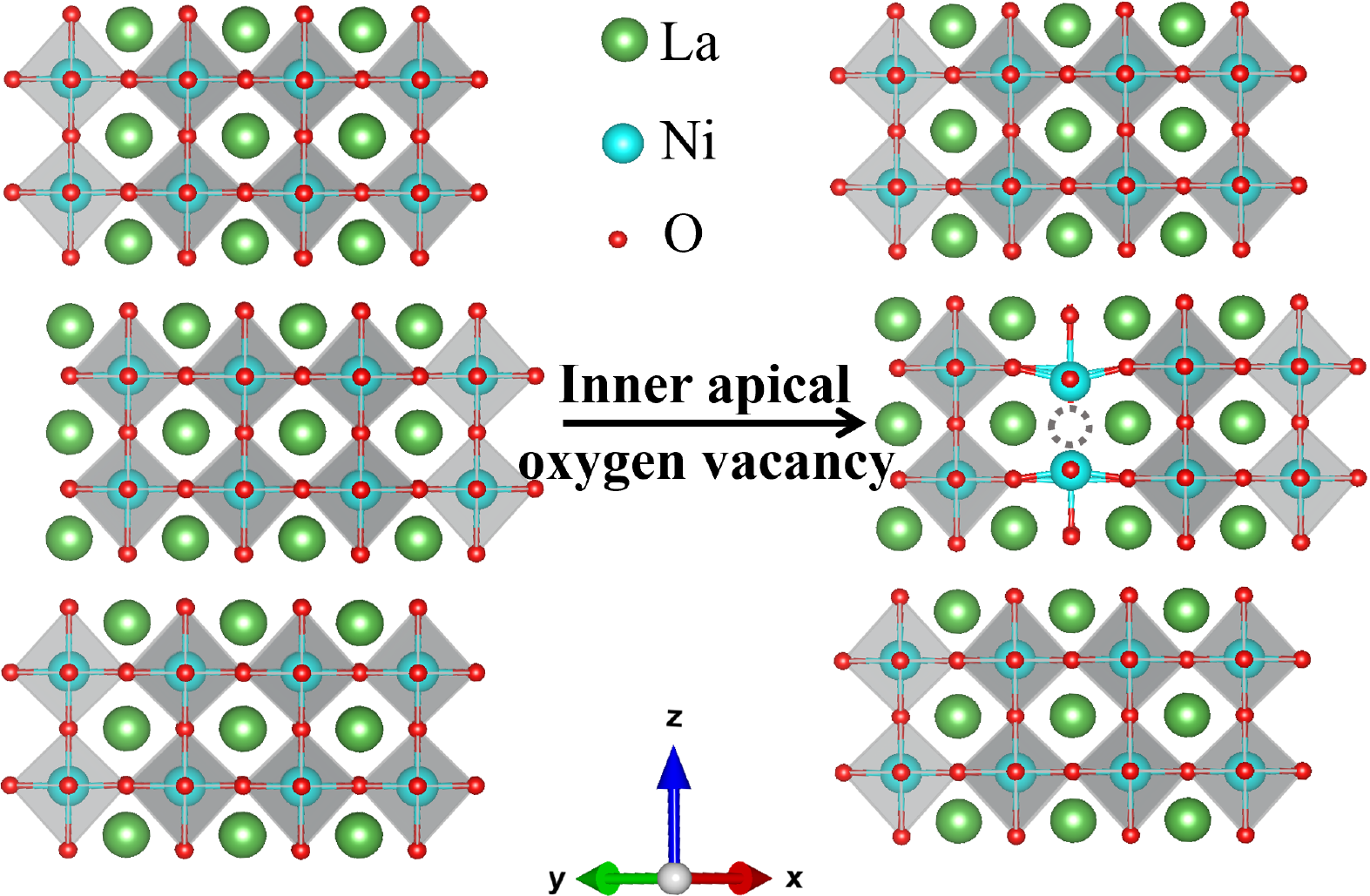}
\caption{Structural comparison between pristine La$_3$Ni$_2$O$_7$ and oxygen-deficient La$_3$Ni$_2$O$_{7-\delta}$, projected along the [110] axis. 
The left panel displays the regular lattice structure of La$_3$Ni$_2$O$_7$ without oxygen vacancy, composed of alternating layers of NiO$_2$ and LaO planes. 
The right panel illustrates La$_3$Ni$_2$O$_{7-\delta}$ with an oxygen vacancy at the inner apical position, indicated by a black dashed circle. 
Figure adapted from Ref.~\cite{lu2025impact}.}
\label{fig:apicalOxygen}
\end{figure}

Furthermore, the role of structural disorder, particularly apical oxygen non-stoichiometry, is highly critical and detrimental to the superconducting state. 
As schematically shown in Fig.~\ref{fig:apicalOxygen}, an apical oxygen vacancy  completely removes the inner oxygen atom located between the upper and lower NiO$_2$ planes within a specific bilayer rung.
At this defect site, the interlayer hopping $t{\perp}^z$ of the $3d_{z^2}$ orbital is effectively destroyed due to the absence of the mediating oxygen $2p_z$ orbital. 
Naturally, this destruction of the interlayer kinetic coupling directly eliminates the local interlayer AFM superexchange $J_{\perp}^x\sim J_{\perp}^z$ that serves as the pairing glue.

Real-space numerical simulations of random apical oxygen vacancies \cite{lu2025impact} reveal a severe suppression of macroscopic superconducting properties.
Microscopically, the removal of an apical oxygen atom physically severs the vertical Ni-O-Ni bond, driving the local superexchange $J_{\perp}$ precisely to zero. 
This disruption effectively eliminates the local electron pairing channel. Simultaneously, the missing oxygen alters the local crystal field, which increases the on-site electron energy and substantially reduces the local DOS near the Fermi level at the defect site.

The presence of apical oxygen vacancies not only suppresses the local rung-pairing amplitude $\Delta_{z}$, but also severely diminishes the global superfluid density. 
Physically, these vacancies act as strong, pair-breaking scattering centers.
They significantly decrease the diamagnetic current density while introducing spatial randomness that dramatically enhances the opposing paramagnetic current density.
This microscopic mechanism provides a clear theoretical explanation for the filamentary SC and extremely low diamagnetic volume fractions experimentally observed in oxygen-deficient samples.
It reveals that pristine inner apical oxygen bonds are an important prerequisite for establishing macroscopic phase coherence and robust high-$T_c$ SC in bilayer nickelates.

\subsection{Electric-Field Tuning Superconductivity in Thin Films}

The specific pairing nature within the bilayer models critically depends on both the effective magnetic coupling strength and the orbital doping levels. 
Under intrinsic conditions, the bilayer structure is perfectly symmetric, resulting in an equal carrier concentration across both NiO$_2$ layers. 
However, the recent successful growth of La$_3$Ni$_2$O$_7$ series ultrathin films on suitable substrates, which exhibit SC around $40$K at ambient pressure, opens up a new experimental avenue for manipulating these strongly correlated materials. 
For such thin-film samples, a highly promising method to independently control the doping levels in each layer is the application of an external perpendicular electric field along the $z$-direction, typically implemented through a gate voltage for thin-film samples ~\cite{shao2026possible}.

\begin{figure}[t!]
\centering
\includegraphics[width=0.95\linewidth]{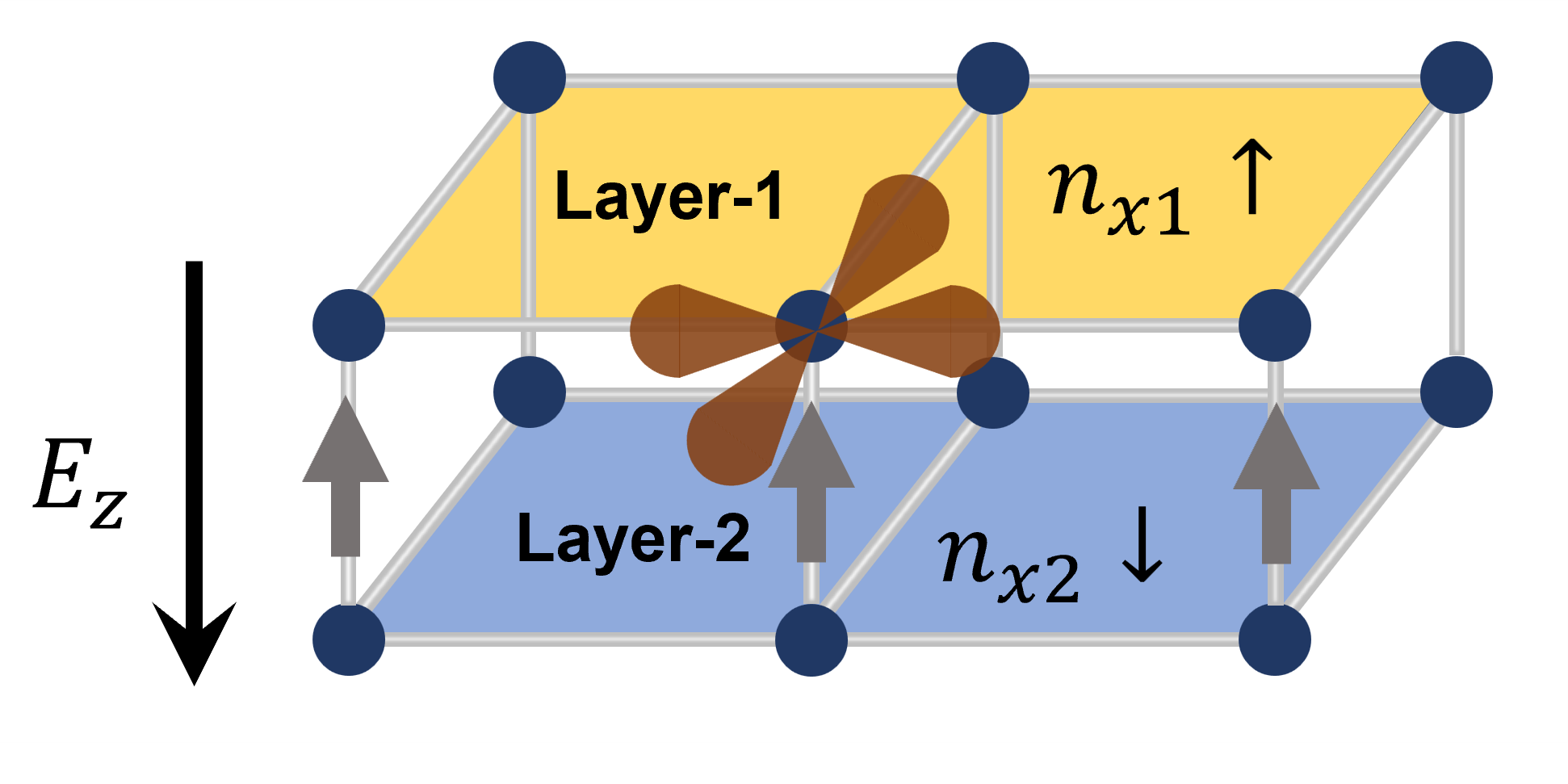}
\caption{Schematic diagram of the electron redistribution induced by a perpendicular electric field, $E_z$. 
Driven by the external field, electrons naturally migrate in the direction opposite to $E_z$ toward the lower-potential layer. 
This macroscopic charge transfer selectively increases the filling of the itinerant $3d_{x^2-y^2}$ orbitals, resulting in a higher electron density $n_{x1}$ in layer $1$ compared to $n_{x2}$ in layer $2$.}
\label{fig:ElectricField}
\end{figure}

Applying this perpendicular electric field explicitly break the spatial inversion symmetry of the two layers, inducing a macroscopic redistribution of the electron density, as depicted in Fig.~\ref{fig:ElectricField}.
Electrons are naturally driven from the layer with higher potential energy into the layer with lower potential energy. 
This interlayer charge transfer exhibits a strong orbital selectivity governed by local electron correlations. 
Because the nearly half-filled $3d_{z^2}$ orbitals sit in close proximity to a Mott insulating state, they cannot accommodate additional electrons due to massive on-site Hubbard repulsion, namely, they keep nearly half-filled under electric field tuning.
The transferred electrons are physically forced to selectively fill the more itinerant $3d_{x^2-y^2}$ orbitals in the lower-potential layer. 
One layer is effectively electron doped, while the other layer is hole doped.

This electric-field-driven orbital-selective filling profoundly alters the superconducting pairing landscape. 
The disparity in the $3d_{x^2-y^2}$ filling fractions between the top and bottom layers shifts their relative Fermi levels.
This shifts creates a direct Fermi surface mismatch between the two layers.
Physically, this mismatch acts as a "pseudo-Zeeman field" operating on the layer index.
It forcefully suppresses the intrinsic interlayer $s$-wave pairing, 
much like a real magnetic field destroys spin-singlet pairs.

As the interlayer $s$-wave pairing is suppressed, the heavily electron-doped layer becomes increasingly analogous to an optimally doped single-layer cuprate. 
Within this specific layer, A robust intralayer $d$-wave superconducting pairing is strongly induced. 
Numerical simulations, including both SBMF theory and DMRG calculations, firmly corroborate this intuitively physical picture.
The results demonstrate a relatively modest and experimentally accessible gate voltage approximately $0.1$ to $0.2$ V is sufficient to drive the transition.
This voltage can elevate the intralayer $d$-wave critical temperature beyond the boiling point of liquid nitrogen ($>77$K) in a single bilayer film at ambient pressure.

Intriguingly, comprehensive two-orbital studies reveal that this electric-field-driven transition stabilizes an exotic, mixed-symmetry quantum state. 
While the itinerant $3d_{x^2-y^2}$ orbitals host the macroscopic intralayer $d$-wave SC, the localized $3d_{z^2}$ orbitals are unaffected by the electric field and still maintain an interlayer $s$-wave pseudo-gap. 
These two distinct pairing channels coexist, naturally with a relative phase mixing ratio of $1:i$.
This complex phase difference forms a chiral $d+is$ state that spontaneously breaks time-reversal symmetry. 
Ultimately, utilizing a perpendicular electric field offers a clean, disorder-free approach to effectively electron-dope the system.
It thus provides a highly feasible and promising route to realize and manipulate liquid-nitrogen-temperature SC in nickelate devices at ambient pressure.

\subsection{Unified understanding of the $T_c$ controlling experiments}

Recent theoretical efforts have also demonstrated the capability to provide a unified physical picture for various structural tuning experiments in La$_3$Ni$_2$O$_7$ \cite{chen2026unified}.
Experiments found evolution of $T_c$ under pressure \cite{li2024pressure}, strain \cite{ko2024sign} and Sm/Nd substitution \cite{li2026bulk,qiu2025interlayer,zhong2025evolution}.
The application of high pressure, the introduction of compressive strain in thin films, and rare-earth element substitutions (such as Sm or Nd for La) all act to modulate the effective interlayer superexchange interaction, $J_{\perp}$ \cite{chen2026unified}. 
For instance, both compressive strain and the reduced $c$-axis lattice constant resulting from Sm/Nd substitution enhance interlayer hopping, thereby significantly amplifying $J_{\perp}$ and elevating the SC critical temperature ($T_c$). 
Furthermore, the pressure dependence of $J_{\perp}$ is found to exhibit a dome-shaped evolution due to competing structural and orbital energy shifts \cite{yi2024nature,yi2025unifying}. 
This trend captures the experimentally observed right-triangle-like SC phase diagram \cite{li2024pressure}, where $T_c$ sharply maximizes around $18$ GPa before gradually declining under further compression.

Beyond structural manipulations, this theoretical framework also captures the system's sensitivity to variations in carrier concentration \cite{chen2026unified}. 
Experimental reports indicate that hole doping, whether induced via over-oxidation or alkaline earth metal substitution, could suppress SC \cite{dong2025interstitial,hao2025sc,zhou2026sc,liu2026sc}.
This are naturally explained by a reduction in the filling fraction of the active $3d_{x^2-y^2}$ orbital. 
the carrier depletion directly lowers the DOS at the Fermi level, leading to a rapid decay of the pairing amplitude and $T_c$. 
This mechanism implies a highly asymmetric doping phase diagram and leads to a compelling theoretical prediction.
Introducing electron doping into the system, potentially through substitution with higher-valence elements, should increase the $3d_{x^2-y^2}$ filling fraction and its corresponding density of states.
This suggests that controlled electron doping presents a viable and promising pathway to further enhance $T_c$ in bilayer nickelates under appropriate structural conditions.

\section{Interplay between the two orbitals}
\label{sec:extended}

While the reduced single $3d_{x^2-y^2}$-orbital model successfully captures the essential physics of the SC,
particularly the interlayer Cooper pairing of the $3d_{x^2-y^2}$ electrons, 
a complete description of the pairing mechanism necessitates the inclusion of the $3d_{z^2}$ orbital.
A comprehensive understanding of high-$T_c$ SC in La$_3$Ni$_2$O$_7$ must be grounded in a strong-coupling bilayer two-orbital model that treats both active $E_g$ orbital degrees of freedom on equal footing from the outset.
This approach is essential to capture the intricate interplay between the two orbitals and the resulting electronic properties, 
which is ultimately critical for the emergence of SC in this system.
Furthermore, to definitively confirm the theoretical proposal that Hund's coupling mediates the effective interlayer pairing for the $3d_{x^2-y^2}$ orbital, 
a direct mathematical analysis incorporating the full Hund's coupling terms is important.

In a subsequent work ~\cite{lu2024interplay}, a strong-coupling bilayer two-orbital model was investigated to clarify the individual roles of each $E_g$ orbital and their profound interplay. 
Within this framework, since the highly localized $3d_{z^2}$ orbital are strongly oriented along the vertical rung direction, they predominantly exhibit a direct strong interlayer AFM superexchange.
Conversely, the itinerant $3d_{x^2-y^2}$ electrons experience two distinct magnetic interactions.
First, they undergo an intralayer AFM exchange arising from direct in-plane superexchange mechanism. 
Second, they experience an effective interlayer AFM exchange, which is transmitted from the $3d_{z^2}$ via the on-site Hund's coupling as depicted previously.

The resulting strong-coupling two-orbital $t$-$J$ model accurately incorporates these distinct magnetic interactions.
It explicitly defines both the intralayer and interlayer AFM exchange for the $3d_{x^2-y^2}$ orbital, alongside the dominant interlayer exchange for the $3d_{z^2}$ orbital:
\begin{equation}
\begin{aligned}
H_{J,x^2}&=J_{\parallel} \sum_{\langle i,j\rangle \alpha} \bm{S}_{x^2\alpha}(i) \cdot \bm{S}_{x^2\alpha}(j) \\
& +J_{\perp} \sum_{i} \bm{S}_{x^21}(i) \cdot \bm{S}_{x^22}(i),\\
H_{J,z^2}&=J_{\perp} \sum_{i} \bm{S}_{z^21}(i) \cdot \bm{S}_{z^22}(i).
\end{aligned}
\end{equation}
Here, the kinetic part $H_0$ remains identical to the tight-binding Hamiltonian previously defined in Eq.~\ref{eq:ham0}.
To solve this strongly correlated two-orbital model, SBMF theory is applied.
In this approach, auxiliary slave particles (charged holons and neutral spinons) are introduced independently for each orbital ~\cite{lu2024interplay}. 
For mathematical simplicity and to clearly isolate the dominant pairing mechanism, only the interlayer pairing channels are considered when decomposing the effective AFM exchange interactions.

\subsection{Orbital-Selective Band Renormalization}

\begin{figure}[t]
\includegraphics[width=1.0\linewidth]{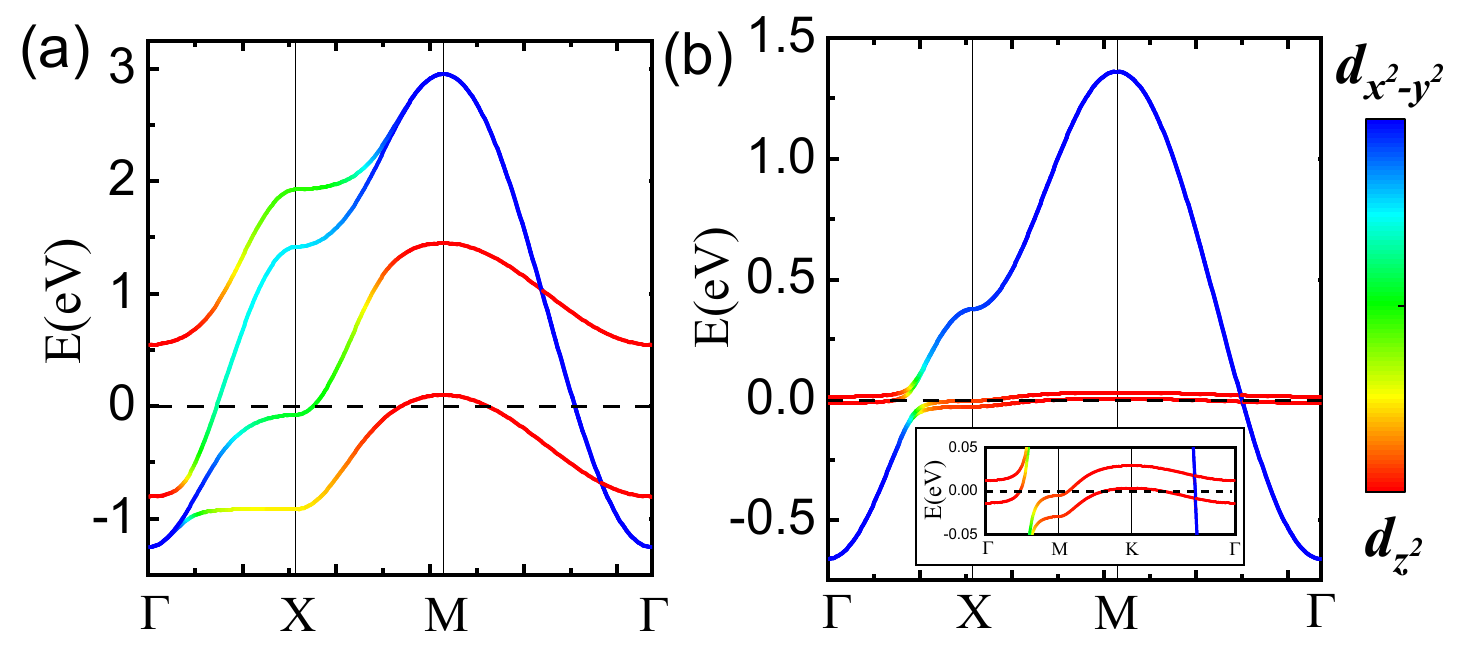}
\caption{($a$) Non-interacting tight-binding band structure. (Note that the tight-binding band structure was originally published in Ref.~\cite{luo2023high} and is presented here for comparison.)
($b$) Strongly correlated spinon band structure derived from SBMF theory.
The color scale on the right indicates the orbital character, with red and blue representing the $3d_{z^2}$ and $3d_{x^2-y^2}$ orbitals, respectively.
The system features four bands corresponding to the two orbitals and their bonding/antibonding combinations.
In the strong-coupling Mott limit, electronic correlations dramatically flatten the $3d_{z^2}$-dominated bands and significantly reduce the bonding-antibonding splitting.
High-symmetry momentum points in the Brillouin zone are marked as $\Gamma(0,0)$, $X=(\pi,0)$, $M=(\pi, \pi)$.
The inset in ($b$) provides a zoomed-in view of the highly renormalized spinon dispersion near the Fermi level.
Figure adapted from Ref.~\cite{lu2024interplay}.
}
\label{fig:BandSpectrum}
\end{figure}

The complex multi-orbital nature of La$_3$Ni$_2$O$_7$ is profoundly shaped by strong electronic correlations, which manifest in a highly orbital-selective manner. 
Theoretical investigations utilizing strong-coupling models have unveiled a pronounced orbital-selective band renormalization in the pressurized superconducting phase. 
While both the $3d_{x^2-y^2}$ and $3d_{z^2}$ orbitals experience correlation-induced band flattening, the magnitude of this effect differs vastly between the two active $E_g$ orbitals.
The contrasting non-interacting and strongly correlated band structures are presented in Fig.~\ref{fig:BandSpectrum} ~\cite{lu2024interplay}.

Numerical calculations based on this two-orbital model using SBMF theory clearly demonstrate this contrast.
The nearly half-filled $3d_{z^2}$ orbital is subjected to overwhelmingly strong correlation effects. 
This intense electron correlation pushes the $3d_{z^2}$ band toward the proximity of a Mott-like localized state, resulting in a drastically reduced quasiparticle weight and a massive enhancement of its effective mass. 
In stark contrast, the nearly quarter-filled $3d_{x^2-y^2}$ orbital experiences a significantly weaker mass enhancement. 
The $3d_{x^2-y^2}$ electrons retain a largely itinerant character and remain primarily responsible for macroscopic charge transport. 
This clear theoretical differentiation indicates that the $3d_{z^2}$ electrons are heavily restricted by Mott physics, while the $3d_{x^2-y^2}$ electrons remain highly mobile, a picture entirely consistent with the pronounced band renormalization observed in experimental measurements.

This pronounced orbital differentiation forms the fundamental physical basis for the unique superconducting mechanism in bilayer nickelates. 
The heavily renormalized, nearly localized $3d_{z^2}$ electrons are responsible for establishing the strong interlayer AFM superexchange.
They effectively act as the stationary magnetic glue and background for the system. Simultaneously, the relatively light and highly mobile $3d_{x^2-y^2}$ electrons interact with these localized spins via the on-site Hund's rule coupling. 
This multi-orbital interplay allows the itinerant $3d_{x^2-y^2}$ carriers to dynamically acquire the necessary pairing interactions from the $3d_{z^2}$ sector, while still maintaining the kinetic energy required to establish long-range macroscopic phase coherence. 
Therefore, the orbital-selective band renormalization is not merely a normal-state anomaly, but a vital electronic prerequisite for realizing high-temperature SC in this complex multi-orbital system.

\subsection{Orbital-dependent pseudogap phase and superconductivity}

While the localized $3d_{z^2}$ orbital does not directly drive the macroscopic superconducting condensate, it plays a vital role in generating the pairing interaction at high temperatures. 
As mentioned previously, the robust interlayer hybridization of the $3d_{z^2}$ orbitals, mediated by the inner apical oxygen, generates an exceptionally strong interlayer AFM superexchange $J_{\perp}^z$. 
This dominant magnetic interaction tightly binds the $3d_{z^2}$ spins into preformed interlayer rung singlets and this orbital exhibits a high pairing temperature $T_{\text{pair}}^z$.
However, the inherently localized nature of the $3d_{z^2}$ orbital prevents these local rung pairs from acquiring the kinetic mobility necessary to establish long-range phase coherence.
As a result, their BEC condensation temperature $T_{\text{BEC}}^z$ is much lower than $T_{\text{pair}}^z$.
Because these pairs cannot move coherently across the lattice, they do not condense into a true superconducting state at the physical superconducting critical temperature $T_c$.
Instead, this localized pairing manifests as an interlayer $s$-wave pseudogap phase.
In this state, a gap in the electronic DOS opens up at temperatures well above the actual $T_c$.

The emergence of true high-$T_c$ SC relies on a critical bridging mechanism between the localized and itinerant electronic sectors: the intra-atomic ferromagnetic Hund's rule coupling. 
Within each nickel atom, this strong Hund's coupling energetically forces the spins of the itinerant $3d_{x^2-y^2}$ electrons to dynamically align with the localized $3d_{z^2}$ spins. 
Because of this local spin alignment, the pre-established robust interlayer pairing correlation of the $3d_{z^2}$ orbitals is efficiently transmitted onto the $3d_{x^2-y^2}$ orbitals.

Through this Hund-assisted mechanism, a potent effective interlayer pairing interaction ($J_{\perp}^x$) is dynamically induced for the $3d_{x^2-y^2}$ electrons. 
Endowed with this transmitted magnetic glue, the $3d_{x^2-y^2}$ electrons can form interlayer rung singlet below their own pairing temperature $T_{\text{pair}}^x$.
At the same time, the highly mobile nature of the $3d_{x^2-y^2}$ band ensures a much higher condensation temperature, such that $T_{\text{BEC}}^x>T_{\text{pair}}^x$. Because their kinetic mobility is sufficiently high, these $3d_{x^2-y^2}$ rung pairs can successfully establish long-range phase coherence simultaneously and transition into a truly macroscopic superconducting state at $T_{\text{pair}}^x$ when the pairing is formed.

Therefore, a compelling, orbital-selective physical picture is established for La$_3$Ni$_2$O$_7$.
The localized $3d_{z^2}$ orbital acts as the stationary magnetic glue generator, responsible for forming the high-temperature pseudogap.
Meanwhile, the highly itinerant $3d_{x^2-y^2}$ orbital serves as the mobile engine that acquires this pairing force, 
driving the long-range phase coherence necessary for the macroscopic superconducting state.

\subsection{Phase Diagram and Temperature Evolution: BEC vs. BCS Pairing}

\begin{figure}[t]
\includegraphics[width=0.85\linewidth]{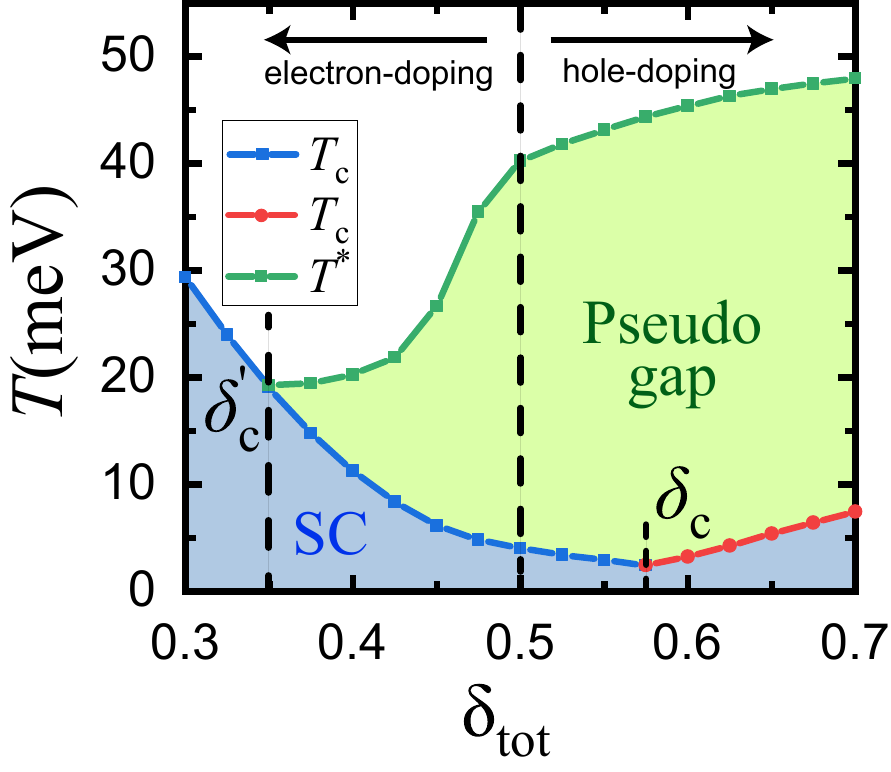}
\caption{Superconducting critical temperature $T_c$ and pseudogap temperature $T^*$ as a function of the total hole doping level $\delta_{\text{tot}}$.
Here, $\delta_{\text{tot}}>0.5$ corresponds to hole doping, while $\delta_{\text{tot}}<0.5$ indicates electron doping relative to the intrinsic stoichiometry (bare $\delta_{\text{tot}}=0.5$ for La$_3$Ni$_2$O$_7$).
The upper curve (green line)marks the onset of the pseudogap phase originating from the localized $3d_{z^2}$ orbital.
This pseudogap phase merges into the macroscopic superconducting phase at the critical doping level $\delta_{c}^{\prime}$.
The lower curve marks the onset of true superconductivity (SC), which exhibits a local minimum at $\delta_c$.
Figure adapted from Ref.~\cite{lu2024interplay}.
}
\label{fig:dopingPhaseDiagram}
\end{figure}

Based on the above physical picture, the global phase diagram of the superconducting La$_3$Ni$_2$O$_7$ system is fundamentally governed by the competition between the spinon pairing temperature ($T_{\text{pair}}$) and the holon condensation temperature ($T_{\text{BEC}}$) within the slave-boson framework. 
Within the slave-boson framework, the onset of SC for each specific orbital is defined by the smaller of its two characteristic energy scales: $\min(T{\text{pair}}, T_{\text{BEC}})$.
This mathematical condition represents the critical point where both local Cooper pairing and macroscopic phase coherence are simultaneously established for a given orbital.
The overall macroscopic critical temperature, $T_c$, of the entire system is then determined by the higher of these two orbital-specific onset temperatures. The simulated results for this complex phase diagram are summarized in Fig.~\ref{fig:dopingPhaseDiagram} ~\cite{lu2024interplay}.

The temperature evolution exhibits a highly orbital-selective behavior, directly stemming from the distinct electronic characters of the two active $E_g$ orbitals.
Driven by the strong interlayer superexchange $J_{\perp}^z$ and its nearly half-filled nature, the $3d_{z^2}$ electrons exhibit an exceptionally high $T_{\text{pair}}^z$. 
However, because the $3d_{z^2}$ orbital is heavily localized and possesses a low effective hole density (resulting in a large effective mass), its phase coherence temperature $T_{\text{BEC}}^z$ is significantly suppressed. 
The condition $T_{\text{BEC}}^z \ll T_{\text{pair}}^z$ is strictly satisfied for the $3d_{z^2}$ orbital.
Physically, this means that while interlayer rung singlets form at high temperatures, they fail to achieve long-range coherence.
On the contrary, they generate a high-temperature pseudogap phase rather than a true superconducting state in this sector.

In sharp contrast, the nearly quarter-filled $3d_{x^2-y^2}$ orbital possesses a much higher degree of itinerancy and a significantly larger condensed holon density.
This results in a $T_{\text{BEC}}^x$ that far exceeds its intrinsic pairing scale $T_{\text{pair}}^x$. 
Within this itinerant sector, the condition $T_{\text{BEC}}^x \gg T_{\text{pair}}^x$ holds.
This implies that for the $3d_{x^2-y^2}$ electrons, their superconducting onset $T_c$ is strictly limited by their pairing temperature $T_{\text{pair}}^x$, placing them on the weak-coupling BCS side of the crossover.

The physical synergy between these two distinct orbitals is mediated by the on-site Hund's rule coupling.
This crucial interaction effectively transmits the high-energy magnetic pairing glue from the localized $3d_{z^2}$ spins to the itinerant $3d_{x^2-y^2}$ charge carriers. 
As the hole doping $\delta$ increases, meaning the band filling $x$ moves further away from half-filling, the macroscopic $T_c$ initially rises because the effective pairing strength is dynamically enhanced.
However, upon further doping, $T_c$ eventually exhibits a dome-like suppression.
This dome-shaped phase diagram is a signature characteristic of many unconventional, strongly correlated superconductors.

Across this wide doping range, the phase coherence temperature $T_{\text{BEC}}^z$ of the $3d_{z^2}$ orbital remains deeply suppressed.
This stark thermodynamic contrast strongly reinforces a compelling, orbital-selective physical picture for the superconducting $T_c$ and doping dependence.
The heavily localized $3d_{z^2}$ orbital acts solely as the stationary source of high-temperature incoherent pairing.
Physically, this manifests physically as a robust pseudogap phase that persists up to high temperatures, particularly in the intrinsic and hole-doped regimes.
In sharp contrast, the highly itinerant $3d_{x^2-y^2}$ orbital dynamically acquires the transmitted interlayer coupling $J_{\perp}^x$ and successfully forms mobile interlayer pairs.
Endowed with this magnetic pairing force, the $3d_{x^2-y^2}$ band serves as the primary mobile engine driving the $80$ K macroscopic superconducting transition, yet it does not intrinsically exhibit a pseudogap phase itself.

This leads to a profound physical conclusion: the true superconducting condensate and the high-temperature pseudogap phenomenon are orbitally decoupled in this bilayer system. 
This represents a fundamental and critical distinction from cuprate superconductors. 
In the single-layer cuprates, the low-energy physics is overwhelmingly dominated by a single $3d_{x^2-y^2}$ orbital. 
The exact same electrons are responsible for forming both the high-temperature pseudogap (due to strong preformed pairing) and the low-temperature macroscopic superconducting state (upon achieving phase coherence). 
In bilayer nickelates, however, the strongly correlated multi-orbital nature physically separates these roles. 
A clear "division of labor" emerges: the $3d_{z^2}$ orbital generates the pairing glue and hosts the pseudogap, while the $3d_{x^2-y^2}$ orbital harnesses this glue to form pair and establish macroscopic phase coherence due to mobile nature.

\subsection{Hund's Coupling-Induced Interlayer Pairing Mechanism}

To confirm the microscopic pairing mechanism in pressurized La$_3$Ni$——2$O$_7$, it is essential to employ a two-orbital framework that explicitly incorporates the intra-atomic Hund's rule coupling ~\cite{qu2023roles,kaneko2023pair}. 
In these effective multi-orbital models, such as the two-orbital Hubbard ladder or bilayer $t-J$ extensions, the local electronic interactions are typically governed by a Hamiltonian that explicitly defines the Hund's coupling term as $H_{\text{Hund}} = -J_H \sum_{i} \bm{S}_{i, d_{x^2-y^2}} \cdot \bm{S}_{i, d_{z^2}}$. 
This ferromagnetic interaction energetically penalizes anti-aligned spins within the same nickel atom, strongly favoring the formation of a local spin-triplet configuration when both $E_g$ orbitals are singly occupied. 
Unlike simple single-orbital effective theories, this explicit treatment of Hund coupling reveals that the macroscopic superconducting state is not driven by an isolated band, but is instead a highly cooperative phenomenon fundamentally orchestrated by the strong $J_H$.

The physical picture of this Hund-assisted SC centers on the dynamic transfer of magnetic pairing glue between the localized and itinerant electronic sectors. 
Due to the nearly $180^\circ$ vertical Ni-O-Ni bond angle under high pressure, the $3d_{z^2}$ orbitals exhibit a massive interlayer hopping $t_\perp^z$, which generates an exceptionally strong interlayer AFM superexchange interaction. 
However, because the $3d_{z^2}$ electrons are heavily correlated and nearly localized, they cannot establish macroscopic phase coherence on their own. 
Here, the local Hund's coupling $J_H$ acts as an indispensable quantum bridge. 
By ferromagnetically locking the spins of the itinerant $3d_{x^2-y^2}$ electrons to the localized $3d_{z^2}$ spins on the same site, the pre-established vertical spin-singlet correlations of the $3d_{z^2}$ rungs are dynamically imprinted onto the $3d_{x^2-y^2}$ orbital. 
This synchronization mechanism effectively induces a potent interlayer pairing interaction for the highly mobile $3d_{x^2-y^2}$ charge carriers, allowing them to bind into robust interlayer $s$-wave Cooper pairs.

The validity and robustness of this interactive‌ pairing mechanism have been rigorously confirmed through unbiased numerical simulations, including DMRG and advanced tensor network calculations ~\cite{qu2023roles,kaneko2023pair}. 
For instance, numerical evaluations on two-orbital Hubbard ladders reveal that the singlet pairing correlation functions of the $3d_{x^2-y^2}$ orbital exhibit a slow algebraic (power-law) decay while spin correlations decay exponentially, which is a definitive hallmark of a Luther-Emery liquid state and quasi-long-range superconducting order. 
Furthermore, global phase diagram analyses demonstrate that the highest transition temperatures emerge in a hybrid-Hund synergistic regime. 
In this optimal parameter space, the interorbital hybridization establishes the necessary kinetic communication to mobilize pairs, while the Hund's rule coupling acts as the indispensable magnetic glue, collaboratively driving the system into a high-temperature superconducting phase that cannot be realized without this multiorbital interplay.

\subsection{Mixed Spin-1 and Spin-1/2 Pairing Mechanism in Bilayer Nickelates}

\begin{figure}[t!]
\centering
\includegraphics[width=1\linewidth]{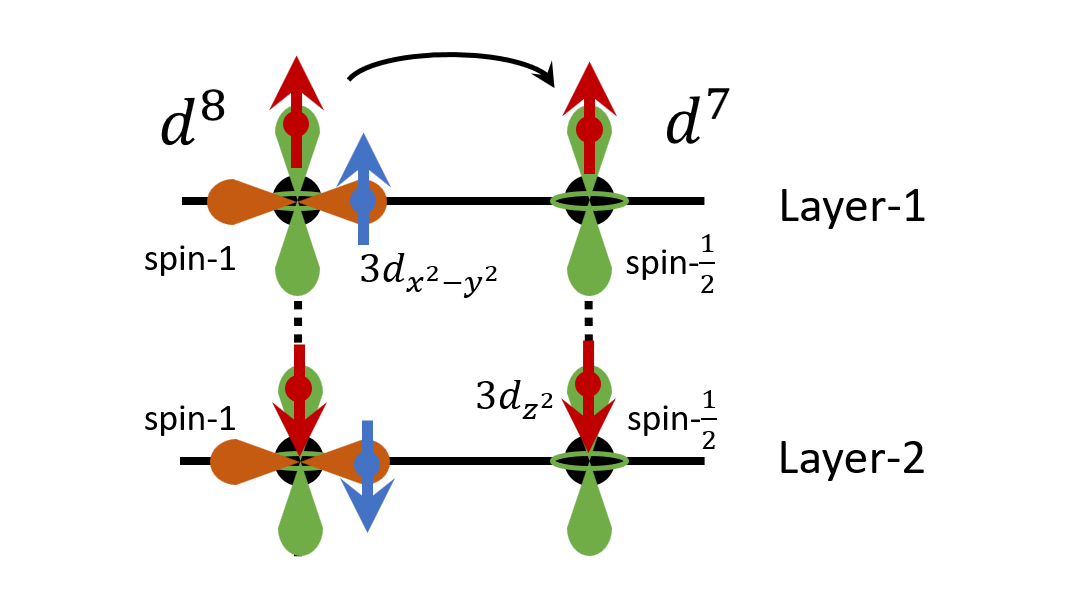}
\caption{Schematic illustration of the two distinct types of interlayer singlet bonds formed on a vertical bilayer rung, driven by strong interlayer superexchange. 
Left: A spin-$1$ singlet bond formed between two local spin-triplets. 
This state is characteristic of the undoped $3d^8$ electronic configuration. 
Right: A spin-$1/2$ singlet bond formed between two localized spin-$1/2$ moments (predominantly $3d_{z^2}$ electrons. 
This configuration naturally emerges upon hole doping the system towards the $3d^7$ state.
The figure is adapted from Ref.~\cite{ji2025strong}.} 
\label{fig:Spin1singlet}
\end{figure}

Further theoretical works explicitly explore the extreme strong-coupling limit,
treating the on-site Hund's coupling as a dominant energy scale.
This framework assumes that the strict no-double-occupancy constraint, driven by a large Hubbard $U$, is already enforced, and the Hilbert space of each nicke ion is limited.
In this regime, the ferromagnetic Hund's coupling tightly binds the electrons in the two active $E_g$ orbitals into local spin-$1$ triplet states, as depicted in Fig.~\ref{fig:Spin1singlet} ~\cite{ji2025strong}.
Each local nickel ion with a nominal $d^8$ configuration electronic configuration acts as an effective, localized spin-$1$ magnetic moment.
Subsequently, virtual hopping processes including both the intralayer and interlayer ones could induce effective AFM superexchange interactions between these local spin-$1$ states.
Because the interlayer hopping integral is naturally larger than the intralayer components, the resulting interlayer AFM interaction overwhelmingly dominates.
In this limit, the system prefers to form interlayer spin-$1$ rung singlets, 
a physical situation that generalizes the familiar concept of spin-$\frac{1}{2}$ rung singlets to a higher-spin manifold.
To establish macroscopic SC, these localized spin-$1$ singlets must eventually acquire kinetic mobility and establish global phase coherence.

To capture such strong-coupling physics of La$_3$Ni$_2$O$_7$ and explicitly account for the Hund's rule induced local spin-$1$ states, a novel mixed spin-$1$ and spin-$1/2$ bilayer $t$-$J$ model has been constructed and numerically simulated ~\cite{ji2025strong}. 
This theoretical approach directly incorporates the nominal $3d^8$ configuration as the underlying insulating parent state.
In this parent state, the strong on-site Hund's coupling enforces the formation of local spin-triplet states. 
Then, the dominant interlayer superexchange strongly couples the two spin-$1$ states along a vertical rung, driving the formation of robust spin-$1$ rung singlets, as illustrated in Fig.~\ref{fig:Spin1singlet}.
The uniform $3d^8$ parent system is thus characterized as an interlayer spin-$1$ valence-bond solid (VBS).
In this highly entangled VBS state, the four $E_g$ electrons residing on each rung are tightly bound into an interlayer singlet.

Superconductivity is then intuitively conceptualized as doping this insulating spin-$1$ VBS state. 
As holes are introduced into the system, the local spin-$1$ triplets are partially broken down into spin-$1/2$ singlons, primarily occupying the $3d_{z^2}$ orbitals, as the $3d_{x^2-y^2}$ electrons are removed.
This doping process drives a fundamental transition from a pure spin-$1$ regime to a highly complex, mixed-spin manifold where both spin-$1$ and spin-$1/2$ singlets coexist and interact.
In the physical La$_3$Ni$_2$O$_7$ system, the intrinsic electronic configuration resides at an average $3d^{7.5}$ valence.
This naturally places the material deep within this entangled, mixed-spin regime.

\begin{figure}[t!]
\centering
\includegraphics[width=1\linewidth]{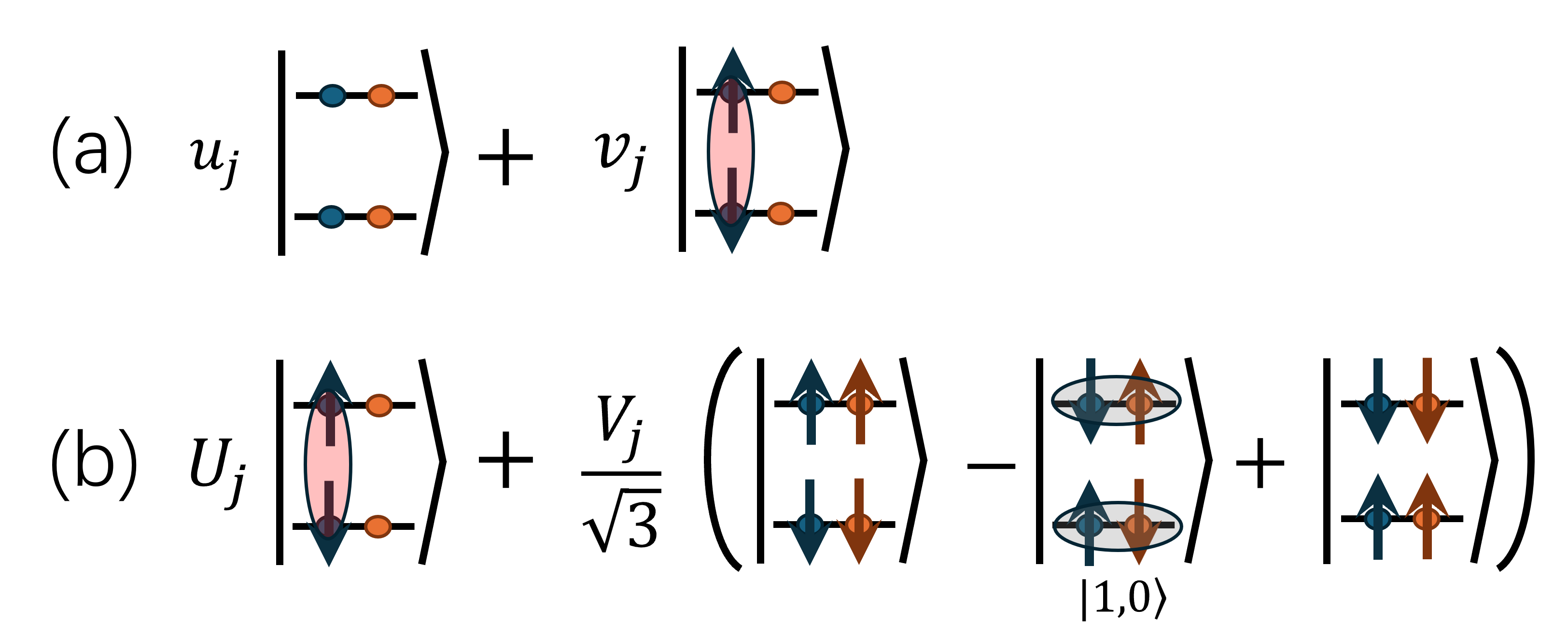}
\caption{Schematic diagrams comparing (a) conventional BCS pairing and (b) the strongly correlated mixed-spin pairing.
The conventional BCS state is represented as a coherent superposition of a fully paired electron state and the vacuum state.
The correlated pairing state is a coherent superposition of two distinct valence bond singlets: one composed of a pair of spin-$1$ states and the other composed of a pair of spin-$1/2$ states.
The blue circles represent the $3d_{z^2}$ orbitals, while the orange circles represent the $3d_{x^2-y^2}$ orbitals. 
The notation $|1,0\rangle$ indicates a local triplet state with a total spin $S=1$ and a total $z$-component spin $S^z=0$.
The figure is adapted from Ref.~\cite{ji2025strong}.
} 
\label{fig:Spin1BCS}
\end{figure}

A defining feature of this theoretical proposal is the exotic nature of its superconducting pairing ~\cite{ji2025strong}.
This composite pairing structure represents a significant departure from standard BCS theory.
The "paired" state is not a simple direct product of independent electron singlets.
Instead, it forms an unusual quantum entangled state mixing different spin manifolds, as schematically shown in Fig.~\ref{fig:Spin1BCS}.
As illustrated in Fig.~\ref{fig:Spin1BCS} (a), ta conventional BCS wavefunction is constructed as a coherent superposition of a fully paired electron state and an empty vacuum state.
In stark contrast, the pairing mechanism in this strong Hund's scenario involves a coherent quantum superposition of two distinct types of local valence bonds, shown in Fig.~\ref{fig:Spin1BCS}(b).
These are the spin-$1/2$ rung singlets formed by the localized $3d_{z^2}$ singlons and the spin-$1$ rung singlets formed by the triplet doublons across the vertical rungs.
When a mobile hole of $3d_{x^2-y^2}$ orbital hops within the plane, a spin-$1$ rung singlet is locally reduced to a spin-$\frac{1}{2}$ rung singlet.
This dynamic interconversion allows the localized pairs to acquire kinetic mobility, effectively establishing macroscopic phase coherence.
Numerical results from both SBMF theory and DMRG simulations consistently corroborate this physical picture.
They demonstrate that the interlayer spin-singlet pairing within the itinerant $3d_{x^2-y^2}$ orbital constitutes the overwhelmingly dominant superconducting channel.
Meanwhile, all triplet pairing channels are exponentially suppressed and remain physically negligible ~\cite{ji2025strong}.

The physical interplay between Hund's coupling and carrier doping plays a decisive role in tuning the superconducting strength within this framework. 
Specifically, the on-site Hund's rule coupling facilitates SC by promoting the local combination of $3d_{z^2}$ and $3d_{x^2-y^2}$ singlons into spin-$1$ triplet doublons.
This process effectively transfers the robust interlayer pairing correlation from the localized $3d_{z^2}$ sector directly to the itinerant $3d_{x^2-y^2}$ electrons. 
Furthermore, the macroscopic pairing amplitude exhibits a marked asymmetry with respect to the type of chemical dopants.
Theoretical simulations reveal that electron doping significantly strengthens the superconducting order, consistent with previous effective single-orbital and two-orbital $t$-$J$ models.
It does so by increasing the effective carrier density and the local doublon occupancy.
Conversely, hole doping is found to be detrimental to the pairing strength.
These findings also emphasize that the high transition temperature in superonducting La$_3$Ni$_2$O$_7$ is intrinsically linked to its orbital-selective behavior and the unique magnetic glue provided by the doped bilayer VBS structure.

\section{Related model}
\label{sec:comp}

Several theoretical models have also proposed strong-coupling mechanism for SC in bilayer La$_3$Ni$_2$O$_7$.
These approaches place distinct emphases on the roles of the Ni $3d_{z^2}$ and $3d_{x^2-y^2}$ orbitals, their hybridization, interlayer bonding, and pairing symmetries such as $s$-wave or $d$-wave. 
Collectively, these models highlight different pathways to SC, ranging from single-orbital, cuprate-like physics to multi-orbital effects driven by strong correlations, self-doping, or Hund's coupling, enriching our understanding for the bilayer SC.

Since the discovery of high-$T_c$ SC in La$_3$Ni$_2$O$_7$, whether its electronic structure and pairing mechanism are akin to those of cuprates has attracted considerable attention \cite{jiang2024pressure,fan2024sc}.
Based on the similarity of their Fermi surfaces, a scenario proposed in Ref.~\cite{fan2024sc} draws a direct parallel between bilayer nickelate and cuprate superconductors with strong bilayer coupling.
This study establishes an effective bilayer $t$-$t{\perp}$-$J$ model, suggesting that the SC is governed by the hybridized Ni $3d_{x^2-y^2}$ and in-plane O $2p$ orbitals.
The analysis also argues that the Ni $3d_{z^2}$ orbitals are decoupled from the active pairing mechanism.
Instead, the pressure-induced $\gamma$-band, which possesses a $3d_{z^2}$ character, functions as a charge reservoir. 
This reservoir extracts holes from the main $\alpha$ and $\beta$ bands, thereby tuning the $\beta$ band toward an optimal doping level. 
The SC pairing originates from intralayer AFM interactions within the $3d_{x^2-y^2}$ orbitals, which intrinsically favors a $d$-wave pairing. 
However, due to the strong bilayer splitting, the $\alpha$ band becomes heavily overdoped, which could introduce an additional $s$-wave component, ultimately leading to a mixed ($d+is$)-wave gap symmetry.

The interlayer coupling might also play a critical role in shaping the electronic structure.
In the strong interlayer coupling limit, the bilayer system naturally forms molecular bonding and anti-bonding bands.
Based on this, Ref.~\cite{jiang2023high,wang2025self,wang2026orbital,wang2026jahn} proposes a self-doped molecular Mott insulator mechanism. 
In this scenario, a large Hubbard $U$ drives two nearly degenerate antisymmetric molecular orbitals (derived from $3d_{z^2}$ and $3d_{x^2-y^2}$) into a two-orbital molecular Mott insulator limit.
The unique aspect of this model is its self-doping mechanism, namely, a small number of electrons transfer to the slightly higher-energy symmetric $3d_{x^2-y^2}$ orbital.
This partial occupation naturally leaves doped holes in the strongly correlated antisymmetric orbitals.
Superconducting pairing emerges from the electron correlations within this hole-doped, two-orbital background.
While intra-orbital interactions naturally support a $d$-wave pairing similar to cuprates, the model demonstrates that strong inter-orbital hopping can instead promote an extended $s$-wave pairing, highlighting the importance of multi-orbital physics.

Theoretical investigations presented in Refs.~\cite{cao2023flat,yang2023interlayer,qin2023high,wang2025fermi} proposes a two-component theory that highlights the distinct roles of the Ni $3d_{z^2}$ and $3d_{x^2-y^2}$ orbitals.
In this scenario, strong on-site Coulomb repulsion localizes the $3d_{z^2}$ electrons, driving the formation of local interlayer spin singlets via a strong $c$-axis superexchange interaction.
High-$T_c$ SC emerges when these localized singlets hybridize with the nearly quarter-filled, itinerant in-plane $3d_{x^2-y^2}$ electrons,
which mobilizes the pairs and establishes macroscopic phase coherence.
Furthermore, this inter-orbital hybridization plays a dual role, namely, it promotes global phase coherence but simultaneously competes with the local singlet pairing strength. 
This intrinsic competition naturally accounts for the experimentally observed non-monotonic variation of $T_c$ under pressure, a mechanism that has been further validated by auxiliary field Monte Carlo simulations of the minimal effective model ~\cite{qin2023high}.

There also exists a series of studies using dynamical mean-field theory ~\cite{ouyang2023hund,tian2023correlation,chen2024nonfermi,ouyang2025phase} propose a distinct multi-orbital picture. 
This approach identifies the primary pairing mechanism as magnetically mediated, driven by local inter-orbital spin coupling and Hund's rule effects. 
This model explicitly includes both the Ni $3d_{x^2-y^2}$ and $3d_{z^2}$ orbitals, which are intrinsically coupled in the superconducting state.
By focusing on a bilayer two-orbital Hubbard model, it suggests that a separate $\gamma$-band is not necessary for SC. 
A key finding is that the SC state exhibits $s_{\pm}$-wave symmetry for both orbitals, where the orthogonal yet correlated orbitals transition simultaneously to either a Mott insulator or a superconductor. 
This illustrates the crucial role of multi-orbital interactions and Hund's coupling in the pairing.

Finally, it should be emphasized that the efforts mentioned above represent only a selected subset of the extensive theoretical investigations into bilayer nickelate superconductors. 
The theoretical landscape of this field is exceptionally rich and diverse. 
Numerous other valuable works have approached the pairing mechanism from a wide array of perspectives, focusing on alternative microscopic interactions, orbital entanglements, and spin-charge dynamics, etc. 
While a unified consensus on the definitive pairing mechanism has not yet been reached, this multitude of theoretical viewpoints underscores a highly vibrant and flourishing research community. 
These continuous explorations are actively driving the field forward, revealing the remarkably complex and fascinating physics inherent in this novel superconducting family.

\section{Discussions}
\label{sec:disc}

The discovery of high-$T_c$ SC in La$_3$Ni$_2$O$_7$ marks a major breakthrough, establishing bilayer nickelates as a novel platform for understanding strongly correlated SC.
A central question in the field is identifying both the fundamental similarities and the distinct differences between these nickelates and the well-known cuprates.
Experimentally, bilayer nickelates exhibit several hallmarks of strong electronic correlations.
These include strange-metal behavior, pseudogap phenomena, charge and spin density waves, and an anomalous normal-state Hall coefficient, etc. 
However, due to the extreme challenges of high-pressure experimental techniques, the exact nature of these behavior remains an open question. 
The successful growth of superconducting thin films at ambient pressure provides a highly promising new avenue for experimental probing, though understanding how these films compare to pressurized bulk samples introduces further complexity. 
Furthermore, the exact pairing symmetry, specifically, whether it is an interlayer $s$-wave or a cuprate-like $d$-wave, remains a subject of intense debate. 
Ultimately, the unique bilayer lattice structure, the active two-orbital degrees of freedom, the strong Coulomb correlations, and the crucial role of Hund's coupling all intertwine, highlighting the uniquely complex and rich physical nature of bilayer nickelates.

In this review, we have specifically emphasized the critical importance of the on-site Hund's coupling $J_H$.
Physically, $J_H$ acts as a vital magnetic bridge. 
By energetically aligning the spins of the localized $3d_{z^2}$ and the itinerant $3d_{x^2-y^2}$ electrons on the same nickel atom, $J_H$ effectively transfers the robust interlayer AFM superexchange $J_{\perp}^z$ from the localized sector to the itinerant sector.
This dynamic transfer generates a massive effective interlayer coupling $J_{\perp}^x$ for the metallic $3d_{x^2-y^2}$ band. 
Consequently, this mechanism elegantly stabilizes a robust interlayer $s$-wave pairing state, which can uniquely survive even in the highly overdoped regime.

A comprehensive analysis of this mechanism reveals a strict division of labor between the two $E_g$ orbitals.
The $3d_{z^2}$ orbital, characterized by strong interlayer spatial hybridization and a proximity to half-filling Mott state, serves as the interaction mediator.
While these localized electrons form strong pre-formed interlayer singlets at high temperatures, their lack of mobility prevents them from establishing macroscopic phase coherence, naturally resulting in a pseudogap phase. 
Conversely, $3d_{x^2-y^2}$ orbital acts as the active superconducting carrier.
Driven by the magnetic pairing strength inherited from the $3d_{z^2}$ channel, it possesses the necessary kinetic mobility to establish long-range phase coherence.
This orbital-selective physical picture highlights that La$_3$Ni$_2$O$_7$ is a fundamentally multi-orbital system where the inter-orbital quantum interplay is just as critical as the intra-orbital transport.

From a broader perspective, the strong-coupling theory of SC highlights the indispensable role of local magnetic superexchange in the Mott regime. 
In this framework, superexchange serves as the pairing glue that binds electrons into pre-formed singlets. 
Furthermore, the high-temperature pseudogap phase is naturally understood as a state where local pairing exists without global phase coherence. 
These physical concepts are broadly shared with the cuprates. 
However, the specific bilayer geometry and the multi-orbital nature of the nickelates introduce profound differences and much richer microscopic details. 
Ultimately, bilayer nickelates provide an exciting new frontier in condensed matter physics. 
Unraveling the complete mysteries of their high-$T_c$ SC will continue to demand intense, collaborative efforts from both theory and experiment.

\section*{Acknowledgment}
C.W. is supported by the National Natural Science Foundation of China under the Grants Nos. 12234016 and 12174317.  
F.Y. is supported by the national natural science foundation of China under the Grant Nos. 12574141 and 12074031.
Z.P. is supported by the National Natural Science Foundation of China under Grant No. 12504219 and the startup funding from Xiamen University. 
C.L. is supported by the National Natural Science Foundation
of China under the Grants No. 12304180. 
This work has been supported by the New Cornerstone Science Foundation.

\twocolumngrid
\bibliography{references}

@article{anderson1987resonating,
  title={The resonating valence bond state in La2CuO4 and superconductivity},
  author={Anderson, Philip W},
  journal={Science},
  volume={235},
  number={4793},
  pages={1196--1198},
  year={1987},
  publisher={American Association for the Advancement of Science}, 
  url={https://www.science.org/doi/abs/10.1126/science.235.4793.1196}
}

@article{sun2023lno,
  title={Signatures of superconductivity near $80$K in a nickelate under high pressure},
  author={Sun, Hualei and Huo, Mengwu and Hu, Xunwu and Li, Jingyuan and Liu, Zengjia and Han, Yifeng and Tang, Lingyun and Mao, Zhongquan and Yang, Pengtao and Wang, Bosen and Cheng, Jinguang and Yao, Dao-Xin and Zhang, Guang-Ming and Wang, Meng},
  journal={Nature},
  volume={621},
  number={7979},
  pages={493--498},
  year={2023},
  publisher={Nature Publishing Group UK London}, 
  url={https://www.nature.com/articles/s41586-023-06408-7}
}

@article{chen2024electronic,
  title={Electronic and magnetic excitations in {L}a$_3${N}i$_2${O}$_7$},
  author={Chen, Xiaoyang and Choi, Jaewon and Jiang, Zhicheng and Mei, Jiong and Jiang, Kun and Li, Jie and Agrestini, Stefano and Garcia-Fernandez, Mirian and Sun, Hualei and Huang, Xing and Shen, Dawei and Wang, Meng and Hu, Jiangping and Lu, Yi and Zhou, Ke-Jin and Feng, Donglai},
  journal={Nat. Commun.},
  volume={15},
  number={1},
  pages={9597},
  year={2024},
  publisher={Nature Publishing Group UK London}, 
  url={https://www.nature.com/articles/s41467-024-53863-5}
}

@article{xie2024strong,
  title={Strong interlayer magnetic exchange coupling in {L}a$_3${N}i$_2${O}$_{7-\delta}$ revealed by inelastic neutron scattering},
  author={Xie, Tao and Huo, Mengwu and Ni, Xiaosheng and Shen, Feiran and Huang, Xing and Sun, Hualei and Walker, Helen C and Adroja, Devashibhai and Yu, Dehong and Shen, Bing and others},
  journal={Sci. Bull.},
  volume={69},
  number={20},
  pages={3221--3227},
  year={2024},
  publisher={Elsevier}, 
  url={https://www.sciencedirect.com/science/article/pii/S2095927324005164}
}

@article{lu2024interlayer,
  title={Interlayer-coupling-driven high-temperature superconductivity in {L}a$_3${N}i$_2${O}$_{7}$ under pressure},
  author={Lu, Chen and Pan, Zhiming and Yang, Fan and Wu, Congjun},
  journal={Phys. Rev. Lett.},
  volume={132},
  number={14},
  pages={146002},
  year={2024},
  publisher={APS}, 
  url={https://journals.aps.org/prl/abstract/10.1103/PhysRevLett.132.146002}
}

@article{qu2024bilayer,
  title={Bilayer $t$-$J$-$J_{\perp}$ model and magnetically mediated pairing in the pressurized nickelate {L}a$_3${N}i$_2${O}$_{7}$},
  author={Qu, Xing-Zhou and Qu, Dai-Wei and Chen, Jialin and Wu, Congjun and Yang, Fan and Li, Wei and Su, Gang},
  journal={Phys. Rev. Lett.},
  volume={132},
  number={3},
  pages={036502},
  year={2024},
  publisher={APS}, 
  url={https://journals.aps.org/prl/abstract/10.1103/PhysRevLett.132.036502}
}

@article{fan2024sc,
  title={Superconductivity in nickelate and cuprate superconductors with strong bilayer coupling},
  author={Fan, Zhen and Zhang, Jian-Feng and Zhan, Bo and Lv, Dingshun and Jiang, Xing-Yu and Normand, Bruce and Xiang, Tao},
  journal={Phys. Rev. B},
  volume={110},
  number={2},
  pages={024514},
  year={2024},
  publisher={APS}, 
  url={https://journals.aps.org/prb/abstract/10.1103/PhysRevB.110.024514}
}

@article{jiang2024pressure,
  title={Pressure driven fractionalization of ionic spins results in cupratelike high-$T_c$ superconductivity in La 3 Ni 2 O 7},
  author={Jiang, Ruoshi and Hou, Jinning and Fan, Zhiyu and Lang, Zi-Jian and Ku, Wei},
  journal={Phys. Rev. Lett.},
  volume={132},
  number={12},
  pages={126503},
  year={2024},
  publisher={APS}, 
  url={https://journals.aps.org/prl/abstract/10.1103/PhysRevLett.132.126503}
}

@article{li2019sc,
  title={Superconductivity in an infinite-layer nickelate},
  author={Li, Danfeng and Lee, Kyuho and Wang, Bai Yang and Osada, Motoki and Crossley, Samuel and Lee, Hye Ryoung and Cui, Yi and Hikita, Yasuyuki and Hwang, Harold Y},
  journal={Nature},
  volume={572},
  number={7771},
  pages={624--627},
  year={2019},
  publisher={Nature Publishing Group UK London}, 
  url={https://www.nature.com/articles/s41586-019-1496-5}
}

@article{nomura2022sc,
  title={Superconductivity in infinite-layer nickelates},
  author={Nomura, Yusuke and Arita, Ryotaro},
  journal={Rep. Prog. Phys.},
  volume={85},
  number={5},
  pages={052501},
  year={2022},
  publisher={IOP Publishing},
  url={https://iopscience.iop.org/article/10.1088/1361-6633/ac5a60/}
}

@article{tsuei2000pairing,
  title={Pairing symmetry in cuprate superconductors},
  author={Tsuei, CC and Kirtley, JR},
  journal={Rev. Mod. Phys.},
  volume={72},
  number={4},
  pages={969},
  year={2000},
  publisher={APS}, 
  url={https://journals.aps.org/rmp/abstract/10.1103/RevModPhys.72.969}
}

@article{lee2006doping,
  title={Doping a Mott insulator: Physics of high-temperature superconductivity},
  author={Lee, Patrick A and Nagaosa, Naoto and Wen, Xiao-Gang},
  journal={Rev. Mod. Phys.},
  volume={78},
  number={1},
  pages={17--85},
  year={2006},
  publisher={APS}, 
  url={https://journals.aps.org/rmp/abstract/10.1103/RevModPhys.78.17}
}

@article{keimer2015quantum,
  title={From quantum matter to high-temperature superconductivity in copper oxides},
  author={Keimer, Bernhard and Kivelson, Steven A and Norman, Michael R and Uchida, Shinichi and Zaanen, J},
  journal={Nature},
  volume={518},
  number={7538},
  pages={179--186},
  year={2015},
  publisher={Nature Publishing Group UK London}, 
  url={https://www.nature.com/articles/nature14165}
}

@article{proust2019remarkable,
  title={The remarkable underlying ground states of cuprate superconductors},
  author={Proust, Cyril and Taillefer, Louis},
  journal={Annu. Rev. Condens. Matter Phys.},
  volume={10},
  number={1},
  pages={409--429},
  year={2019},
  publisher={Annual Reviews}, 
  url={https://www.annualreviews.org/content/journals/10.1146/annurev-conmatphys-031218-013210}
}

@article{bhatt2025resolving,
  title={Resolving Structural Origins for Superconductivity in Strain-Engineered La$_3$Ni$_2$O$_7$ Thin Films},
  author={Bhatt, Lopa and Jiang, Abigail Y and Ko, Eun Kyo and Schnitzer, Noah and Pan, Grace A and Segedin, Dan Ferenc and Liu, Yidi and Yu, Yijun and Zhao, Yi-Feng and Morales, Edgar Abarca and others},
  journal={arXiv:2501.08204},
  year={2025},
  url={https://arxiv.org/abs/2501.08204}
}

@article{zhong2025epitaxial,
  title={Epitaxial strain tuning of electronic and spin excitations in La$_3$Ni$_2$O$_7$ thin films},
  author={Zhong, Hengyang and Wei, Yuan and Zhang, Zhijia and Liu, Ruixian and Huang, Xinru and Ni, Xiao-Sheng and Cantarino, Marli dos Reis and Cao, Kun and Nie, Yuefeng and Schmitt, Thorsten and others},
  journal={aarXiv:2502.03178},
  year={2025},
  url={https://arxiv.org/abs/2502.03178}
}

@article{ko2024sign,
  title={Signatures of ambient pressure superconductivity in thin film {L}a$_3${N}i$_2${O}$_7$},
  author={Ko, Eun Kyo and Yu, Yijun and Liu, Yidi and Bhatt, Lopa and Li, Jiarui and Thampy, Vivek and Kuo, Cheng-Tai and Wang, Bai Yang and Lee, Yonghun and Lee, Kyuho and others},
  journal={Nature},
  volume={638},
  pages={935--940},
  year={2025},
  publisher={Nature Publishing Group UK London},
  url={https://www.nature.com/articles/s41586-024-08525-3}
}

@article{zhou2025ambient,
  title={Ambient-pressure superconductivity onset above 40 K in (La, Pr) 3Ni2O7 films},
  author={Zhou, Guangdi and Lv, Wei and Wang, Heng and Nie, Zihao and Chen, Yaqi and Li, Yueying and Huang, Haoliang and Chen, Wei-Qiang and Sun, Yu-Jie and Xue, Qi-Kun and others},
  journal={Nature},
  volume={640},
  number={8059},
  pages={641--646},
  year={2025},
  publisher={Nature Publishing Group UK London},
  url={https://www.nature.com/articles/s41586-025-08755-z}
}

@article{wang2024pressure,
  title={Pressure-induced superconductivity in polycrystalline {L}a$_3${N}i$_2${O}$_{7-\delta}$},
  author={Wang, Gang and Wang, NN and Shen, XL and Hou, J and Ma, L and Shi, LF and Ren, ZA and Gu, YD and Ma, HM and Yang, PT and others},
  journal={Phys. Rev. X},
  volume={14},
  number={1},
  pages={011040},
  year={2024},
  publisher={APS},
  url={https://journals.aps.org/prx/abstract/10.1103/PhysRevX.14.011040}
}

@article{wang2024bulk,
  title={Bulk high-temperature superconductivity in pressurized tetragonal {L}a$_2${P}r{N}i$_2${O}$_7$},
  author={Wang, Ningning and Wang, Gang and Shen, Xiaoling and Hou, Jun and Luo, Jun and Ma, Xiaoping and Yang, Huaixin and Shi, Lifen and Dou, Jie and Feng, Jie and others},
  journal={Nature},
  volume={634},
  number={8034},
  pages={579--584},
  year={2024},
  publisher={Nature Publishing Group UK London},
  url={https://www.nature.com/articles/s41586-024-07996-8}
}

@article{liu2025sc,
  title={Superconductivity and normal-state transport in compressively strained La$_2$PrNi$_2$O$_7$ thin films},
  author={Liu, Yidi and Ko, Eun Kyo and Tarn, Yaoju and Bhatt, Lopa and Li, Jiarui and Thampy, Vivek and Goodge, Berit H and Muller, David A and Raghu, Srinivas and Yu, Yijun and others},
  journal={Nat. Mater.},
  volume={24},
  pages={1221–1227},
  year={2025},
  publisher={Nature Publishing Group UK London},
  url={https://www.nature.com/articles/s41563-025-02258-y}
}

@article{zhang2024high,
  title={High-temperature superconductivity with zero resistance and strange-metal behaviour in {L}a$_3${N}i$_2${O}$_{7-\delta}$},
  author={Zhang, Yanan and Su, Dajun and Huang, Yanen and Shan, Zhaoyang and Sun, Hualei and Huo, Mengwu and Ye, Kaixin and Zhang, Jiawen and Yang, Zihan and Xu, Yongkang and others},
  journal={Nat. Phys.},
  volume={20},
  number={8},
  pages={1269--1273},
  year={2024},
  publisher={Nature Publishing Group UK London},
  url={https://www.nature.com/articles/s41567-024-02515-y}
}

@article{yue2025correlated,
  author = {Yue, Changming and Miao, Jian-Jian and Huang, Haoliang and Hua, Yichen and Li, Peng and Li, Yueying and Zhou, Guangdi and Lv, Wei and Yang, Qishuo and Yang, Fan and Sun, Hongyi and Sun, Yu-Jie and Lin, Junhao and Xue, Qi-Kun and Chen, Zhuoyu and Chen, Wei-Qiang},
  title = {Correlated electronic structures and unconventional superconductivity in bilayer nickelate heterostructures},
  journal = {Nat. Sci. Rev.},
  pages = {nwaf253},
  year = {2025},
  month = {06},
  issn = {2095-5138},
  doi = {10.1093/nsr/nwaf253},
  url = {https://doi.org/10.1093/nsr/nwaf253}
}

@article{wang2025mottness,
  title={The Mottness and the Anderson localization in bilayer nickelate La3Ni2O7},
  author={Wang, Yuxin and Chen, Ziyan and Zhang, Yi and Jiang, Kun and Hu, Jiangping},
  journal={arXiv:2501.08536},
  year={2025},
  url={https://arxiv.org/abs/2501.08536}
}

@article{li2025angle,
  author = {Li, Peng and Zhou, Guangdi and Lv, Wei and Li, Yueying and Yue, Changming and Huang, Haoliang and Xu, Lizhi and Shen, Jianchang and Miao, Yu and Song, Wenhua and Nie, Zihao and Chen, Yaqi and Wang, Heng and Chen, Weiqiang and Huang, Yaobo and Chen, Zhen-Hua and Qian, Tian and Lin, Junhao and He, Junfeng and Sun, Yu-Jie and Chen, Zhuoyu and Xue, Qi-Kun},
  title = {Angle-resolved photoemission spectroscopy of superconducting (La,Pr)$_3$Ni$_2$O$_7$/SrLaAlO$_4$ heterostructures},
  journal = {Nat. Sci. Rev.},
  pages = {nwaf205},
  year = {2025},
  month = {05},
  issn = {2095-5138},
  doi = {10.1093/nsr/nwaf205},
  url = {https://doi.org/10.1093/nsr/nwaf205}
}

@article{shao2025band,
  title = {Band structure and pairing nature of La$_{3}$Ni$_{2}$O$_{7}$ thin film at ambient pressure},
  author = {Shao, Zhi-Yan and Liu, Yu-Bo and Liu, Min and Yang, Fan},
  journal = {Phys. Rev. B},
  volume = {112},
  issue = {2},
  pages = {024506},
  numpages = {11},
  year = {2025},
  month = {Jul},
  publisher = {American Physical Society},
  url = {https://link.aps.org/doi/10.1103/9t6n-jqr5}
}

@article{li2026bulk,
  title={Bulk superconductivity up to 96 K in pressurized nickelate single crystals},
  author={Li, Feiyu and Xing, Zhenfang and Peng, Di and Dou, Jie and Guo, Ning and Ma, Liang and Zhang, Yulin and Wang, Lingzhen and Luo, Jun and Yang, Jie and others},
  journal={Nature},
  volume={649},
  pages={871--878},
  year={2026},
  publisher={Nature Publishing Group},
  url={https://www.nature.com/articles/s41586-025-09954-4}
}

@article{liu2024origin,
  title = {Origin of the diagonal double-stripe spin density wave and potential superconductivity in bulk ${\mathrm{La}}_{3}{\mathrm{Ni}}_{2}{\mathrm{O}}_{7}$ at ambient pressure},
  author = {Liu, Yu-Bo and Sun, Hongyi and Zhang, Ming and Liu, Qihang and Chen, Wei-Qiang and Yang, Fan},
  journal = {Phys. Rev. B},
  volume = {112},
  issue = {1},
  pages = {014510},
  numpages = {13},
  year = {2025},
  month = {Jul},
  publisher = {American Physical Society},
  url = {https://link.aps.org/doi/10.1103/24f4-349n}
}

@article{li2025direct,
  title={Direct Visualization of an Incommensurate Unidirectional Charge Density Wave in La$_4$Ni$_3$O$_{10}$},
  author={Li, Mingzhe and Gong, Jiashuo and Zhu, Yinghao and Chen, Ziyuan and Zhang, Jiakang and Zhang, Enkang and Li, Yuanji and Yin, Ruotong and Wang, Shiyuan and Zhao, Jun and others},
  journal={arXiv:2501.18885},
  year={2025},
  url={https://arxiv.org/abs/2501.18885}
}

@article{liu2023evidence,
  title={Evidence for charge and spin density waves in single crystals of {L}a$_3${N}i$_2${O}$_7$ and {L}a$_3${N}i$_2${O}$_6$},
  author={Liu, Zengjia and Sun, Hualei and Huo, Mengwu and Ma, Xiaoyan and Ji, Yi and Yi, Enkui and Li, Lisi and Liu, Hui and Yu, Jia and Zhang, Ziyou and Chen, Zhiqiang and Liang, Feixiang and Dong, Hongliang and Guo, Hanjie and Zhong, Dingyong and Shen, Bing and Li, Shiliang and Wang, Meng},
  journal={Sci. China-Phys. Mech. Astron.},
  volume={66},
  number={1},
  pages={217411},
  year={2023},
  publisher={Springer},
  url = {https://link.springer.com/article/10.1007/s11433-022-1962-4}
}

@article{zhang1994synthesis,
  title={Synthesis, structure, and properties of the layered perovskite La3Ni2O7-$\delta$},
  author={Zhang, Z and Greenblatt, M and Goodenough, JB},
  journal={Journal of Solid State Chemistry},
  volume={108},
  number={2},
  pages={402--409},
  year={1994},
  publisher={Elsevier},
  url={https://www.sciencedirect.com/science/article/pii/S0022459684710590}
}

@article{liu2024ele,
  title={Electronic correlations and partial gap in the bilayer nickelate {L}a$_3${N}i$_2${O}$_7$},
  author={Liu, Zhe and Huo, Mengwu and Li, Jie and Li, Qing and Liu, Yuecong and Dai, Yaomin and Zhou, Xiaoxiang and Hao, Jiahao and Lu, Yi and Wang, Meng and Wen, Hai-Hu},
  journal={Nat. Commun.},
  volume={15},
  number={1},
  pages={7570},
  year={2024},
  publisher={Nature Publishing Group UK London},
  url={https://www.nature.com/articles/s41467-024-52001-5}
}

@article{hou2023emergence,
   author = {Jun Hou and Peng-Tao Yang and Zi-Yi Liu and Jing-Yuan Li and Peng-Fei Shan and Liang Ma and Gang Wang and Ning-Ning Wang and Hai-Zhong Guo and Jian-Ping Sun and Yoshiya Uwatoko and Meng Wang and Guang-Ming Zhang and Bo-Sen Wang and Jin-Guang Cheng},
   title = {Emergence of High-Temperature Superconducting Phase in Pressurized La$_{3}$Ni$_{2}$O$_7$ Crystals},
   publisher = {Chin. Phys. Lett.},
   year = {2023},
   journal = {Chin. Phys. Lett.},
   volume = {40},
   number = {11},
   eid = {117302},
   pages = {117302},
   url = {https://cpl.iphy.ac.cn/EN/abstract/article_116425.shtml},
   doi = {10.1088/0256-307X/40/11/117302}
}

@article{yang2023arpes,
  title={Orbital-dependent electron correlation in double-layer nickelate {L}a$_3${N}i$_2${O}$_7$},
  author={Yang, Jiangang and Sun, Hualei and Hu, Xunwu and Xie, Yuyang and Miao, Taimin and Luo, Hailan and Chen, Hao and Liang, Bo and Zhu, Wenpei and Qu, Gexing and others},
  journal={Nat. Commun.},
  volume={15},
  number={1},
  pages={4373},
  year={2024},
  publisher={Nature Publishing Group UK London},
  url={https://www.nature.com/articles/s41467-024-48701-7}
}

@article{zhang2023pressure,
  title={Effects of pressure and doping on Ruddlesden-Popper phases {L}a$_{n+1}${N}i$_n${O}$_{3n+1}$},
  author={Zhang, Mingxin and Pei, Cuiying and Wang, Qi and Zhao, Yi and Li, Changhua and Cao, Weizheng and Zhu, Shihao and Wu, Juefei and Qi, Yanpeng},
  journal={Journal of Materials Science \& Technology},
  volume={185},
  pages={147--154},
  year={2024},
  publisher={Elsevier},
  url={https://www.sciencedirect.com/science/article/pii/S1005030223009829}
}

@article{wang2023la2prnio7,
  title={Observation of high-temperature superconductivity in the high-pressure tetragonal phase of {L}a$_2${P}r{N}i$_2${O}$_{7-\delta}$}, 
  author={Gang Wang and Ningning Wang and Yuxin Wang and Lifen Shi and Xiaoling Shen and Jun Hou and Hanming Ma and Pengtao Yang and Ziyi Liu and Hua Zhang and Xiaoli Dong and Jianping Sun and Bosen Wang and Kun Jiang and Jiangping Hu and Yoshiya Uwatoko and Jinguang Cheng},
  journal={arXiv:2311.08212},
  url = {https://arxiv.org/abs/2311.08212},
  year={2023}
}

@article{wang2023structure,
  title={Structure Responsible for the Superconducting State in {L}a$_3${N}i$_2${O}$_7$ at High-Pressure and Low-Temperature Conditions},
  author={Wang, Luhong and Li, Yan and Xie, Sheng-Yi and Liu, Fuyang and Sun, Hualei and Huang, Chaoxin and Gao, Yang and Nakagawa, Takeshi and Fu, Boyang and Dong, Bo and others},
  journal={J. Am. Chem. Soc.},
  volume={146},
  number={11},
  pages={7506--7514},
  year={2024},
  publisher={ACS Publications},
  url={https://pubs.acs.org/doi/abs/10.1021/jacs.3c13094}
}

@article{zhang2026spin,
  title={Spin-density-wave transition in monolayer-trilayer La3Ni2O7 single crystals},
  author={Zhang, Mingxin and Dou, Jie and Peng, Di and Pei, Cuiying and Wang, Qi and Zhao, Yi and Xiong, Chao and Li, Shuo and Luo, Jun and Wu, Juefei and others},
  journal={arXiv:2601.13090},
  year={2026},
  url = {https://arxiv.org/abs/2601.13090}
}

@article{zhang2025ident,
  title={Identifying the structure of La3Ni2O7 in the pressurized superconducting state},
  author={Zhang, Hengyuan and Zhang, Jielong and Huo, Mengwu and Chen, Junfeng and Hu, Deyuan and Yao, Dao-Xin and Sun, Hualei and Cao, Kun and Wang, Meng},
  journal={arXiv:2511.15265},
  year={2025},
  url = {https://arxiv.org/abs/2511.15265}
}

@article{liu2025optimally,
  title={Optimally Tensile Strained La3Ni2O7 Films as Candidate High-Temperature Superconductors on Designer Ba1-xSrxO (001) and SrO-SrTiO3 Substrates},
  author={Liu, Liangliang and Peng, Junhao and Qiao, Zhuangzhuang and Cai, Shuo and Dong, Huafeng and Jia, Yu and Zhang, Zhenyu},
  journal={arXiv:2509.13820},
  year={2025},
  url = {https://arxiv.org/abs/2509.13820}
}

@article{cao2025direct,
  title={Direct Observation of $d$-Wave Superconducting Gap Symmetry in Pressurized La$_3$Ni$_2$O$_7$-delta Single Crystals},
  author={Cao, Zi-Yu and Peng, Di and Choi, Seokmin and Lan, Fujun and Yu, Lan and Zhang, Enkang and Xing, Zhenfang and Liu, Yuxin and Zhang, Feiyang and Luo, Tao and others},
  journal={arXiv:2509.12606},
  year={2025},
  url = {https://arxiv.org/abs/2509.12606}
}

@article{zhou2023evidence,
  title={Investigations of key issues on the reproducibility of high-$T_c$ superconductivity emerging from compressed {L}a$_3${N}i$_2${O}$_7$},
  author={Zhou, Yazhou and Guo, Jing and Cai, Shu and Sun, Hualei and Li, Chengyu and Zhao, Jinyu and Wang, Pengyu and Han, Jinyu and Chen, Xintian and Chen, Yongjin and others},
  journal={Matter and Radiation at Extremes},
  volume={10},
  number={2},
  year={2025},
  publisher={AIP Publishing},
  url={https://pubs.aip.org/aip/mre/article/10/2/027801/3331819}
}

@article{cui2023strain,
  title={Strain-mediated phase crossover in Ruddlesden--Popper nickelates},
  author={Cui, Ting and Choi, Songhee and Lin, Ting and Liu, Chen and Wang, Gang and Wang, Ningning and Chen, Shengru and Hong, Haitao and Rong, Dongke and Wang, Qianying and others},
  journal={Commun. Mater.},
  volume={5},
  number={1},
  pages={32},
  year={2024},
  publisher={Nature Publishing Group UK London},
  url={https://www.nature.com/articles/s43246-024-00478-4}
}

@article{sui2023rno,
  title={Electronic properties of the bilayer nickelates {R}$_3${N}i$_2${O}$_7$ with oxygen vacancies ({R}= {L}a or {C}e)},
  author={Sui, Xuelei and Han, Xiangru and Jin, Heng and Chen, Xiaojun and Qiao, Liang and Shao, Xiaohong and Huang, Bing},
  journal={Phys. Rev. B},
  volume={109},
  number={20},
  pages={205156},
  year={2024},
  publisher={APS},
  url={https://journals.aps.org/prb/abstract/10.1103/PhysRevB.109.205156}
}

@article{luo2023bilayer,
  title={Bilayer two-orbital model of {L}a$_3${N}i$_2${O}$_7$ under pressure},
  author={Luo, Zhihui and Hu, Xunwu and Wang, Meng and W{\'u}, W{\'e}i and Yao, Dao-Xin},
  journal={Phys. Rev. Lett.},
  volume={131},
  number={12},
  pages={126001},
  year={2023},
  publisher={APS},
  url={https://journals.aps.org/prl/abstract/10.1103/PhysRevLett.131.126001}
}

@article{zhang2023electronic,
  title = {Electronic structure, dimer physics, orbital-selective behavior, and magnetic tendencies in the bilayer nickelate superconductor {L}a$_3${N}i$_2${O}$_7$ under pressure},
  author = {Zhang, Yang and Lin, Ling-Fang and Moreo, Adriana and Dagotto, Elbio},
  journal = {Phys. Rev. B},
  volume = {108},
  issue = {18},
  pages = {L180510},
  numpages = {5},
  year = {2023},
  month = {Nov},
  publisher = {American Physical Society},
  doi = {10.1103/PhysRevB.108.L180510},
  url = {https://link.aps.org/doi/10.1103/PhysRevB.108.L180510}
}

@article{yang2023possible,
  title = {Possible ${S}_{\pm}$-wave superconductivity in {L}a$_3${N}i$_2${O}$_7$},
  author = {Yang, Qing-Geng and Wang, Da and Wang, Qiang-Hua},
  journal = {Phys. Rev. B},
  volume = {108},
  issue = {14},
  pages = {L140505},
  numpages = {5},
  year = {2023},
  month = {Oct},
  publisher = {American Physical Society},
  doi = {10.1103/PhysRevB.108.L140505},
  url = {https://link.aps.org/doi/10.1103/PhysRevB.108.L140505}
}

@article{lechermann2023,
  title = {Electronic correlations and superconducting instability in {L}a$_3${N}i$_2${O}$_7$ under high pressure},
  author = {Lechermann, Frank and Gondolf, Jannik and B\"otzel, Steffen and Eremin, Ilya M.},
  journal = {Phys. Rev. B},
  volume = {108},
  issue = {20},
  pages = {L201121},
  numpages = {6},
  year = {2023},
  month = {Nov},
  publisher = {American Physical Society},
  doi = {10.1103/PhysRevB.108.L201121},
  url = {https://link.aps.org/doi/10.1103/PhysRevB.108.L201121}
}

@article{sakakibara2024possible,
  title = {Possible High ${T}_{c}$ Superconductivity in {L}a$_3${N}i$_2${O}$_7$ under High Pressure through Manifestation of a Nearly Half-Filled Bilayer Hubbard Model},
  author = {Sakakibara, Hirofumi and Kitamine, Naoya and Ochi, Masayuki and Kuroki, Kazuhiko},
  journal = {Phys. Rev. Lett.},
  volume = {132},
  issue = {10},
  pages = {106002},
  numpages = {6},
  year = {2024},
  month = {Mar},
  publisher = {American Physical Society},
  doi = {10.1103/PhysRevLett.132.106002},
  url = {https://link.aps.org/doi/10.1103/PhysRevLett.132.106002}
}

@article{gu2023effective,
  title = {Effective model and pairing tendency in the bilayer Ni-based superconductor ${\mathrm{La}}_{3}{\mathrm{Ni}}_{2}{\mathrm{O}}_{7}$},
  author = {Gu, Yuhao and Le, Congcong and Yang, Zhesen and Wu, Xianxin and Hu, Jiangping},
  journal = {Phys. Rev. B},
  volume = {111},
  issue = {17},
  pages = {174506},
  numpages = {7},
  year = {2025},
  month = {May},
  publisher = {American Physical Society},
  url = {https://link.aps.org/doi/10.1103/PhysRevB.111.174506}
}

@article{shen2023effective,
  title={Effective Bi-Layer Model Hamiltonian and Density-Matrix Renormalization Group Study for the High-${T}_c$ Superconductivity {L}a$_3${N}i$_2${O}$_7$ under High Pressure},
  author={Shen, Yang and Qin, Mingpu and Zhang, Guang-Ming},
  journal={Chin. Phys. Lett.},
  volume={40},
  number={12},
  pages={127401},
  year={2023},
  publisher={Chinese Physical Society},
  url={https://iopscience.iop.org/article/10.1088/0256-307X/40/12/127401}
}

@article{christiansson2023correlated,
  title = {Correlated Electronic Structure of {L}a$_3${N}i$_2${O}$_7$ under Pressure},
  author = {Christiansson, Viktor and Petocchi, Francesco and Werner, Philipp},
  journal = {Phys. Rev. Lett.},
  volume = {131},
  issue = {20},
  pages = {206501},
  numpages = {6},
  year = {2023},
  month = {Nov},
  publisher = {American Physical Society},
  doi = {10.1103/PhysRevLett.131.206501},
  url = {https://link.aps.org/doi/10.1103/PhysRevLett.131.206501}
}

@article{shilenko2023correlated,
  title = {Correlated electronic structure, orbital-selective behavior, and magnetic correlations in double-layer {L}a$_3${N}i$_2${O}$_7$ under pressure},
  author = {Shilenko, D. A. and Leonov, I. V.},
  journal = {Phys. Rev. B},
  volume = {108},
  issue = {12},
  pages = {125105},
  numpages = {9},
  year = {2023},
  month = {Sep},
  publisher = {American Physical Society},
  doi = {10.1103/PhysRevB.108.125105},
  url = {https://link.aps.org/doi/10.1103/PhysRevB.108.125105}
}

@article{wu2024superexchange,
  title={Superexchange and charge transfer in the nickelate superconductor {L}a$_3${N}i$_2${O}$_7$ under pressure},
  author={W{\'u}, W{\'e}i and Luo, Zhihui and Yao, Dao-Xin and Wang, Meng},
  journal={Sci. China-Phys. Mech. Astron.},
  volume={67},
  number={11},
  pages={117402},
  year={2024},
  publisher={Springer},
  url={https://link.springer.com/article/10.1007/s11433-023-2300-4}
}

@article{cao2023flat,
  title = {Flat bands promoted by Hund's rule coupling in the candidate double-layer high-temperature superconductor {L}a$_3${N}i$_2${O}$_7$ under high pressure},
  author = {Cao, Yingying and Yang, Yi-feng},
  journal = {Phys. Rev. B},
  volume = {109},
  issue = {8},
  pages = {L081105},
  numpages = {6},
  year = {2024},
  month = {Feb},
  publisher = {American Physical Society},
  doi = {10.1103/PhysRevB.109.L081105},
  url = {https://link.aps.org/doi/10.1103/PhysRevB.109.L081105}
}

@article{chen2023critical,
  title={Charge and spin instabilities in superconducting {L}a$_3${N}i$_2${O}$_7$},
  author={Chen, Xuejiao and Jiang, Peiheng and Li, Jie and Zhong, Zhicheng and Lu, Yi},
  journal={Phys. Rev. B},
  volume={111},
  number={1},
  pages={014515},
  year={2025},
  publisher={APS},
  url={https://journals.aps.org/prb/abstract/10.1103/PhysRevB.111.014515}
}

@article{liu2023spm,
  title = {s$^{\pm}$-Wave Pairing and the Destructive Role of Apical-Oxygen Deficiencies in {L}a$_3${N}i$_2${O}$_7$ under Pressure},
  author = {Liu, Yu-Bo and Mei, Jia-Wei and Ye, Fei and Chen, Wei-Qiang and Yang, Fan},
  journal = {Phys. Rev. Lett.},
  volume = {131},
  issue = {23},
  pages = {236002},
  numpages = {6},
  year = {2023},
  month = {Dec},
  publisher = {American Physical Society},
  doi = {10.1103/PhysRevLett.131.236002},
  url = {https://link.aps.org/doi/10.1103/PhysRevLett.131.236002}
}

@article{oh2023type2,
  title = {Type-{II} $t$-${J}$ model and shared superexchange coupling from Hund's rule in superconducting {L}a$_3${N}i$_2${O}$_7$},
  author = {Oh, Hanbit and Zhang, Ya-Hui},
  journal = {Phys. Rev. B},
  volume = {108},
  issue = {17},
  pages = {174511},
  numpages = {8},
  year = {2023},
  month = {Nov},
  publisher = {American Physical Society},
  doi = {10.1103/PhysRevB.108.174511},
  url = {https://link.aps.org/doi/10.1103/PhysRevB.108.174511}
}

@article{zhang2023structural,
  title={Structural phase transition, $s_{\pm}$-wave pairing, and magnetic stripe order in bilayered superconductor La3Ni2O7 under pressure},
  author={Zhang, Yang and Lin, Ling-Fang and Moreo, Adriana and Maier, Thomas A and Dagotto, Elbio},
  journal={Nat. Commun.},
  volume={15},
  number={1},
  pages={2470},
  year={2024},
  publisher={Nature Publishing Group UK London},
  url={https://www.nature.com/articles/s41467-024-46622-z}
}

@article{liao2023electron,
  title={Electron correlations and superconductivity in {L}a$_3${N}i$_2${O}$_7$ under pressure tuning},
  author={Liao, Zhiguang and Chen, Lei and Duan, Guijing and Wang, Yiming and Liu, Changle and Yu, Rong and Si, Qimiao},
  journal={Phys. Rev. B},
  volume={108},
  number={21},
  pages={214522},
  year={2023},
  publisher={APS},
  url={https://journals.aps.org/prb/abstract/10.1103/PhysRevB.108.214522}
}

@article{yang2023interlayer,
  title = {Interlayer valence bonds and two-component theory for high-${T}_{c}$ superconductivity of {L}a$_3${N}i$_2${O}$_7$ under pressure},
  author = {Yang, Yi-feng and Zhang, Guang-Ming and Zhang, Fu-Chun},
  journal = {Phys. Rev. B},
  volume = {108},
  issue = {20},
  pages = {L201108},
  numpages = {6},
  year = {2023},
  month = {Nov},
  publisher = {American Physical Society},
  doi = {10.1103/PhysRevB.108.L201108},
  url = {https://link.aps.org/doi/10.1103/PhysRevB.108.L201108}
}

@article{jiang2023high,
  title={High temperature superconductivity in {L}a$_3${N}i$_2${O}$_7$},
  author={Jiang, Kun and Wang, Ziqiang and Zhang, Fu-Chun},
  journal={Chin. Phys. Lett.},
  year={2023},
  url={https://iopscience.iop.org/article/10.1088/0256-307X/41/1/017402}
}

@article{zhang2023trends,
  title = {Trends in electronic structures and $s_{\pm}$-wave pairing for the rare-earth series in bilayer nickelate superconductor ${R}_3${N}i$_2${O}$_7$},
  author = {Zhang, Yang and Lin, Ling-Fang and Moreo, Adriana and Maier, Thomas A. and Dagotto, Elbio},
  journal = {Phys. Rev. B},
  volume = {108},
  issue = {16},
  pages = {165141},
  numpages = {8},
  year = {2023},
  month = {Oct},
  publisher = {American Physical Society},
  doi = {10.1103/PhysRevB.108.165141},
  url = {https://link.aps.org/doi/10.1103/PhysRevB.108.165141}
}

@article{huang2023impurity,
  title = {Impurity and vortex states in the bilayer high-temperature superconductor {L}a$_3${N}i$_2${O}$_7$},
  author = {Huang, Junkang and Wang, Z. D. and Zhou, Tao},
  journal = {Phys. Rev. B},
  volume = {108},
  issue = {17},
  pages = {174501},
  numpages = {7},
  year = {2023},
  month = {Nov},
  publisher = {American Physical Society},
  doi = {10.1103/PhysRevB.108.174501},
  url = {https://link.aps.org/doi/10.1103/PhysRevB.108.174501}
}

@article{qin2023high,
  title = {High-${T}_{c}$ superconductivity by mobilizing local spin singlets and possible route to higher ${T}_{c}$ in pressurized {L}a$_3${N}i$_2${O}$_7$},
  author = {Qin, Qiong and Yang, Yi-Feng},
  journal = {Phys. Rev. B},
  volume = {108},
  issue = {14},
  pages = {L140504},
  numpages = {6},
  year = {2023},
  month = {Oct},
  publisher = {American Physical Society},
  doi = {10.1103/PhysRevB.108.L140504},
  url = {https://link.aps.org/doi/10.1103/PhysRevB.108.L140504}
}

@article{tian2023correlation,
  title = {Correlation effects and concomitant two-orbital ${s}_{\pm}$-wave superconductivity in {L}a$_3${N}i$_2${O}$_7$ under high pressure},
  author = {Tian, Yi-Heng and Chen, Yin and Wang, Jia-Ming and He, Rong-Qiang and Lu, Zhong-Yi},
  journal = {Phys. Rev. B},
  volume = {109},
  issue = {16},
  pages = {165154},
  numpages = {6},
  year = {2024},
  month = {Apr},
  publisher = {American Physical Society},
  doi = {10.1103/PhysRevB.109.165154},
  url = {https://link.aps.org/doi/10.1103/PhysRevB.109.165154}
}

@article{lu2023sc,
  title={Superconductivity from Doping Symmetric Mass Generation Insulators: Application to {L}a$_3${N}i$_2${O}$_7$ under Pressure},
  author={Lu, Da-Chuan and Li, Miao and Zeng, Zhao-Yi and Hou, Wanda and Wang, Juven and Yang, Fan and You, Yi-Zhuang},
  journal={arXiv:2308.11195},
  year={2023},
  url = {https://arxiv.org/abs/2308.11195}
}

@article{kitamine2023,
  title = {Designing nickelate superconductors with ${d}^{8}$ configuration exploiting mixed-anion strategy},
  author = {Kitamine, Naoya and Ochi, Masayuki and Kuroki, Kazuhiko},
  journal = {Phys. Rev. Res.},
  volume = {2},
  issue = {4},
  pages = {042032},
  numpages = {7},
  year = {2020},
  month = {Nov},
  publisher = {American Physical Society},
  doi = {10.1103/PhysRevResearch.2.042032},
  url = {https://link.aps.org/doi/10.1103/PhysRevResearch.2.042032}
}

@article{luo2023high,
  title={High-$T_c$ superconductivity in La3Ni2O7 based on the bilayer two-orbital tJ model},
  author={Luo, Zhihui and Lv, Biao and Wang, Meng and W{\'u}, W{\'e}i and Yao, Dao-Xin},
  journal={npj Quantum Mater.},
  volume={9},
  number={1},
  pages={61},
  year={2024},
  publisher={Nature Publishing Group UK London},
  url={https://www.nature.com/articles/s41535-024-00668-w}
}

@article{zhang2023strong,
  title={Strong pairing originated from an emergent $\mathbb{Z}_2$ Berry phase in {L}a$_3${N}i$_2${O}$_7$},
  author={Zhang, Jia-Xin and Zhang, Hao-Kai and You, Yi-Zhuang and Weng, Zheng-Yu},
  journal={Phys. Rev. Lett.},
  volume={133},
  number={12},
  pages={126501},
  year={2024},
  publisher={APS},
  url={https://journals.aps.org/prl/abstract/10.1103/PhysRevLett.133.126501}
}

@article{pan2023rno,
  title={Effect of rare-earth element substitution in superconducting R3Ni2O7 under pressure},
  author={Pan, Zhiming and Lu, Chen and Yang, Fan and Wu, Congjun},
  journal={Chin. Phys. Lett.},
  volume={41},
  number={8},
  pages={087401},
  year={2024},
  publisher={IOP Publishing},
  url={https://iopscience.iop.org/article/10.1088/0256-307X/41/8/087401}
}

@article{sakakibara2023La4Ni3O10,
  title = {Theoretical analysis on the possibility of superconductivity in the trilayer Ruddlesden-Popper nickelate {L}a$_4${N}i$_3${O}$_{10}$ under pressure and its experimental examination: Comparison with {L}a$_3${N}i$_2${O}$_7$},
  author = {Sakakibara, Hirofumi and Ochi, Masayuki and Nagata, Hibiki and Ueki, Yuta and Sakurai, Hiroya and Matsumoto, Ryo and Terashima, Kensei and Hirose, Keisuke and Ohta, Hiroto and Kato, Masaki and Takano, Yoshihiko and Kuroki, Kazuhiko},
  journal = {Phys. Rev. B},
  volume = {109},
  issue = {14},
  pages = {144511},
  numpages = {10},
  year = {2024},
  month = {Apr},
  publisher = {American Physical Society},
  doi = {10.1103/PhysRevB.109.144511},
  url = {https://link.aps.org/doi/10.1103/PhysRevB.109.144511}
}

@article{lange2023mixedtj,
  title={Pairing dome from an emergent Feshbach resonance in a strongly repulsive bilayer model},
  author={Lange, Hannah and Homeier, Lukas and Demler, Eugene and Schollw{\"o}ck, Ulrich and Bohrdt, Annabelle and Grusdt, Fabian},
  journal={Phys. Rev. B},
  volume={110},
  number={8},
  pages={L081113},
  year={2024},
  publisher={APS},
  url={https://journals.aps.org/prb/abstract/10.1103/PhysRevB.110.L081113}
}

@article{geisler2023structural,
  title={Structural transitions, octahedral rotations, and electronic properties of ${A}_3${N}i$_2${O}$_7$ rare-earth nickelates under high pressure},
  author={Geisler, Benjamin and Hamlin, James J and Stewart, Gregory R and Hennig, Richard G and Hirschfeld, PJ},
  journal={npj Quantum Mater.},
  volume={9},
  number={1},
  pages={38},
  year={2024},
  publisher={Nature Publishing Group UK London},
  url={https://www.nature.com/articles/s41535-024-00648-0}
}

@article{yang2023strong,
  title={Strong pairing from a small Fermi surface beyond weak coupling: Application to {L}a$_3${N}i$_2${O}$_7$},
  author={Yang, Hui and Oh, Hanbit and Zhang, Ya-Hui},
  journal={Phys. Rev. B},
  volume={110},
  number={10},
  pages={104517},
  year={2024},
  publisher={APS},
  url={https://journals.aps.org/prb/abstract/10.1103/PhysRevB.110.104517}
}

@article{rhodes2023structural,
  title = {Structural routes to stabilize superconducting {L}a$_3${N}i$_2${O}$_7$ at ambient pressure},
  author = {Rhodes, Luke C. and Wahl, Peter},
  journal = {Phys. Rev. Mater.},
  volume = {8},
  issue = {4},
  pages = {044801},
  numpages = {9},
  year = {2024},
  month = {Apr},
  publisher = {American Physical Society},
  doi = {10.1103/PhysRevMaterials.8.044801},
  url = {https://link.aps.org/doi/10.1103/PhysRevMaterials.8.044801}
}

@article{lange2023feshbach,
  title={Feshbach resonance in a strongly repulsive ladder of mixed dimensionality: A possible scenario for bilayer nickelate superconductors},
  author={Lange, Hannah and Homeier, Lukas and Demler, Eugene and Schollw{\"o}ck, Ulrich and Grusdt, Fabian and Bohrdt, Annabelle},
  journal={Phys. Rev. B},
  volume={109},
  number={4},
  pages={045127},
  year={2024},
  publisher={APS},
  url={https://journals.aps.org/prb/abstract/10.1103/PhysRevB.109.045127}
}

@article{kumar2023softening,
  title = {Softening of $\mathit{dd}$ excitation in the resonant inelastic x-ray scattering spectra as a signature of Hund's coupling in nickelates},
  author = {Kumar, Umesh and Melnick, Corey and Kotliar, Gabriel},
  journal = {Phys. Rev. Res.},
  volume = {7},
  issue = {1},
  pages = {L012066},
  numpages = {7},
  year = {2025},
  month = {Mar},
  publisher = {American Physical Society},
  doi = {10.1103/PhysRevResearch.7.L012066},
  url = {https://link.aps.org/doi/10.1103/PhysRevResearch.7.L012066}
}

@article{kaneko2023pair,
  title = {Pair correlations in the two-orbital Hubbard ladder: Implications for superconductivity in the bilayer nickelate {L}a$_3${N}i$_2$O$_7$},
  author = {Kaneko, Tatsuya and Sakakibara, Hirofumi and Ochi, Masayuki and Kuroki, Kazuhiko},
  journal = {Phys. Rev. B},
  volume = {109},
  issue = {4},
  pages = {045154},
  numpages = {5},
  year = {2024},
  month = {Jan},
  publisher = {American Physical Society},
  doi = {10.1103/PhysRevB.109.045154},
  url = {https://link.aps.org/doi/10.1103/PhysRevB.109.045154}
}

@article{ryee2023critical,
  title={Quenched pair breaking by interlayer correlations as a key to superconductivity in {L}a$_3${N}i$_2${O}$_7$},
  author={Ryee, Siheon and Witt, Niklas and Wehling, Tim O},
  journal={Phys. Rev. Lett.},
  volume={133},
  number={9},
  pages={096002},
  year={2024},
  publisher={APS},
  url={https://journals.aps.org/prl/abstract/10.1103/PhysRevLett.133.096002}
}

@article{zhang2023la3ni2o6,
  title = {Electronic structure, magnetic correlations, and superconducting pairing in the reduced Ruddlesden-Popper bilayer {L}a$_3${N}i$_2${O}$_6$ under pressure: Different role of $d_{3z^2-r^2}$ orbital compared with {L}a$_3${N}i$_2${O}$_7$},
  author = {Zhang, Yang and Lin, Ling-Fang and Moreo, Adriana and Maier, Thomas A. and Dagotto, Elbio},
  journal = {Phys. Rev. B},
  volume = {109},
  issue = {4},
  pages = {045151},
  numpages = {10},
  year = {2024},
  month = {Jan},
  publisher = {American Physical Society},
  doi = {10.1103/PhysRevB.109.045151},
  url = {https://link.aps.org/doi/10.1103/PhysRevB.109.045151}
}

@article{grusdt2023lno03349,
  title={Superconductivity in the pressurized nickelate La$_3$Ni$_2$O$_7$ in the vicinity of a BEC--BCS crossover},
  author={Schl{\"o}mer, Henning and Schollw{\"o}ck, Ulrich and Grusdt, Fabian and Bohrdt, Annabelle},
  journal={Commun. Phys.},
  volume={7},
  number={1},
  pages={366},
  year={2024},
  publisher={Nature Publishing Group UK London},
  url={https://www.nature.com/articles/s42005-024-01854-9}
}

@article{chen2023iPEPS,
  title={Orbital-selective superconductivity in the pressurized bilayer nickelate {L}a$_3${N}i$_2${O}$_7$: An infinite projected entangled-pair state study},
  author={Chen, Jialin and Yang, Fan and Li, Wei},
  journal={Phys. Rev. B},
  volume={110},
  number={4},
  pages={L041111},
  year={2024},
  publisher={APS},
  url={https://journals.aps.org/prb/abstract/10.1103/PhysRevB.110.L041111}
}

@article{liu2023dxy,
  title={Sensitive dependence of pairing symmetry on Ni-eg crystal field splitting in the nickelate superconductor {L}a$_3${N}i$_2${O}$_7$},
  author={Xia, Chengliang and Liu, Hongquan and Zhou, Shengjie and Chen, Hanghui},
  journal={Nat. Commun.},
  volume={16},
  number={1},
  pages={1054},
  year={2025},
  publisher={Nature Publishing Group UK London},
  url={https://www.nature.com/articles/s41467-025-56206-0}
}

@article{ouyang2023hund,
  title = {Hund electronic correlation in {L}a$_3${N}i$_2${O}$_7$ under high pressure},
  author = {Ouyang, Zhenfeng and Wang, Jia-Ming and Wang, Jing-Xuan and He, Rong-Qiang and Huang, Li and Lu, Zhong-Yi},
  journal = {Phys. Rev. B},
  volume = {109},
  issue = {11},
  pages = {115114},
  numpages = {7},
  year = {2024},
  month = {Mar},
  publisher = {American Physical Society},
  doi = {10.1103/PhysRevB.109.115114},
  url = {https://link.aps.org/doi/10.1103/PhysRevB.109.115114}
}

@article{qu2023roles,
  title = {Hund's rule, interorbital hybridization, and high-${T}_{c}$ superconductivity in the bilayer nickelate ${\mathrm{La}}_{3}{\mathrm{Ni}}_{2}{\mathrm{O}}_{7}$},
  author = {Qu, Xing-Zhou and Qu, Dai-Wei and Yi, Xin-Wei and Li, Wei and Su, Gang},
  journal = {Phys. Rev. B},
  volume = {112},
  issue = {16},
  pages = {L161101},
  numpages = {8},
  year = {2025},
  month = {Oct},
  publisher = {American Physical Society},
  url = {https://link.aps.org/doi/10.1103/171w-6kjw}
}

@article{zheng2023twoorbital,
  title = {${s}_{\pm}$-wave superconductivity in the bilayer two-orbital Hubbard model},
  author = {Zheng, Yao-Yuan and W\'u, W\'ei},
  journal = {Phys. Rev. B},
  volume = {111},
  issue = {3},
  pages = {035108},
  numpages = {7},
  year = {2025},
  month = {Jan},
  publisher = {American Physical Society},
  doi = {10.1103/PhysRevB.111.035108},
  url = {https://link.aps.org/doi/10.1103/PhysRevB.111.035108}
}

@article{kakoi2023pair,
  title={Pair correlations of the hybridized orbitals in a ladder model for the bilayer nickelate {L}a$_3${N}i$_2${O}$_7$},
  author={Kakoi, Masataka and Kaneko, Tatsuya and Sakakibara, Hirofumi and Ochi, Masayuki and Kuroki, Kazuhiko},
  journal={Phys. Rev. B},
  volume={109},
  number={20},
  pages={L201124},
  year={2024},
  publisher={APS},
  url={https://journals.aps.org/prb/abstract/10.1103/PhysRevB.109.L201124}
}

@article{heier2023competing,
  title = {Competing ${d}_{xy}$ and ${s}_{\pm}$ pairing symmetries in superconducting {L}a$_3${N}i$_2${O}$_7$: $\mathrm{LDA}+\mathrm{FLEX}$ calculations},
  author = {Heier, Griffin and Park, Kyungwha and Savrasov, Sergey Y.},
  journal = {Phys. Rev. B},
  volume = {109},
  issue = {10},
  pages = {104508},
  numpages = {9},
  year = {2024},
  month = {Mar},
  publisher = {American Physical Society},
  doi = {10.1103/PhysRevB.109.104508},
  url = {https://link.aps.org/doi/10.1103/PhysRevB.109.104508}
}

@article{zhao2024pressure,
  title={Pressure-enhanced spin-density-wave transition in double-layer nickelate La3Ni2O7-$\delta$},
  author={Zhao, Dan and Zhou, Yanbing and Huo, Mengwu and Wang, Yu and Nie, Linpeng and Yang, Ye and Ying, Jianjun and Wang, Meng and Wu, Tao and Chen, Xianhui},
  journal={Science Bulletin},
  year={2025},
  publisher={Elsevier},
  url={https://www.sciencedirect.com/science/article/pii/S2095927325001811}
}

@article{abadi2024electronic,
  title = {Electronic Structure of the Alternating Monolayer-Trilayer Phase of ${\mathrm{La}}_{3}{\text{Ni}}_{2}{\mathrm{O}}_{7}$},
  author = {Abadi, Sebastien and Xu, Ke-Jun and Lomeli, Eder G. and Puphal, Pascal and Isobe, Masahiko and Zhong, Yong and Fedorov, Alexei V. and Mo, Sung-Kwan and Hashimoto, Makoto and Lu, Dong-Hui and Moritz, Brian and Keimer, Bernhard and Devereaux, Thomas P. and Hepting, Matthias and Shen, Zhi-Xun},
  journal = {Phys. Rev. Lett.},
  volume = {134},
  issue = {12},
  pages = {126001},
  numpages = {7},
  year = {2025},
  month = {Mar},
  publisher = {American Physical Society},
  doi = {10.1103/PhysRevLett.134.126001},
  url = {https://link.aps.org/doi/10.1103/PhysRevLett.134.126001}
}

@article{li2024electronic,
  title={Electronic correlation and pseudogap-like behavior of high-temperature superconductor La3Ni2O7},
  author={Li, Yidian and Du, Xian and Cao, Yantao and Pei, Cuiying and Zhang, Mingxin and Zhao, Wenxuan and Zhai, Kaiyi and Xu, Runzhe and Liu, Zhongkai and Li, Zhiwei and others},
  journal={Chin. Phys. Lett.},
  volume={41},
  number={8},
  pages={087402},
  year={2024},
  publisher={IOP Publishing},
  url={https://iopscience.iop.org/article/10.1088/0256-307X/41/8/087402/}
}

@article{zhang2024doping,
  title={Doping evolution of the normal state magnetic excitations in pressurized La3Ni2O7},
  author={Zhang, Hai-Yang and Bai, Yu-Jie and Kong, Fan-Jie and Wu, Xiu-Qiang and Xing, Yu-Heng and Xu, Ning},
  journal={New J. Phys.},
  volume={26},
  number={12},
  pages={123027},
  year={2024},
  publisher={IOP Publishing},
  url={https://iopscience.iop.org/article/10.1088/1367-2630/ada0d4/meta}
}

@article{ren2025resolving,
  title={Resolving the electronic ground state of La3Ni2O7-$\delta$ films},
  author={Ren, Xiaolin and Sutarto, Ronny and Wu, Xianxin and Zhang, Jianfeng and Huang, Hai and Xiang, Tao and Hu, Jiangping and Comin, Riccardo and Zhou, Xingjiang and Zhu, Zhihai},
  journal={Commun. Phys.},
  volume={8},
  number={1},
  pages={52},
  year={2025},
  publisher={Nature Publishing Group UK London},
  url={https://www.nature.com/articles/s42005-025-01971-z}
}

@article{li2024dis,
  title={Distinguishing Electronic Band Structure of Single-layer and Bilayer Ruddlesden-Popper Nickelates Probed by in-situ High Pressure X-ray Absorption Near-edge Spectroscopy},
  author={Li, Mingtao and Wang, Yiming and Pei, Cuiying and Zhang, Mingxin and Li, Nana and Guan, Jiayi and Amboage, Monica and Adama, N and Kong, Qingyu and Qi, Yanpeng and others},
  journal={arXiv:2410.04230},
  year={2024},
  url={https://arxiv.org/abs/2410.04230}
}

@article{zhou2024revealing,
  title={Revealing nanoscale structural phase separation in La3Ni2O7-$\delta$ single crystal via scanning near-field optical microscopy},
  author={Zhou, Xiaoxiang and He, Weihong and Zhou, Zijian and Ni, Kaipeng and Huo, Mengwu and Hu, Deyuan and Zhu, Yinghao and Zhang, Enkang and Jiang, Zhicheng and Zhang, Shuaikang and others},
  journal={arXiv:2410.06602},
  year={2024},
  url={https://arxiv.org/abs/2410.06602}
}

@article{su2024strongly,
  title={Strongly Anisotropic Charge Dynamics in La3Ni2O7 with Coherent-to-Incoherent Crossover of Interlayer Charge Dynamics},
  author={Su, Bo and Huang, Chaoxin and Zhao, Jianzhou and Huo, Mengwu and Luo, Jianlin and Wang, Meng and Chen, Zhi-Guo},
  journal={arXiv:2411.10786},
  year={2024},
  url={https://arxiv.org/abs/2411.10786}
}

@article{geisler2024fermi,
  title={Fermi surface reconstruction in strained La3Ni2O7 on LaAlO$_3$(001) and SrTiO$_3$(001)},
  author={Geisler, Benjamin and Hamlin, James J and Stewart, Gregory R and Hennig, Richard G and Hirschfeld, PJ},
  journal={arXiv:2411.14600},
  year={2024},
  url={https://arxiv.org/abs/2411.14600}
}

@article{mijit2024local,
  title={Local electronic properties of La3Ni2O7 under pressure},
  author={Mijit, Emin and Ma, Peiyue and Sahle, Christoph J and Rosa, Angelika D and Hu, Zhiwei and De Angelis, Francesco and Lopez, Alberto and Amatori, Simone and Tchoudinov, Georghii and Joly, Yves and others},
  journal={arXiv:2412.08269},
  year={2024},
  url={https://arxiv.org/abs/2412.08269}
}

@article{chen2024unveiling,
  title = {Unveiling the multiband metallic nature of the normal state in the nickelate ${\mathrm{La}}_{3}{\mathrm{Ni}}_{2}{\mathrm{O}}_{7}$},
  author = {Chen, Bowen and Zhang, Hengyuan and Li, Jingyuan and Hu, Deyuan and Huo, Mengwu and Wang, Shuyang and Xi, Chuanying and Wang, Zhaosheng and Sun, Hualei and Wang, Meng and Shen, Bing},
  journal = {Phys. Rev. B},
  volume = {111},
  issue = {5},
  pages = {054519},
  numpages = {7},
  year = {2025},
  month = {Feb},
  publisher = {American Physical Society},
  doi = {10.1103/PhysRevB.111.054519},
  url = {https://link.aps.org/doi/10.1103/PhysRevB.111.054519}
}

@article{xu2025com,
  title = {Incommensurate spin fluctuations and competing pairing symmetries in La$_3$Ni$_2$O$_7$},
  author = {Xu, Han-Xiang and Guterding, Daniel},
  journal = {Phys. Rev. B},
  volume = {112},
  issue = {17},
  pages = {174519},
  numpages = {8},
  year = {2025},
  month = {Nov},
  publisher = {American Physical Society},
  url = {https://link.aps.org/doi/10.1103/f6nj-34th}
}

@article{shi2025pre,
  title={Prerequisite of superconductivity: SDW rather than tetragonal structure in double-layer La3Ni2O7-x},
  author={Shi, Mengzhu and Peng, Di and Li, Yikang and Xing, Zhenfang and Wang, Yuzhu and Fan, Kaibao and Li, Houpu and Wu, Rongqi and Zeng, Zhidan and Zeng, Qiaoshi and others},
  journal={arXiv:2501.14202},
  year={2025},
  url={https://arxiv.org/abs/2501.14202}
}

@article{huo2025low,
  title={Low volume fraction of high-Tc superconductivity in La3Ni2O7 at 80 K and ambient pressure},
  author={Huo, Mengwu and Ma, Peiyue and Huang, Chaoxin and Huang, Xing and Sun, Hualei and Wang, Meng},
  journal={arXiv:2501.15929},
  year={2025},
  url={https://arxiv.org/abs/2501.15929}
}

@article{huo2025first,
  title={First-principles study of the Pr-doped bilayer nickelate L a 3 N i 2 O 7},
  author={Huo, Zihao and Zhang, Peng and Shi, Haoliang and Yan, Xiaochun and Duan, Defang and Cui, Tian},
  journal={Phys. Rev. B},
  volume={111},
  number={19},
  pages={195118},
  year={2025},
  publisher={APS},
  url={https://journals.aps.org/prb/abstract/10.1103/PhysRevB.111.195118}
}

@article{zhang2026strong,
title = {Strong oxidizing annealing of bilayer La$_3$Ni$_2$O$_{7-\delta}$ results in suppression of superconductivity under high pressure},
author = {Yulin Zhang and Cuiying Pei and Ning Guo and Feiyu Li and Longlong Fan and Mingxin Zhang and Lingzhen Wang and Gongting Zhang and Yunong Wang and Chao Ma and Wenyong Cheng and Shanpeng Wang and Qiang Zheng and Yanpeng Qi and Junjie Zhang},
journal = {Journal of Solid State Chemistry},
volume = {355},
pages = {125757},
year = {2026},
issn = {0022-4596},
url = {https://www.sciencedirect.com/science/article/pii/S002245962500581X}
}

@article{bhatta2025structural,
  title={Structural and Electronic Evolution of Bilayer Nickelates Under Biaxial Strain},
  author={Bhatta, HC and Zhang, Xiaoliang and Zhong, Yong and Jia, Chunjing},
  journal={arXiv:2502.01624},
  year={2025},
  url={https://arxiv.org/abs/2502.01624}
}

@article{zhang2022review,
  title={A review on strain study of cuprate superconductors},
  author={Zhang, Jian and Wu, Haiyan and Zhao, Guangzhen and Han, Lu and Zhang, Jun},
  journal={Nanomaterials},
  volume={12},
  number={19},
  pages={3340},
  year={2022},
  publisher={MDPI},
  url={https://www.mdpi.com/2079-4991/12/19/3340}
}

@article{armitage2010progress,
  title={Progress and perspectives on electron-doped cuprates},
  author={Armitage, NP and Fournier, P and Greene, RL},
  journal={Rev. Mod. Phys.},
  volume={82},
  number={3},
  pages={2421--2487},
  year={2010},
  publisher={APS},
  url={https://journals.aps.org/rmp/abstract/10.1103/RevModPhys.82.2421}
}

@article{taillefer2010scattering,
  title={Scattering and pairing in cuprate superconductors},
  author={Taillefer, Louis},
  journal={Annu. Rev. Condens. Matter Phys.},
  volume={1},
  number={1},
  pages={51--70},
  year={2010},
  publisher={Annual Reviews},
  url={https://www.annualreviews.org/content/journals/10.1146/annurev-conmatphys-070909-104117}
}

@article{damascelli2003angle,
  title={Angle-resolved photoemission studies of the cuprate superconductors},
  author={Damascelli, Andrea and Hussain, Zahid and Shen, Zhi-Xun},
  journal={Rev. Mod. Phys.},
  volume={75},
  number={2},
  pages={473},
  year={2003},
  publisher={APS},
  url={https://journals.aps.org/rmp/abstract/10.1103/RevModPhys.75.473}
}

@article{shi2025effect,
  title={The effect of Carrier Doping and Thickness on the Electronic Structures of {L}a$_3${N}i$_2${O}$_7$ Thin Films},
  author={Shi, Haoliang and Huo, Zihao and Li, Guanlin and Ma, Hao and Cui, Tian and Yao, Dao-Xin and Duan, Defang},
  journal={arXiv:2502.04255},
  year={2025},
  url={https://arxiv.org/abs/2502.04255}
}

@article{pardo2011dft,
  title={Metal-insulator transition in layered nickelates {L}a$_3${N}i$_2${O}$_{7-\delta}$ ($\delta$= 0.0, 0.5, 1)},
  author={Pardo, Victor and Pickett, Warren E},
  journal = {Phys. Rev. B},
  volume={83},
  number={24},
  pages={245128},
  year={2011},
  publisher = {American Physical Society},
  doi = {10.1103/PhysRevB.83.245128},
  url = {https://link.aps.org/doi/10.1103/PhysRevB.83.245128}
}

@article{nakata2017finite,
  title={Finite-energy spin fluctuations as a pairing glue in systems with coexisting electron and hole bands},
  author={Nakata, Masahiro and Ogura, Daisuke and Usui, Hidetomo and Kuroki, Kazuhiko},
  journal={Phys. Rev. B},
  volume={95},
  number={21},
  pages={214509},
  year={2017},
  publisher={APS},
  url = {https://link.aps.org/doi/10.1103/PhysRevB.95.214509}
}

@article{bednorz1986LBCO,
  title={Possible high {T}$_c$ superconductivity in the {B}a-{L}a-{C}u-{O} system},
  author={Bednorz, J George and M{\"u}ller, K Alex},
  journal={Zeitschrift f{\"u}r Physik B Condensed Matter},
  volume={64},
  number={2},
  pages={189--193},
  year={1986},
  publisher={Springer},
  doi = {10.1007/BF01303701},
  url = {https://link.springer.com/article/10.1007/BF01303701}
}

@article{lee2004infinite,
  title={Infinite-layer {L}a{N}i{O}$_2$: {N}i$^{1+}$ is not {C}u$^{2+}$},
  author={Lee, K-W and Pickett, WE},
  journal={Phys. Rev. B},
  volume={70},
  number={16},
  pages={165109},
  year={2004},
  publisher={APS}
}

@article{lu2024interplay,
  title={Interplay of two $E_g$ orbitals in superconducting La 3 Ni 2 O 7 under pressure},
  author={Lu, Chen and Pan, Zhiming and Yang, Fan and Wu, Congjun},
  journal={Phys. Rev. B},
  volume={110},
  number={9},
  pages={094509},
  year={2024},
  publisher={APS},
  url={https://journals.aps.org/prb/abstract/10.1103/PhysRevB.110.094509}
}

@article{botana2021nickelate,
  title={Nickelate superconductors: an ongoing dialog between theory and experiments},
  author={Botana, Antia S and Bernardini, Fabio and Cano, Andr{\'e}s},
  journal={Journal of Experimental and Theoretical Physics},
  volume={132},
  pages={618--627},
  year={2021},
  publisher={Springer},
  doi={10.1134/S1063776121040026},
  url={https://link.springer.com/article/10.1134/S1063776121040026}
}

@article{anisimov1999electronic,
  title = {Electronic structure of possible nickelate analogs to the cuprates},
  author = {Anisimov, V. I. and Bukhvalov, D. and Rice, T. M.},
  journal = {Phys. Rev. B},
  volume = {59},
  issue = {12},
  pages = {7901--7906},
  numpages = {0},
  year = {1999},
  month = {Mar},
  publisher = {American Physical Society},
  doi = {10.1103/PhysRevB.59.7901},
  url = {https://link.aps.org/doi/10.1103/PhysRevB.59.7901}
}

@article{labollita2024electronic,
  title = {Electronic structure and magnetic tendencies of trilayer {L}a$_4${N}i$_3${O}$_{10}$ under pressure: Structural transition, molecular orbitals, and layer differentiation},
  author = {LaBollita, Harrison and Kapeghian, Jesse and Norman, Michael R. and Botana, Antia S.},
  journal = {Phys. Rev. B},
  volume = {109},
  issue = {19},
  pages = {195151},
  numpages = {12},
  year = {2024},
  month = {May},
  publisher = {American Physical Society},
  doi = {10.1103/PhysRevB.109.195151},
  url = {https://link.aps.org/doi/10.1103/PhysRevB.109.195151}
}

@article{li2024pressure,
  author = {Li, Jingyuan and Peng, Di and Ma, Peiyue and Zhang, Hengyuan and Xing, Zhenfang and Huang, Xing and Huang, Chaoxin and Huo, Mengwu and Hu, Deyuan and Dong, Zixian and Chen, Xiang and Xie, Tao and Dong, Hongliang and Sun, Hualei and Zeng, Qiaoshi and Mao, Ho-kwang and Wang, Meng},
  title = {Identification of superconductivity in bilayer nickelate La$_3$Ni$_2$O$_7$ under high pressure up to 100 GPa},
  journal = {Natl. Sci. Rev.},
  volume = {12},
  number = {10},
  pages = {nwaf220},
  year = {2025},
  month = {05},
  issn = {2095-5138},
  doi = {10.1093/nsr/nwaf220},
  url = {https://academic.oup.com/nsr/article/12/10/nwaf220/8152906},
}

@article{li2024ultrafast,
  title={Distinct ultrafast dynamics of bilayer and trilayer nickelate superconductors regarding the density-wave-like transitions},
  author={Li, Yidian and Cao, Yantao and Liu, Liangyang and Peng, Pai and Lin, Hao and Pei, Cuiying and Zhang, Mingxin and Wu, Heng and Du, Xian and Zhao, Wenxuan and others},
  journal={Science Bulletin},
  volume={70},
  number={2},
  pages={180--186},
  year={2025},
  publisher={Elsevier},
  url={https://www.sciencedirect.com/science/article/pii/S2095927325001811}
}

@article{ji2025signatures,
  title={Signatures of spin-glass superconductivity in nickelate (La, Pr, Sm)$_3$Ni$_2$O$_7$ films}, 
  author={Haoran Ji and Zheyuan Xie and Yaqi Chen and Guangdi Zhou and Longxin Pan and Heng Wang and Haoliang Huang and Jun Ge and Yi Liu and Guang-Ming Zhang and Ziqiang Wang and Qi-Kun Xue and Zhuoyu Chen and Jian Wang},
  year={2025},
  journal={arXiv:2508.16412},
  url={https://arxiv.org/abs/2508.16412}
}

@article{wu2025ultrafast,
  title={Ultrafast Optical Evidence of Split Density Waves in Bilayer Nickelate La$_3$Ni$_2$O$_7$},
  author={Wu, Qi-Yi and Hu, De-Yuan and Zhang, Chen and Huo, Mengwu and Liu, Hao and Chen, Bo and Zhou, Ying and Fu, Zhong-Tuo and Lv, Chun-Hui and Xu, Zi-Jie and others},
  journal={arXiv:2508.09436},
  year={2025},
  url={https://arxiv.org/abs/2508.09436}
}

@article{dong2025interstitial,
  title={Interstitial oxygen order and its competition with superconductivity in La$_2$PrNi$_2$O$_{7+\delta}$}, 
  author={Zehao Dong and Gang Wang and Ningning Wang and Wen-Han Dong and Lin Gu and Yong Xu and Jinguang Cheng and Zhen Chen and Yayu Wang},
  year={2025},
  journal={arXiv:2508.03414},
  url={https://arxiv.org/abs/2508.03414}
}

@article{wang2025fermi,
  title={Fermi liquid and isotropic superconductivity of Hund scenario for bilayer nickelates}, 
  author={Jiangfan Wang and Yi-feng Yang},
  year={2025},
  journal={arXiv:2507.19301},
  url={https://arxiv.org/abs/2507.19301}
}

@article{cao2025strain,
  title={Strain-Engineered Electronic Structure and Superconductivity in La$_3$Ni$_2$O$_7$ Thin Films}, 
  author={Yu-Han Cao and Kai-Yue Jiang and Hong-Yan Lu and Da Wang and Qiang-Hua Wang},
  year={2025},
  journal={arXiv:2507.13694},
  url={https://arxiv.org/abs/2507.13694}
}

@article{zhu2025quantum,
  title={Quantum phase transition driven by competing intralayer and interlayer hopping of Ni-$d_{3z^2-r^2}$ orbitals in bilayer nickelates}, 
  author={Xiaoyu Zhu and Wei Qin and Ping Cui and Zhenyu Zhang},
  year={2025},
  journal={arXiv:2507.11169},
  url={https://arxiv.org/abs/2507.11169}
}

@article{li2026enhanced,
  title={Enhanced superconductivity in the compressively strained bilayer nickelate thin films by pressure},
  author={Li, Qing and Sun, Jianping and B{\"o}tzel, Steffen and Ou, Mengjun and Xiang, Zhe-Ning and Lechermann, Frank and Wang, Bosen and Wang, Yi and Zhang, Ying-Jie and Cheng, Jinguang and others},
  journal={Nat. Commun.},
  year={2026},
  publisher={Nature Publishing Group UK London},
  url={https://www.nature.com/articles/s41467-026-69660-1}
}

@article{yang2025evolution,
  title={Evolution from intralayer to interlayer superconductivity in a bilayer $t$-$J$ model}, 
  author={Yuan Yang and Xin Lu and Yuan Wan and Wei-Qiang Chen and Shou-Shu Gong},
  year={2025},
  journal={arXiv:2507.07545},
  url={https://arxiv.org/abs/2507.07545}, 
}

@article{sun2025observation,
  title={Observation of superconductivity-induced leading-edge gap in Sr-doped $\mathrm{La}_{3}\mathrm{Ni}_{2}\mathrm{O}_{7}$ thin films}, 
  author={Wenjie Sun and Zhicheng Jiang and Bo Hao and Shengjun Yan and Hongyi Zhang and Maosen Wang and Yang Yang and Haoying Sun and Zhengtai Liu and Dianxiang Ji and Zhengbin Gu and Jian Zhou and Dawei Shen and Donglai Feng and Yuefeng Nie},
  year={2025},
  journal={arXiv:2507.07409},
  url={https://arxiv.org/abs/2507.07409}, 
}

@article{wang2025self,
    author = {Wang, Zhan and Zhang, Heng-Jia and Jiang, Kun and Zhang, Fu-Chun},
    title = {Self-doped molecular Mott insulator for bilayer high-temperature superconducting La$_3$Ni$_2$O$_7$},
    journal = {Nat. Sci. Rev.},
    pages = {nwaf353},
    year = {2025},
    month = {08},
    issn = {2095-5138},
    doi = {10.1093/nsr/nwaf353},
    url = {https://doi.org/10.1093/nsr/nwaf353}
}

@article{chen2024nonfermi,
  title = {Non-Fermi liquid and antiferromagnetic correlations with hole doping in the bilayer two-orbital Hubbard model of {La}$_{3}${Ni}$_{2}${O}$_{7}$ at zero temperature},
  author = {Chen, Yin and Tian, Yi-Heng and Wang, Jia-Ming and He, Rong-Qiang and Lu, Zhong-Yi},
  journal = {Phys. Rev. B},
  volume = {110},
  issue = {23},
  pages = {235119},
  numpages = {7},
  year = {2024},
  month = {Dec},
  publisher = {American Physical Society},
  doi = {10.1103/PhysRevB.110.235119},
  url = {https://link.aps.org/doi/10.1103/PhysRevB.110.235119}
}

@article{ouyang2025phase,
  title = {Phase diagrams and two key factors to superconductivity of Ruddlesden-Popper nickelates},
  author = {Ouyang, Zhenfeng and He, Rong-Qiang and Lu, Zhong-Yi},
  journal = {Phys. Rev. B},
  volume = {112},
  issue = {4},
  pages = {045127},
  numpages = {10},
  year = {2025},
  month = {Jul},
  publisher = {American Physical Society},
  doi = {10.1103/1412-nfzm},
  url = {https://link.aps.org/doi/10.1103/1412-nfzm}
}

@article{wang2025electronic,
  title={Electronic structure of compressively strained thin film La$_2$PrNi$_2$O$_7$}, 
  author={Bai Yang Wang and Yong Zhong and Sebastien Abadi and Yidi Liu and Yijun Yu and Xiaoliang Zhang and Yi-Ming Wu and Ruohan Wang and Jiarui Li and Yaoju Tarn and Eun Kyo Ko and Vivek Thampy and Makoto Hashimoto and Donghui Lu and Young S. Lee and Thomas P. Devereaux and Chunjing Jia and Harold Y. Hwang and Zhi-Xun Shen},
  year={2025},
  journal={arXiv:2504.16372},
  url={https://arxiv.org/abs/2504.16372}, 
}

@article{shen2025nodeless,
  title={Nodeless superconducting gap and electron-boson coupling in (La,Pr,Sm)$_{3}$Ni$_2$O$_7$ films}, 
  author={Jianchang Shen and Guangdi Zhou and Yu Miao and Peng Li and Zhipeng Ou and Yaqi Chen and Zechao Wang and Runqing Luan and Hongxu Sun and Zikun Feng and Xinru Yong and Yueying Li and Lizhi Xu and Wei Lv and Zihao Nie and Heng Wang and Haoliang Huang and Yu-Jie Sun and Qi-Kun Xue and Junfeng He and Zhuoyu Chen},
  year={2025},
  journal={arXiv:2502.17831},
  url={https://arxiv.org/abs/2502.17831}
}

@article{shu2026cont,
  title={Contrasting Momentum-Selective Spin-Density-Wave Gaps in Bilayer and Trilayer Nickelates},
  author={Shu, Jun and Shen, Jun and Zhou, Xiaoxiang and Zhu, Yinghao and Wang, Qingsong and Wang, Dengjing and He, Weihong and Yuan, Jie and Jin, Kui and Shen, Dawei and others},
  year={2026},
  journal={arXiv:2602.02174},
  url={https://arxiv.org/abs/2602.02174}
}

@article{ouyang2024absence,
  title={Absence of electron-phonon coupling superconductivity in the bilayer phase of La3Ni2O7 under pressure},
  author={Ouyang, Zhenfeng and Gao, Miao and Lu, Zhong-Yi},
  journal={npj Quantum Materials},
  volume={9},
  number={1},
  pages={80},
  year={2024},
  publisher={Nature Publishing Group UK London},
  url={https://www.nature.com/articles/s41535-024-00689-5}
}

@article{talantsev2024debye,
  author={E.F. Talantsev and V.V. Chistyakov},
  title={Debye temperature, electron-phonon coupling constant, and three-dome shape of crystalline strain as a function of pressure in highly compressed La$_3$Ni$_2$O$_{7-\delta}$},
  publisher={Letters on Materials},
  year={2024},
  journal={Letters on Materials},
  volume={14},
  number={3},
  pages={262-268},
  url={https://lettersonmaterials.com/en/Readers/Article.aspx?aid=44771}
}

@article{zhu2025magnetic,
  title={Magnetic phases and electron--phonon coupling in La3Ni2O7 under pressure},
  author={Zhu, Cong and Li, Bin and Fan, Yuxiang and Yin, Chuanhui and Zhai, Junjie and Cheng, Jie and Liu, Shengli and Shi, Zhixiang},
  journal={Computational Materials Science},
  volume={250},
  pages={113676},
  year={2025},
  publisher={Elsevier},
  url={https://www.sciencedirect.com/science/article/pii/S0927025625000199}
}

@article{shao2026possible,
  title={Possible liquid-nitrogen-temperature superconductivity driven by perpendicular electric field in the single-bilayer film of La$_3$Ni$_2$O$_7$ at ambient pressure},
  author={Shao, Zhi-Yan and Ji, Jia-Heng and Wu, Congjun and Yao, Dao-Xin and Yang, Fan},
  journal={Nat. Commun.},
  volume={17},
  pages={1120},
  year={2026},
  publisher={Nature Publishing Group UK London},
  url={https://www.nature.com/articles/s41467-025-67880-5}
}

@article{lu2025impact,
  title={Impact of pressure and apical oxygen vacancies on superconductivity in La3Ni2O7},
  author={Lu, Chen and Zhang, Ming and Pan, Zhiming and Wu, Congjun and Yang, Fan},
  journal={Commun. Phys.},
  volume={8},
  number={1},
  pages={354},
  year={2025},
  publisher={Nature Publishing Group UK London},
  url={https://www.nature.com/articles/s42005-025-02266-z}
}

@article{ji2025strong,
  title = {Strong-coupling study of the pairing mechanism in pressurized La$_3$Ni$_2$O$_7$},
  author = {Ji, Jia-Heng and Lu, Chen and Shao, Zhi-Yan and Pan, Zhiming and Yang, Fan and Wu, Congjun},
  journal = {Phys. Rev. B},
  volume = {112},
  issue = {21},
  pages = {214515},
  numpages = {18},
  year = {2025},
  month = {Dec},
  publisher = {American Physical Society},
  url = {https://link.aps.org/doi/10.1103/f6sr-t6js}
}

@article{sharma2026struct,
  title={Structural stability, electronic structure, and magnetic properties of the single-layer trilayer La 3 Ni 2 O 7 polymorph},
  author={Sharma, Shekhar and Zhao, Yi-Feng and Botana, Antia S},
  journal={Phys. Rev. B},
  volume={113},
  number={4},
  pages={045139},
  year={2026},
  publisher={APS},
  url={https://link.aps.org/doi/10.1103/f3wc-hrqb}
}

@article{liu2025evidence,
  title={Evidence for the Meissner Effect in the Nickelate Superconductor La 3 Ni 2 O 7-$\delta$ Single Crystal Using Diamond Quantum Sensors},
  author={Liu, Lin and Guo, Jianning and Hu, Deyuan and Yan, Guizhen and Chen, Yuzhi and Yu, Lunxuan and Wang, Meng and Liu, Xiao-Di and Huang, Xiaoli},
  journal={Phys. Rev. Lett.},
  volume={135},
  number={9},
  pages={096001},
  year={2025},
  publisher={APS},
  url={https://link.aps.org/doi/10.1103/yvj7-htb4}
}

@article{li2026machine,
  title={Machine learning accelerated search for superconductors in BCN based compounds and R3Ni2O7-type nickelates},
  author={Li, Xiaoying and Tu, Wenqian and Lv, Run and Liu, Li'e and Shao, Dingfu and Sun, Yuping and Lu, Wenjian},
  journal={Phys. Rev. B},
  volume={113},
  number={5},
  pages={054521},
  year={2026},
  publisher={APS},
  url={https://link.aps.org/doi/10.1103/wn98-9yc7}
}

@article{haque2025dft,
  title={DFT exploration of pressure dependent physical properties of the recently discovered La3Ni2O7 superconductor},
  author={Haque, Md Enamul and Ali, Ruman and Masum, MA and Hassan, Jahid and Naqib, SH},
  journal={arXiv:2504.15853},
  year={2025},
  url={https://arxiv.org/abs/2504.15853}
}

@article{geisler2026elect,
  title={Electronic reconstruction and interface engineering of emergent spin fluctuations in compressively strained La 3 Ni 2 O 7 on SrLaAlO 4 (001)},
  author={Geisler, Benjamin and Hamlin, James J and Stewart, Gregory R and Hennig, Richard G and Hirschfeld, PJ},
  journal={Phys. Rev. B},
  volume={113},
  number={5},
  pages={054516},
  year={2026},
  publisher={APS},
  url={https://link.aps.org/doi/10.1103/v4r2-xsnq}
}

@article{li2025orbital,
  title={Orbital Signatures of Density Wave Transition in La3Ni2O7-delta and La2PrNi2O7-delta RP-Nickelates Probed via in-situ X-ray Absorption Near-edge Spectroscopy},
  author={Li, Mingtao and Zhang, Mingxin and Wang, Yiming and Guan, Jiayi and Li, Nana and Pei, Cuiying and Adama, N and Kong, Qingyu and Qi, Yanpeng and Yang, Wenge and others},
  journal={arXiv:2502.10962},
  year={2025},
  url={https://arxiv.org/abs/2502.10962}
}

@article{ma2022doping,
  title={Doping-driven antiferromagnetic insulator-superconductor transition: A quantum Monte Carlo study},
  author={Ma, Tianxing and Wang, Da and Wu, Congjun},
  journal={Phys. Rev. B},
  volume={106},
  number={5},
  pages={054510},
  year={2022},
  publisher={APS},
  url={https://journals.aps.org/prb/abstract/10.1103/PhysRevB.106.054510}
}

@article{ma2025parameter,
  title={Parameter-Dependent Superconducting Transition Temperature in a Sign-Problem-Free Bilayer Model},
  author={Ma, Runyu and Fan, Zenghui and Ma, Tianxing and Wu, Congjun},
  journal={Chin. Phys. Lett.},
  volume={42},
  number={11},
  pages={110705},
  year={2025},
  publisher={Chinese Physical Society and IOP Publishing Ltd},
  url={https://iopscience.iop.org/article/10.1088/0256-307X/42/11/110705/}
}

@article{scalapino1998so,
  title={$SO(5)$ symmetric ladder},
  author={Scalapino, D and Zhang, Shou-Cheng and Hanke, W},
  journal={Phys. Rev. B},
  volume={58},
  number={1},
  pages={443},
  year={1998},
  publisher={APS},
  url={https://journals.aps.org/prb/abstract/10.1103/PhysRevB.58.443}
}

@article{wu2005sufficient,
  title={Sufficient condition for absence of the sign problem in the fermionic quantum Monte Carlo algorithm},
  author={Wu, Congjun and Zhang, Shou-Cheng},
  journal={Phys. Rev. B},
  volume={71},
  number={15},
  pages={155115},
  year={2005},
  publisher={APS},
  url={https://journals.aps.org/prb/abstract/10.1103/PhysRevB.71.155115}
}

@article{wu2003exact,
  title={Exact $SO(5)$ symmetry in the spin-$3/2$ fermionic system},
  author={Wu, Congjun and Hu, Jiang-ping and Zhang, Shou-cheng},
  journal={Phys. Rev. Lett.},
  volume={91},
  number={18},
  pages={186402},
  year={2003},
  publisher={APS},
  url={https://journals.aps.org/prl/abstract/10.1103/PhysRevLett.91.186402}
}

@article{capponi2004current,
  title={Current carrying ground state in a bilayer model of strongly correlated systems},
  author={Capponi, Sylvain and Wu, Congjun and Zhang, Shou-Cheng},
  journal={Phys. Rev. B},
  volume={70},
  number={22},
  pages={220505},
  year={2004},
  publisher={APS},
  url={https://journals.aps.org/prb/abstract/10.1103/PhysRevB.70.220505}
}

@article{wu2006hidden,
  title={Hidden symmetry and quantum phases in spin-3/2 cold atomic systems},
  author={Wu, Congjun},
  journal={Mod. Phys. Lett. B},
  volume={20},
  number={27},
  pages={1707--1738},
  year={2006},
  publisher={World Scientific},
  url={https://www.worldscientific.com/doi/abs/10.1142/S0217984906012213}
}

@article{chen2025variation,
  title={Variation Monte Carlo Study on the bilayer $t$-$J_{\parallel}$-$J_{\perp}$ model for La$_3$Ni$_2$O$_7$},
  author={Chen, Zeyu and Liu, Yu-Bo and Yang, Fan},
  journal={arXiv:2510.04224},
  year={2025},
  url={https://arxiv.org/abs/2510.04224}
}

@article{chen2026unified,
  title={A Unified Understanding of the Experimental Controlling of the $T_c$ of Bilayer Nickelates},
  author={Chen, Zeyu and Ji, Jia-Heng and Liu, Yu-Bo and Zhang, Ming and Yang, Fan},
  journal={arXiv:2603.14519},
  year={2026},
  url={https://arxiv.org/abs/2603.14519}
}

@article{qiu2025interlayer,
  title={Interlayer coupling enhanced superconductivity near 100 K in La$_{3-x}$Nd$_x$Ni$_2$O$_7$},
  author={Qiu, Zhengyang and Chen, Junfeng and Semenok, Dmitrii V and Zhong, Qingyi and Zhou, Di and Li, Jingyuan and Ma, Peiyue and Huang, Xing and Huo, Mengwu and Xie, Tao and others},
  journal={arXiv:2510.12359},
  year={2025},
  url={https://arxiv.org/abs/2510.12359}
}

@article{zhong2025evolution,
  title={Evolution of the superconductivity in pressurized La$_{3-x}$Sm$_x$Ni$_2$O$_7$},
  author={Zhong, Qingyi and Chen, Junfeng and Qiu, Zhengyang and Li, Jingyuan and Huang, Xing and Ma, Peiyue and Huo, Mengwu and Dong, Hongliang and Sun, Hualei and Wang, Meng},
  journal={arXiv:2510.13342},
  year={2025},
  url={https://arxiv.org/abs/2510.13342}
}

@article{yi2025unifying,
  title={Unifying strain-and pressure-driven superconductivity in La$_3$Ni$_2$O$_7$: Suppressed charge and spin density waves and enhanced interlayer coupling},
  author={Yi, Xin-Wei and Li, Wei and You, Jing-Yang and Gu, Bo and Su, Gang},
  journal={Phys. Rev. B},
  volume={112},
  number={14},
  pages={L140504},
  year={2025},
  publisher={APS},
  url={https://link.aps.org/doi/10.1103/85qv-ncxb}
}

@article{yi2024nature,
  title={Nature of charge density waves and metal-insulator transition in pressurized La$_3$Ni$_2$O$_7$},
  author={Yi, Xin-Wei and Meng, Ying and Li, Jia-Wen and Liao, Zheng-Wei and Li, Wei and You, Jing-Yang and Gu, Bo and Su, Gang},
  journal={Phys. Rev. B},
  volume={110},
  number={14},
  pages={L140508},
  year={2024},
  publisher={APS},
  url={https://journals.aps.org/prb/abstract/10.1103/PhysRevB.110.L140508}
}

@article{zhou2026sc,
  title={Superconductivity onset above $60$K in ambient-pressure nickelate films},
  author={Zhou, Guangdi and Wang, Heng and Huang, Haoliang and Chen, Yaqi and Peng, Fei and Lv, Wei and Nie, Zihao and Wang, Wei and Jia, Jin-Feng and Xue, Qi-Kun and others},
  journal={Nat. Sci. Rev.},
  pages={nwag151},
  year={2026},
  publisher={Oxford University Press},
  url={https://academic.oup.com/nsr/advance-article-abstract/doi/10.1093/nsr/nwag151/8512895}
}

@article{liu2026sc,
  title={A superconducting half-dome in bilayer nickelates},
  author={Liu, Yidi and Wang, Bai Yang and Li, Jiarui and Tarn, Yaoju and Bhatt, Lopa and Colletta, Michael and Wu, Yi-Ming and Kuo, Cheng-Tai and Lee, Jun-Sik and Goodge, Berit H and others},
  journal={arXiv:2603.12196},
  year={2026},
  url={https://arxiv.org/abs/2603.12196}
}

@article{hao2025sc,
  title={Superconductivity in Sr-doped La$_3$Ni$_2$O$_7$ thin films},
  author={Hao, Bo and Wang, Maosen and Sun, Wenjie and Yang, Yang and Mao, Zhangwen and Yan, Shengjun and Sun, Haoying and Zhang, Hongyi and Han, Lu and Gu, Zhengbin and others},
  journal={Nature Materials},
  volume={24},
  number={11},
  pages={1756--1762},
  year={2025},
  publisher={Nature Publishing Group UK London},
  url={https://www.nature.com/articles/s41563-025-02327-2}
}

@article{zhong2026doping,
  title={Doping evolution of spin excitations in La$_{3-x}$Sr$_{x}$Ni$_2$O$_7$/SrLaAlO$_4$ superconducting thin films},
  author={Zhong, Hengyang and Hao, Bo and Chen, Anni and Huang, Xinru and Li, Chunyi and Zhang, Wenting and Liu, Chang and Kummer, Kurt and Brookes, Nicholas and Nie, Yuefeng and others},
  journal={arXiv:2603.01120},
  year={2026},
  url={https://arxiv.org/abs/2603.01120}
}

@article{wu2026sc,
  title={Superconductivity and magnetism in bilayer nickelates: itinerant perspective},
  author={Wu, Yi-Ming and Wang, Hao-Xin and Smailagi{\'c}, Salahudin V and Helbig, Tobias and Raghu, Srinivas},
  journal={arXiv:2602.20288},
  year={2026},
  url={https://arxiv.org/abs/2602.20288}
}

@article{chen2026electronic,
  title={Electronic structures and superconductivity in Nd-doped La$_3$Ni$_2$O$_7$},
  author={Chen, Cui-Qun and Qiu, Wenyuan and Luo, Zhihui and Wang, Meng and Yao, Dao-Xin},
  journal={Sci. China Phys. Mech. Astron.},
  volume={69},
  number={4},
  pages={247414},
  year={2026},
  publisher={Springer},
  url={https://link.springer.com/article/10.1007/s11433-025-2869-1}
}

@article{chen2025sc,
  title={Superconductivity of bilayer two-orbital Hubbard model for La$_3$Ni$_2$O$_7$ under high pressure},
  author={Chen, Wei-Yang and Chen, Cui-Qun and Wang, Meng and Gong, Shou-Shu and Yao, Dao-Xin},
  journal={arXiv:2511.01801},
  year={2025},
  url={https://arxiv.org/abs/2511.01801}
}

@article{zhang2026sc,
  title={Superconductivity in Ruddlesden-Popper nickelates: a review of recent progress, focusing on thin films},
  author={Yang Zhang and Ling-Fang Lin and Thomas A. Maier and Elbio Dagotto},
  journal={arXiv:2604.18385},
  year={2026},
  url={https://arxiv.org/abs/2604.18385}
}

@article{wu2026pair,
  title={Pairing Mechanism in Bilayer Nickelate La$_3$Ni$_2$O$_7$ Superconductors}, 
  author={Xianxin Wu and Tao Xiang and Jiangping Hu},
  year={2026},
  journal={arXiv:2604.17181},
  url={https://arxiv.org/abs/2604.17181}, 
}

@article{zhang2026interlayer,
  title={Interlayer hybridization enables superconductivity in bilayer nickelates},
  author={Zhang, Shilong and Zhang, Meng and Luo, Qilin and Tao, Zihao and Huang, Hsiao-Yu and Li, Kunhao and Li, Jie and Fu, Junchi and Huang, Di-Jing and Xie, Yanwu and others},
  journal={arXiv:2604.14701},
  year={2026},
  url={https://arxiv.org/abs/2604.14701}
}

@article{kriener2026control,
  title={Controlling the Band Filling and the Band Width in Nickelate Superconductors},
  author={Kriener, M and Terakura, C and Kikkawa, A and Liu, Z and Murayama, H and Nakajima, M and Fujishiro, Y and Sasano, S and Ishikawa, R and Shibata, N and others},
  journal={arXiv:2604.13875},
  year={2026},
  url={https://arxiv.org/abs/2604.13875}
}

@article{bejas2026raman,
  title={Raman response in superconducting multiorbital systems with application to nickelates},
  author={Bejas, Mat{\'\i}as and Zhan, Jun and Wu, Xianxin and Schnyder, Andreas P and Greco, Andr{\'e}s},
  journal={Physical Review B},
  volume={113},
  number={14},
  pages={144504},
  year={2026},
  publisher={APS},
  url={https://link.aps.org/pdf/10.1103/mvc9-3dhq}
}

@article{flavenot2026decoding,
  title={Decoding Superconductivity in La$_3$Ni$_2$O$_{7-delta}$ Thin Films via Ozone-Driven Structure and Oxidation Tuning},
  author={Flavenot, Mathieu and Sahib, Hoshang and Robert, J{\'e}r{\^o}me and Lenertz, Marc and Versini, Gilles and Schlur, Laurent and Gloter, Alexandre and Viart, Nathalie and Preziosi, Daniele},
  journal={arXiv:2604.09807},
  year={2026},
  url={https://arxiv.org/abs/2604.09807}
}

@article{li2026three,
  title={Three-Dimensional Electronic Structures in Superconducting Ruddlesden-Popper Bilayer Nickelate Films},
  author={Li, Yueying and Xu, Lizhi and Lv, Wei and Nie, Zihao and Wang, Zechao and Miao, Yu and Shen, Jianchang and Zhou, Guangdi and Song, Wenhua and Wang, Heng and others},
  journal={arXiv:2604.08430},
  year={2026},
  url={https://arxiv.org/abs/2604.08430}
}

@article{wang2026orbital,
  title={Orbital-Selective $d$-wave Superconductivity in the Two-Band $t$-$J$ Model: Possible Applications to La $_3$Ni$_2$O$_7$},
  author={Wang, Zhan and Jiang, Kun and Zhang, Fu-Chun and Jin, Hui-Ke},
  journal={arXiv:2604.08319},
  year={2026},
  url={https://arxiv.org/abs/2604.08319}
}

@article{stepanov2026co,
  title={Co-operating multiorbital and nonlocal correlations in bilayer nickelate},
  author={Stepanov, Evgeny A and B{\"o}tzel, Steffen and Eremin, Ilya M and Lechermann, Frank},
  journal={arXiv:2604.08221},
  year={2026},
  url={https://arxiv.org/abs/2604.08221}
}

@article{han2026granular,
  title={Granular Superconductivity in La$_2$PrNi$_2$O$_{7-\delta}$ Thin Films},
  author={Han, Ziao and Xiang, Lifen and Zhou, XJ and Zhu, Zhihai},
  journal={arXiv:2604.07807},
  year={2026},
  url={https://arxiv.org/abs/2604.07807}
}

@article{wei2026perp,
  title={Perpendicular electric field induced $s^{\pm}$-wave to $d$-wave superconducting transition in thin film La$_3$Ni$_2$O$_7$},
  author={Wei, Yongping and Liu, Xun and Yang, Fan and Jiang, Mi},
  journal={arXiv:2604.07185},
  year={2026},
  url={https://arxiv.org/abs/2604.07185}
}

@article{cao2026tunable,
  title={Tunable superconductivity and spin density wave in La3Ni2O7/LaAlO3 thin films},
  author={Cao, Yu-Han and Jiang, Kai-Yue and Lu, Hong-Yan and Wang, Da and Wang, Qiang-Hua},
  journal={arXiv:2604.05590},
  year={2026},
  url={https://arxiv.org/abs/2604.05590}
}

@article{wang2026jahn,
  title={Jahn-Teller distortion on strained La$_3$Ni$_2$O$_7$ thin films},
  author={Wang, Yuxin and Wang, Zhan and Zhang, Fu-Chun and Jiang, Kun},
  journal={arXiv:2604.02191},
  year={2026},
  url={https://arxiv.org/abs/2604.02191}
}

@article{chen2026dissecting,
  title={Dissecting superconductivity in the Ruddlesden-Popper nickelates: The role of electron correlation and interlayer magnetic exchange},
  author={Chen, Xiaoyang and Li, Zezhong and Xie, Mei and Hu, Deyuan and Chiu, Yiu-Fung and Agrestini, Stefano and Zhang, Wenliang and Lu, Yi and Wang, Meng and Garcia-Fernandez, Mirian and others},
  journal={arXiv:2604.01902},
  year={2026},
  url={https://arxiv.org/abs/2604.01902}
}

@article{zhan2026detecting,
  title={Detecting pairing symmetry of bilayer nickelates using electronic Raman scattering},
  author={Zhan, Jun and Bejas, Mat{\'\i}as and Schnyder, Andreas P and Greco, Andr{\'e}s and Wu, Xianxin and Hu, Jiangping},
  journal={Chin. Phys. Lett.},
  year={2026},
  volume={43},
  pages={020706},
  url={https://iopscience.iop.org/article/10.1088/0256-307X/43/2/020706/}
}

@article{kumar2026non,
  title={Non-Fermi liquid behavior in La$_3$Ni$_2$O$_7$ thin films under hydrostatic pressure},
  author={Kumar, Deepak and Dans, Jared Z and Avers, Keenan E and Paxson, Ryan and Takeuchi, Ichiro and Paglione, Johnpierre},
  journal={arXiv:2603.26978},
  year={2026},
  url={https://arxiv.org/abs/2603.26978}
}

@article{zhao2026pseudogap,
  title={Pseudogap and Non-Fermi-liquid criticality in double Kondo model for bilayer nickelates},
  author={Zhao, Jing-Yu and Zhang, Ya-Hui},
  journal={arXiv:2603.25742},
  year={2026},
  url={https://arxiv.org/abs/2603.25742}
}

@article{misawa2026polar,
  title={Polar, checkerboard charge order in bilayer nickelate La$_3$Ni$_2$O$_7$},
  author={Misawa, Ryo and Kitou, Shunsuke and Sun, Jian-Ping and Yu, Yingpeng and Koyama, Chihaya and Nakamura, Yuiga and Arima, Taka-hisa and Cheng, Jin-Guang and Hirschberger, Max},
  journal={arXiv:2603.25119},
  year={2026},
  url={https://arxiv.org/abs/2603.25119}
}

@article{wu2026pressure,
  title={Pressure and strain tuning of the alternating bilayer-trilayer Ruddlesden-Popper nickelate: crystal and electronic structure},
  author={Wu, Huan and Zhao, Yi-Feng and Botana, Antia S},
  journal={arXiv:2603.16072},
  year={2026},
  url={https://arxiv.org/abs/2603.16072}
}

@article{watanabe2026hier,
  title={Hierarchical structure of primary and hybridization-induced superconducting correlations in bilayer nickelates},
  author={Watanabe, Hiroshi and Sakakibara, Hirofumi and Kuroki, Kazuhiko},
  journal={arXiv:2603.13604},
  year={2026},
  url={https://arxiv.org/abs/2603.13604}
}

@article{qiu2026progress,
  title={Progress of ambient-pressure superconductivity in bilayer nickelate thin films},
  author={Qiu, Wenyuan and Yao, Dao-Xin},
  journal={arXiv:2603.11235},
  year={2026},
  url={https://arxiv.org/abs/2603.11235}
}

@article{wang2026pauli,
  title={Pauli-limited upper critical field and anisotropic depairing effect of La2. 82Sr0. 18Ni2O7 superconducting thin film},
  author={Wang, Ke and Wang, Maosen and Wei, Wei and Hao, Bo and Liu, Mengqin and Xiang, Qiaochao and Zhou, Xin and Hou, Qiang and Sun, Yue and Zhu, Zengwei and others},
  journal={Materials Today},
  volume={95},
  pages={103269},
  year={2026},
  publisher={Elsevier},
  url={https://www.sciencedirect.com/science/article/pii/S136970212600115X}
}

@article{gao2026enhance,
  title={Enhancement of metallicity by Na doping in La$_3$Ni$_2 $O$_{7+\delta}$},
  author={Gao, Yingying and Zhou, Wei and Guo, WH and Xu, Chunqiang and Chen, HF and Han, ZD and Xu, Xiaofeng and Wu, Yinzhong and Qian, Bin},
  journal={arXiv:2603.08168},
  year={2026},
  url={https://arxiv.org/abs/2603.08168}
}

@article{liu2026triplon,
  title={Triplon-mediated pairing and the superconducting gap structure in bilayer nickelates},
  author={Liu, Huimei and Khaliullin, Giniyat},
  journal={arXiv:2602.23989},
  year={2026},
  url={https://arxiv.org/abs/2602.23989}
}

@article{korolev2026threefold,
  title={Threefold error in the reported zero-field cooled magnetic moment of single crystal La$_2$SmNi$_2$O$_7$},
  author={Korolev, Aleksandr V and Talantsev, Evgeny F},
  journal={arXiv:2602.23240},
  year={2026},
  url={https://arxiv.org/abs/2602.23240}
}

@article{korolev2026nearly,
  title={Nearly twofold overestimation of the superconducting volume fraction in pressurized Ruddlesden-Popper nickelates},
  author={Korolev, Aleksandr V and Talantsev, Evgeny F},
  journal={arXiv:2602.19282},
  year={2026},
  url={https://arxiv.org/abs/2602.19282}
}

@article{lange2026simulating,
  title={Simulating superconductivity in mixed-dimensional $t_{\parallel}$-$J_{\parallel}$-$J_{\perp}$ bilayers with neural quantum states},
  author={Lange, Hannah and Chen, Ao and Georges, Antoine and Grusdt, Fabian and Bohrdt, Annabelle and Roth, Christopher},
  journal={arXiv:2602.10091},
  year={2026},
  url={https://arxiv.org/abs/2602.10091}
}

@article{he2026anisotropic,
  title={Anisotropic Electronic Correlations in the Spin Density Wave State of La$_3$Ni$_2$O$_7$},
  author={He, Ge and Shen, Jun and Xie, Shiyu and Zhang, Haotian and Huo, Mengwu and Shu, Jun and Hu, Deyuan and Zhou, Xiaoxiang and Zhang, Yanmin and Qin, Lei and others},
  journal={arXiv:2602.07998},
  year={2026},
  url={https://arxiv.org/abs/2602.07998}
}

@article{xu2026coex,
  title={Coexistence of Antiferromagnetic Spin Fluctuations and Superconductivity in La$_2$SmNi$_2$O$_7$ Thin Films},
  author={Xu, Minhui and Wang, Yibo and Liu, Jia and Cheng, Long and Li, Shuyin and Yin, Shuaishuai and Zheng, Xu and Yu, Lixin and Zhao, Aidi and Li, Xiaolong and others},
  journal={arXiv:2602.07994},
  year={2026},
  url={https://arxiv.org/abs/2602.07994}
}

@article{gim2026spect,
  title={Spectroscopic Evidence of Competing Diagonal Spin Interactions and Spin Disproportionation in the Bilayer Nickelate La$_3$Ni$_2$O$_7$},
  author={Gim, Dong-Hyeon and Wulferding, Dirk and Zhang, Hengyuan and Wang, Meng and Kim, Kee Hoon},
  journal={arXiv:2602.05365},
  year={2026},
  url={https://arxiv.org/abs/2602.05365}
}

@article{oh2026incom,
  title={Incommensurate pair-density-wave correlations in two-leg ladder $t$-$J$-$J_{\perp}$ model},
  author={Oh, Hanbit and May-Mann, Julian and Zhang, Ya-Hui},
  journal={arXiv:2602.04945},
  year={2026},
  url={https://arxiv.org/abs/2602.04945}
}


\end{document}